\documentclass[prl,
showpacs,aps,nofootinbib,floatfix,amsmath,amssymb]{revtex4}
\usepackage{graphicx}
\usepackage{soul}
\usepackage{color}
\usepackage[usenames,dvipsnames]{xcolor}
\begin{document}

\makeatletter
%Feynman slash
\newbox\slashbox \setbox\slashbox=\hbox{$/$}
\newbox\Slashbox \setbox\Slashbox=\hbox{\large$/$}
\def\pFMslash#1{\setbox\@tempboxa=\hbox{$#1$}
  \@tempdima=0.5\wd\slashbox \advance\@tempdima 0.5\wd\@tempboxa
  \copy\slashbox \kern-\@tempdima \box\@tempboxa}
\def\pFMSlash#1{\setbox\@tempboxa=\hbox{$#1$}
  \@tempdima=0.5\wd\Slashbox \advance\@tempdima 0.5\wd\@tempboxa
  \copy\Slashbox \kern-\@tempdima \box\@tempboxa}
\def\FMslash{\protect\pFMslash}
\def\FMSlash{\protect\pFMSlash}
\def\miss#1{\ifmmode{/\mkern-11mu #1}\else{${/\mkern-11mu #1}$}\fi}
%%%% Uso:  \pFMSlash{p}

%%%%%%%%%%%%%%%%%%%%%%%%%%%%%%%%%%%%%%%%
\newcommand{\psum}[1]{{\sum_{ #1}\!\!\!}'\,}
%%%%%%%%%%%%%%%%%%%%%%%%%%%%%%%%%%%%%%%%%%%%%%%
\makeatother

%\tightenlines
\title{The Standard Model in extra dimensions and its Kaluza-Klein effective Lagrangian}

\author{I. Garc\'ia-Jim\'enez$^{\rm a}$, M. A. L\'opez-Osorio$^{\rm b}$, E. Mart\'inez-Pascual$^{\rm b}$, G. I. N\'apoles-Ca\~nedo$^{\rm c}$, H. Novales-S\'anchez$^{\rm c}$, J. J. Toscano$^{\rm c}$}

\affiliation{$^{\rm a}$Departamento de Ciencias B\' asicas, Instituto Tecnol\' ogico de Oaxaca, Avenida Ing. V\'\i ctor Bravo Ahuja No. 125 Esquina Calzada Tecnol\' ogico, C.P. 68030, Oaxaca de Ju\' arez, Oaxaca, M\' exico.\\$^{\rm b}$Departamento de Ciencias Naturales y Exactas, Centro Universitario de los Valles, Universidad de Guadalajara,
Carretera Guadalajara-Ameca Km. 45.5, CP 46600, Ameca, Jalisco, M\' exico.\\$^{\rm c}$Facultad de Ciencias F\'{\i}sico Matem\'aticas,
Benem\'erita Universidad Aut\'onoma de Puebla, Apartado Postal
1152, Puebla, Puebla, M\'exico.}

\begin{abstract}
An effective theory for the Standard Model with extra dimensions is constructed. We start from a field theory governed by the extra-dimensional Poincar\' e group ${\rm ISO}(1,3+n)$ and by the extended gauge group $G_{\rm SM}({\cal M}^{4+n})=SU_C(3,{\cal M}^{4+n})\times SU_L(2,{\cal M}^{4+n})\times U_Y(1,{\cal M}^{4+n})$, which is characterized by an unknown energy scale $\Lambda$ and is assumed to be valid at energies far below this scale. Assuming that the size of the extra dimensions is much larger than the distance scale at which this theory is valid, an effective theory with symmetry groups ${\rm ISO}(1,3)$ and $G_{\rm SM}({\cal M}^{4})$ is constructed. The transition between such theories is carried out via a canonical transformation that allows us to hide the extended symmetries $\{{\rm ISO}(1,3+n), \, G_{\rm SM}({\cal M}^{4+n})\}$ into the standard symmetries $\{{\rm ISO}(1,3), \, G_{\rm SM}({\cal M}^{4})\}$, and thus endow the Kaluza-Klein gauge fields with mass. Using a set of orthogonal functions $\{f^{(\underline{0})},f^{(\underline{m})}(\bar x)\}$, which is generated by the Casimir invariant $\bar {P}^2$ associated with the translations subgroup $T(n)\subset {\rm ISO}(n)$, the degrees of freedom of $\{{\rm ISO}(1,3+n), G_{\rm SM}({\cal M}^{4+n})\}$ are expanded via a general Fourier series, whose coefficients are the degrees of freedom of $\{{\rm  ISO}(1,3), G({\cal M}^{4})\}$. It is shown that these functions, which correspond to the projection on the coordinates basis $\{|\bar{x} \big >\}$ of the discrete basis $\{|0\big >,|p^{(\underline{m})}\big >\}$ generated by $\bar {P}^2$, play a central role in defining the effective theory. It is shown that those components along the ground state $f^{(\underline{0})}=\big <\bar x|0\big>$ do not receive mass at the compactification scale, so they are identified with the Standard Model fields; but components along excited states $f^{(\underline{m})}=\big <\bar x|p^{(\underline{m})}\big>$ do receive mass at this scale, so they correspond to Kaluza-Klein excitations. In particular, it is shown that associated with any direction $|p^{(\underline{m})}\neq0\big >$ there are a massive gauge field and a pseudo-Goldstone boson. Some resemblances of this mass-generating mechanism with the Englert-Higgs mechanism are stressed and some physical implications are discussed. We perform a comprehensive study of the couplings in all sectors of the effective theory, which includes a full catalog of Lagrangian terms that can be used to calculate Feynman rules.
\end{abstract}


\maketitle

\section{Introduction}
\label{I}Candidates to fundamental theories at the Planck scale, which encompass models that are not theories of fields or particles in Wigner's sense~\cite{Wigner,Scadron}, are commonly formulated in more than four dimensions in order to be self-consistent. The emblematic case is superstring theory~\cite{GS,P,W}. The interest in phenomenological effects produced by new physics involving extra dimensions arose after the works by Antoniadis, Arkani-Hamed, Dimopoulos and Dvali~\cite{A,ADD,AADD}, who argued that relatively large extra dimensions could be detected at the TeV scale. If such large extra dimensions exist, they might affect the dynamics of  known particles. In order to investigate the consequences of extra dimensions at our four-dimensional realm, in this paper we present an extension of the Standard Model (SM) based on the well-known universal extra dimensions (UEDs) approach~\cite{ACD1}. The phenomenological implications of theories in the presence of extra dimensions have been exploited  in various contexts such as dark matter~\cite{UEDDM1,UEDDM2,UEDDM3,UEDDM4,UEDDM5,UEDDM6,UEDDM7,UEDDM8,UEDDM9,UEDDM10,UEDDM11,UEDDM12,UEDDM13,UEDDM14,UEDDM15,UEDDM16,UEDDM17}, neutrino physics~\cite{UEDn1,UEDn2}, Higgs physics~\cite{UEDH1,UEDH2,UEDH3,UEDH4,UEDH5,UEDH6,UEDH7,UEDH8,UEDH9}, flavor physics~\cite{UEDf1,UEDf2,UEDf3,UEDf4,UEDf5}, Hadronic and linear colliders~\cite{UEDc1,UEDc2,UEDc3,UEDc4,UEDc5,UEDc6,UEDc7,UEDc8,UEDc9,UEDc10,UEDc11,UEDc12,UEDc13,UEDc14,UEDc15,UEDc16,UEDc17,UEDc18,UEDc19,UEDc20,UEDc21,UEDc22,UEDc23,UEDc24}, and electroweak gauge couplings~\cite{OPP1,OPP2}; also, theoretical aspects of field theories with extra dimensions have been studied by several authors~\cite{TAED1,TAED2,TAED3,TAED4,TAED5,TAED6,TAED7,TAED8,TAED9,TAED10,TAED11,TAED12,TAED13,TAED14,TAED15}.\\

At sufficiently-low energy, effects of fundamental theories can be described by effective field theories. In this work, we are interested in deriving a SM extension to extra dimensions. Our approach will be based on the notion of {\it hidden symmetry}. The concept of hidden symmetry is a powerful instrument that allows to elegantly describe some subtle and complex scenarios of fundamental physics. Hidden symmetry and spontaneous symmetry breaking (SSB)~\cite{SSB1,SSB2,SSB3,SSB4} are the cornerstone of the SM of particle physics. The merger of both ideas is the very essence of the Englert-Higgs mechanism (EHM)~\cite{EHM1,EHM2,EHM3}. In recent communications~\cite{NoTo,GGNNT} by some of us, the symmetric structure of Yang-Mills theories with extra dimensions has been studied and its resemblance with the EHM stressed. In this work, we follow this approach closely to lay the foundations of a SM extension to extra flat dimensions, which we refer to by the acronym EDSM, for ``extra-dimensional Standard Model''. We consider, as a starting point, an action for a field theory defined on a flat-spacetime manifold of dimension $d=4+n$: ${\mathcal M}^{d}={\mathcal M}^4\times {\mathcal N}^n$, which is the result of the cartesian product of the four-dimensional Minkowski spacetime ${\cal M}^4$ and some $ n $-dimensional noncompact manifold ${\cal N}^n$ that represents a spatial extension. We assume this higher-dimensional theory to be invariant under the Poincar\'e group ${\rm ISO}(1,3+n)$ and the gauge group $G_{\rm SM}({\mathcal M}^{d})\equiv SU_C(3,{\mathcal M}^{d})\times SU_L(2,{\mathcal M}^{d})\times U_Y(1,{\mathcal M}^{d})$, which is a gauge group with all of its group parameters defined on $ \mathcal{M}^{d} $. Since this theory is not renormalizable in the Dyson's sense, it is given by a Lagrangian that includes an infinite number of $\{{\rm ISO}(1,3+n),\, G_{\rm SM}({\cal M}^d)\}$-invariant terms of all possible canonical dimensions. The lowest-dimensional term corresponds to a direct $(4+n)$-dimensional extension of the standard $4$-dimensional theory, while those terms of greater and increasing dimension are suppressed by inverse powers of an unknown energy scale $\Lambda$, which is assumed to be far above the Fermi scale. We assume that the average size of the extra dimensions, $R$, is so large compared with the distance scale at which this theory is valid that it actually can be considered as infinite. At energies far above the compactification scale $R^{-1}$, this theory is governed by the extended groups $\{{\rm ISO}(1,3+n)\,,  G_{\rm SM}({\cal M}^d)\}$. To describe the physical phenomena at much-smaller energies, where the compactness of the extra dimensions becomes apparent, we need to hide the $\{{\rm ISO}(1,3+n),\, G_{\rm SM}({\cal M}^d)\}$ symmetry into $\{{\rm ISO}(1,3),\, G_{\rm SM}({\cal M}^4)\}$. Observe that $G_{\rm SM}({\cal M}^d)$ and $G_{\rm SM}({\cal M}^4)$ coincide as Lie groups, but they differ as gauge groups. It should be noted that the process of hiding a symmetry does not mean moving from one theory to another, but rather focusing on the same theory from another perspective. This means that we must pass from the description based on $\{{\rm ISO}(1,3+n),\, G_{\rm SM}({\cal M}^d)\}$ to that characterized by $\{{\rm ISO}(1,3),\, G_{\rm SM}({\cal M}^4)\}$ through a canonical transformation. As it occurs in theories with SSB, the physical content is a matter of scales. In the SM one uses the groups $SU_C(3,{\cal M}^4)\times U_e(1,{\cal M}^4)$ to describe physical phenomena at energies of the order of the Fermi scale $v$, but at energies far above $v$ the $SU_C(3,{\cal M}^4)\times SU_L(2,{\cal M}^4)\times U_e(1,{\cal M}^4)$ description must be used. In our case, at energies of the order of the compactification scale $R^{-1}$, we use the $\{ {\rm ISO}(1,3), \, G_{\rm SM}({\cal M}^4)\}$ description. However, at energies far above the $R^{-1}$ scale, we use the $\{ {\rm ISO}(1,3+n), \, G_{\rm SM}({\cal M}^d)\}$ description, since at these energies we are exploring distances so small that the compact dimensions would really look infinite.\\

The compactification program comprises a number of nontrivial steps. First, one must define a canonical transformation that maps covariant objets of $\{ {\rm ISO}(1,3+n), \, G_{\rm SM}({\cal M}^d)\}$ into covariant objets of $\{ {\rm ISO}(1,3), \, G_{\rm SM}({\cal M}^4)\}$. This transformation is crucial to hide the extended symmetry into the standard one. Second, since the number of connections of $G_{\rm SM}({\cal M}^4)$ is smaller than that of $G_{\rm SM}({\cal M}^d)$, the difference will appear as tensorial representations of $G_{\rm SM}({\cal M}^4)$. So from the perspective of the SM gauge group, these connections can be endowed with mass. Indeed, it is necessary to endow such connections with mass at the compactification scale $R^{-1}$ because these new-physics effects must be of decoupling nature in accordance with the Appelquist-Carazzone's decoupling theorem~\cite{AC}. This means that some instrument analogous to the SSB must be introduced in order to generate such masses. Any new particle that emerges as a consequence of the process of hiding the symmetry must be endowed with mass through that instrument. The passing from the extended symmetry to the standard one should not spoil the gauge structure of the theory, which means that we must be able to examine the physical phenomenon from both the $\{ {\rm ISO}(1,3), \, G_{\rm SM}({\cal M}^4)\}$ and  $\{ {\rm ISO}(1,3+n), \, G_{\rm SM}({\cal M}^d)\}$ points of view. As it occurs in theories with SSB, the hiding of the gauge symmetry manifests itself through the presence of two types of gauge transformations: the standard gauge transformations (SGTs) of the group $G_{\rm SM}({\cal M}^4)$ and a set of nonstandard gauge transformations (NSGTs). The NSGTs are determined by gauge parameters which do not belong to the $G_{\rm SM}({\cal M}^4)$ gauge group (see Refs.~\cite{OPT1,OPT2,OPT3}).\\

A nontrivial consequence of hiding the extended symmetry into the standard one is the presence of an infinite number of basic fields produced by the canonical map. These fields correspond to the so-called Kaluza-Klein (KK) towers; some of these towers involve a field that can be identified with a SM field on $ \mathcal{M}^{4} $ because its mass is not affected by the compactification mechanism, while all the remaining KK fields receive mass through this mechanism. This aspect is key to ensure the decoupling of the new-physics effects when the compactification scale is much higher than the Fermi scale.  Besides the well-known interactions among the SM fields, new  couplings between the SM fields and the excited KK modes, as well as couplings exclusively involving excited KK modes, arise in this model. The additional terms bring new phenomena in four dimensions induced by the presence of the compact spatial extra dimensions, which motivates us to present a comprehensive  study of the different sectors of the model.\\

As commented, our approach is based on the notion of hidden symmetry. Since this concept is central to the SM, one of our main purposes is to establish a parallelism of the Kaluza-Klein mass generating mechanism with the EHM mechanism, highlighting both coincidences and discrepancies. An important goal is to establish the sequence of hiding symmetries:
\begin{eqnarray}
\label{HS}
\{{\rm ISO}(1,3+n),\, G({\cal M}^d)\} && \xrightarrow{R^{-1}} \{{\rm ISO}(1,3),\, G({\cal M}^4)\} \nonumber \\
 &&\xrightarrow{v} \{{\rm ISO}(1,3),\, SU_C(3,{\cal M}^4)\times U_e(1,{\cal M}^4)\}\, ,
\end{eqnarray}
where the last step corresponds to the EHM. We will see that this step introduces nontrivial new elements at the Fermi scale.

The rest of the paper has been organized as follows. In Sec. \ref{P}, some general features of theories with extra dimensions, which are needed for subsequent developments, are discussed. In Sec. \ref{BS}, we present explicit expressions for the Yang-Mills sector and the Higgs sector of the EDSM. Sec. \ref{FS} is devoted to study the Yukawa sector and the currents sector of the theory. In Sec. \ref{C}, a summary of our results is presented. In~\ref{AA}, we introduce a representation of the higher-dimensional Dirac matrices that is particularly suitable for our calculations; the covariant terms that are present in the currents sector can be found in~\ref{AB}; finally, in~\ref{AC} and~\ref{AD} we gather the different scalar charged currents associated to the $ W $ gauge boson and the scalar neutral currents associated to the $ Z$ gauge boson.

\section{Preliminaries}
\label{P}The starting point is an effective gauge field theory that is governed by the extended groups $\{{\rm ISO}(1,3+n), \,G_{\rm SM}({\mathcal M}^{d})\}$, and whose gauge parameters are defined all over the spacetime $\mathcal{M}^{d}=\mathcal{M}^{4}\times\mathcal{N}^{n} $. The action of the theory is assumed to be a functional on gauge and matter fields, collectively denoted by $\Phi_{A}(x,\bar{x})$ at spacetime points $ (x,\bar{x})\in\mathcal{M}^{d} $. These basic fields of the theory furnish a representation of the Lorentz group $ {\rm SO}(1,3+n) $. Since the theory is not renormalizable in the Dyson's sense, the corresponding action consists of an infinite series of Lorentz and gauge invariant terms of growing canonical dimension, that is,
\begin{equation}\label{AG}
S=\int d^{\,4}x\,d^{\,n}\bar{x} \,\, {\cal L}_{4+n}\left(\Phi_{A}(x,\bar{x}),\partial_M \Phi_{A}(x,\bar{x}) \right)\, ,
\end{equation}
where
\begin{eqnarray}
\label{LG}
{\cal L}_{4+n}&=&{\cal L}_{(4+n)}^{\rm SM}\left(\Phi_{A}(x,\bar{x}),\partial_M \Phi_{A}(x,\bar{x}) \right)\nonumber \\
&&+\sum_{\textbf{d}} \frac{\lambda_{\textbf{d}}}{\Lambda^{\textbf{d}}}\,{\cal L}^{(\textbf{d}+d)}\left(\Phi_{A}(x,\bar{x}),\partial_M \Phi_{A}(x,\bar{x}) \right)\, .
\end{eqnarray}
Here, ${\cal L}^{(\textbf{d}+d)}$ represents gauge- and Lorentz-invariant interactions of canonical dimensions greater than $d$, formulated from fields $ \Phi_{A} $ and its spacetime derivatives $ \partial_{M}\Phi_{A} $, multiplied by unknown coupling constants $ \lambda_{\textbf{d}}/\Lambda^{\textbf{d}} $. It is worth to mention that the fields $ \Phi_{A} $ describe particles of at most spin 1; although in an up-to-bottom approach this expression should incorporate higher-spin particle content, we will not consider such cases here. The first term of the right-hand side of Eq.~\eqref{LG} corresponds to a straightforward extension of the functional structure of the well-known SM Lagrangian from four dimensions to $ d $ dimensions. The terms of increasing dimension are suppressed by inverse powers of $\Lambda$, which is assumed to be far above the Fermi scale. In fact, according to the effective Lagrangian approach, this theory is valid only for energies less than $\Lambda$. We assume that the size of the extra dimensions is so large compared with this energies that it actually can be considered as infinite. This justifies the ${\rm ISO}(1,3+n)$ description of the effective theory~\eqref{LG}. Note that the first term in Eq.~\eqref{LG} does not depend on the $\Lambda$ scale, although it does depend on dimensionful coupling constants. As we see below, this term plays a central role in the $\{{\rm ISO}(1,3), \,G_{\rm SM}({\mathcal M}^{4})\}$ description.\\

Our main goal is to construct a SM extension to extra dimensions. To do this, we need to define canonical maps that allow us to descend towards low-energy regimes in accordance with the pattern of hiding symmetries that we schematized in Eq.~(\ref{HS}). To carry out this program, we stress that $ {\rm SO}(1,3) $ and $ {\rm SO}(n) $ are subgroups of $ {\rm SO}(1,3+n) $. We also have the following remarks, which are useful for the subsequent study: in the case where $ \Phi_{A}(x,\bar{x}) $ represents a scalar field $ \Phi(x,\bar{x}) $, it is also a scalar with respect to $ {\rm SO}(1,3) $ and $ {\rm SO}(n) $; in the case in which $ \Phi_{A}(x,\bar{x})$ represents an $ {\rm SO}(1,3+n) $ vector field, which could be renamed as $  \mathcal{A}_{M}(x,\bar{x})  $ ($ M=0,1,\ldots,3;\,5,\ldots,d\equiv \mu; \,\bar{\mu} $), it can be seen as an $ {\rm SO}(1,3) $-vector field, with components $ \mathcal{A}_{\mu} $, and  $ n $ $ {\rm SO}(1,3) $-scalars denoted by\footnote{We will refer to these scalars as the {\it scalars associated to the vector field} $ \mathcal{A}_{\mu} $.} $ \mathcal{A}_{\bar{\mu}} $; under the ${\rm SO}(n)$ group, the four components of $ \mathcal{A}_{\mu} $ are scalar fields, whereas $ \mathcal{A}_{\bar{\mu}} $ can be seen as a vector. Letting $ d $ be even, the $ 2^{d/2} =2^{\frac{4+n}{2}}$ components of a spinor $ \Psi(x,\bar{x}) $ of $ {\rm SO}(1,3+n) $ can be unfolded into $ 2^{n/2} $ (four-component) spinors of $ {\rm SO}(1,3) $ and $2^{d/2}$ scalars of $ {\rm SO}(n) $. Alternatively, the spinor $ \Psi(x,\bar{x})$ can be unfolded into $4$ ($2^{n/2}$-component) spinors of $ {\rm SO}(n) $ and $2^{d/2}$ scalars of $ {\rm SO}(1,3) $. In the spinorial case, we will adopt the first option, since we are interested in constructing an effective theory at the $R^{-1}$ scale governed by the standard Lorentz group $ {\rm SO}(1,3) $. \\

In the next sections, we implement the procedure outlined in Refs.~\cite{NoTo,GGNNT} in order to obtain the $\{ {\rm SO}(1,3), \, G_{\rm SM}(\mathcal{M}^{4}) \}$-invariant field theory,  $ \mathcal{L}_{\!\textrm{eff}}^{\textbf{d}=4} $, for the whole SM. This work complements the results given in these references on the extra-dimensional Yang-Mills sector with the corresponding Higgs and fermionic sectors. We also point out the phenomenologically relevant aspects involved in $ \mathcal{L}_{\!\textrm{eff}}^{\textbf{d}=4} $.

According to Ref.~\cite{GGNNT}, the effective Lagrangian with terms of canonical dimension $ \textbf{d}=4 $ is derived from the straightforward extension of the SM as follows:
\begin{equation}\label{4DSM}
{\mathcal{L}}_{\rm eff}^{\textbf{d}=4}:=\int d^n\bar{x}\,{\cal L}^{\rm SM}_{(4+n)}\, ,
\end{equation}
where
\begin{equation}
{\cal L}^{\rm SM}_{(4+n)}={\cal L}^{\rm YM}_{(4+n)}+{\cal L}^{\rm H}_{(4+n)}+{\cal L}^{\rm C}_{(4+n)}+{\cal L}^{\rm Y}_{(4+n)}\, ,
\end{equation}
with ${\cal L}^{\rm YM}_{(4+n)}$, ${\cal L}^{\rm H}_{(4+n)}$, ${\cal L}^{\rm C}_{(4+n)}$, and ${\cal L}^{\rm Y}_{(4+n)}$ respectively representing direct $(4+n)$-dimensional generalizations of the Yang-Mills, Higgs, Currents, and Yukawa sectors as we know them in the SM. In what follows we present a comprehensive discussion of each of these sectors and their corresponding compactification.

\section{The bosonic sector}
\label{BS}
In Refs.\cite{NoTo,GGNNT}, a thorough discussion of Yang-Mills theories in more than four dimensions and their subsequent compactification was presented. Here, we introduce notation and conventions, and present only the relevant results. The Yang-Mills sector defined on $ \mathcal{M}^{d} $ is given by
\begin{eqnarray}
\label{ymD}
{\cal L}^{\rm YM}_{(4+n)}&=&-\frac{1}{4}{\cal G}^a_{MN}(x,\bar{x}){\cal G}^{MN}_a(x,\bar{x})-\frac{1}{4}{\cal W}^i_{MN}(x,\bar{x}){\cal W}^{MN}_i(x,\bar{x})\nonumber \\
&&-\frac{1}{4}{\cal B}_{MN}(x,\bar{x}){\cal B}^{MN}(x,\bar{x})\, ,
\end{eqnarray}
where ${\cal G}^a_{MN}(x,\bar{x})$, ${\cal W}^i_{MN}(x,\bar{x})$, and ${\cal B}_{MN}(x,\bar{x})$ are the components of the curvatures valued on the Lie algebras $su_C(3,{\cal M}^d)$, $su_L(2,{\cal M}^d)$, and $u_Y(1,{\cal M}^{d})$, respectively. In terms of the corresponding connection components, they are given by
\begin{subequations}\label{EDCurv}
\begin{align}
{\cal G}^a_{MN}& =\partial_M {\cal G}^a_N-\partial_N {\cal G}^a_M+g_{s\, 4+n}f^{abc}{\cal G}^b_M{\cal G}^c_N\, ,\quad a,b,c=1,\ldots,8\\
{\cal W}^i_{MN}& =\partial_M {\cal W}^i_N-\partial_N {\cal W}^i_M+g_{4+n}\epsilon^{ijk}{\cal W}^j_M{\cal W}^k_N \, ,\quad i,j,k=1,2,3\\
{\cal B}_{MN}&=\partial_M {\cal B}_N-\partial_N {\cal B}_M \, .
\end{align}
\end{subequations}
where $ g_{s\, 4+n} $ and $ g_{4+n} $ are dimensionful coupling constants, while $ f^{abc} $ and $ \epsilon^{ijk} $ are the structure constants that define the Lie algebras $ su(3) $ and $ su(2) $, respectively. The Lagrangian (\ref{ymD}) is invariant under $\{{\rm ISO}(1,3+n), \, G_{\rm SM}({\cal M}^{d})\}$.

Hence after introducing the Fourier decompositions of the curvature components into the Yang-Mills sector at the right-hand side of Eq.~\eqref{ymD}, and integrating out $ \mathcal{N}^{n} $ using the orthogonality properties of $ f^{(\underline{m})}_{E,O} $ (see Appendix B of Ref.~\cite{GGNNT}), we obtain a four-dimensional Yang-Mills Lagrangian that is conveniently decomposed into the quantum-chromodynamics (QCD) and electroweak (EW) sectors, that is,
\begin{equation}
{\cal L}^{\rm YM}_{\rm eff}={\cal L}^{\rm YM\, QCD}_{\rm eff}+{\cal L}^{\rm YM\, EW}_{\rm eff}\, .
\end{equation}
At this level, both Lagrangians can be analyzed in a similar fashion. However, due to the presence of the Higgs sector and its couplings with EW gauge fields, via covariant derivatives, we have decided to separately deal with $ {\cal L}^{\rm YM\, QCD}_{\rm eff} $ and $ {\cal L}^{\rm YM\, EW}_{\rm eff} $.

\subsection{The QCD sector: Gluons}
\label{QCD}
In the notation used for the discussion of extra-dimensional Yang-Mills theories, ${\cal A}^a_M(x,\overline{x})$, written in caligraphic letters, represented extra-dimensional gauge fields, while fields $A^{(\underline{0})a}_\mu$, $A^{(\underline{m})a}_\mu$, and $A^{(\underline{m})a}_{\bar \mu}$, written in italic letters, denoted KK modes that emerged from the extra-dimensional connection. As Eq.~(\ref{ymD}) suggests, we will follow this notation for the rest of the paper. With this in mind, let us comment that the QCD sector ${\cal L}_{\rm eff}^{\rm YM\,QCD}$ contains, after the identification of $ G^{(\underline{0})a}_{\mu} $ with gluon fields of the SM, the well-known four-dimensional SM Yang-Mills-QCD sector plus other terms that contain excited KK modes. Following Ref.~\cite{GGNNT}, we divide $ \mathcal{L}_{\textrm{eff}}^{\textrm{YM QCD}} $ into three different sectors depending on the kind of local interaction, namely
\begin{equation}
\mathcal{L}_{\textrm{eff}}^{\textrm{YM QCD}}=\mathcal{L}_{\textrm{v-v}}^{\textrm{YM QCD}}+\mathcal{L}_{\textrm{v-s}}^{\textrm{YM QCD}}+\mathcal{L}_{\textrm{s-s}}^{\textrm{YM QCD}}\ .
\end{equation}
Using the $ su(3)$-valued fields $ G^{(\underline{m})}_{\bar{\mu}}=G^{(\underline{m})a}_{\bar{\mu}}T^{a} $, the commutator symbol $ [\cdot\, ,\cdot] $, and the normalization convention Tr$(T^{a}T^{b})=\delta^{ab}/2  $, we can write the pure-vector sector ${\cal L}^{\rm YM\,QCD}_{\textrm{v-v}}$ as
\begin{equation}\label{LYMQCDvv}
\mathcal{L}^{\textrm{YM QCD}}_{\textrm{v-v}}:=-\frac{1}{2}\textrm{Tr}\Big\lbrace \mathcal{G}^{(\underline{0})}_{\mu \nu}
\mathcal{G}^{(\underline{0})\mu \nu}+\sum_{(\underline{m})}\mathcal{G}^{(\underline{m})}_{\mu \nu}
\mathcal{G}^{(\underline{m})\mu \nu}\Big\rbrace\ ,
\end{equation}
where the following $ su(3) $-valued objects are defined:
\begin{align}
{\cal G}^{(\underline{0})}_{\mu \nu} & =G^{(\underline{0})}_{\mu \nu}-ig_s\sum_{(\underline{m})}\left[G^{(\underline{m})}_\mu, \, G^{(\underline{m})}_\nu\right]\  ,\\
{\cal G}^{(\underline{m})}_{\mu \nu} & =\mathcal{D}^{(\underline{0})}_{\mu}G^{(\underline{m})}_\nu
- \mathcal{D}^{(\underline{0})}_{\nu}G^{(\underline{m})}_\mu
-ig_s\sum_{(\underline{r}\underline{s})} \Delta_{(\underline{mrs})}\left[G^{(\underline{r})}_\mu, \, G^{(\underline{s})}_\nu\right]\, .
\end{align}
Recall that we have identified $ G^{(\underline{0})a}_{\mu} $ with the SM gluon fields, so note that $ \mathcal{L}_{\textrm{v-v}}^{\textrm{YM QCD}} $ contains the usual $ {\rm SU}(3) $ kinematic Yang-Mills term. Besides this term, we also encounter the kinematical term associated to $ G^{(\underline{m})}_{\mu} $, interactions between the SM gluon fields and its excited KK modes, and interactions only among excited KK modes. As already noted in Ref.~\cite{GGNNT}, the Lagrangian given in Eq.~\eqref{LYMQCDvv} is not affected by the rotation $ \mathcal{R} $, which is introduced in order to determine the physical  scalar fields and pseudo-Goldstone bosons of this sector (see Ref.~\cite{GGNNT}).

From the $\mathcal{L}_{\textrm{v-s}}^{\textrm{YM QCD}}$ Lagrangian, the masses corresponding to the vectorial KK excitations $ G^{(\underline{m})a}_{\mu}$ emerge as a result of the Kaluza-Klein mass-generating mechanism~\cite{GGNNT}. From the same lagrangian term, the interactions of these KK vector excitations with scalar modes $ G^{(\underline{m})a}_{\bar{\mu}} $ are determined as well. Notice that $ \mathcal{G}^{(\underline{m})}_{\mu\bar{\nu}}\equiv G^{(\underline{m})a}_{\mu\bar{\nu}}T^{a} $ does not behave as a vector under the rotation $ \mathcal{R} $; in fact,
\begin{equation}\label{Gmbn-trans}
\mathcal{G}^{(\underline{m})}_{\mu\bar{\nu}}\to\mathcal{R}^{(\underline{m})}_{\bar{\nu}\bar{\nu}'}
\mathcal{D}^{(\underline{0})}_{\mu}
G^{(\underline{m})}_{\bar{\nu}'}-ig_{s}\sum_{(\underline{kr})}
\Delta'_{(\underline{krm})}\mathcal{R}^{(\underline{r})}_{\bar{\nu}\bar{\nu}'}\left[G_{\mu}^{(\underline{k})},
G^{(\underline{r})}_{\bar{\nu}'}\right]+p_{\bar{\nu}}^{(\underline m)}G^{(\underline{m})}_{\mu}\ .
\end{equation}
Then,
\begin{eqnarray}
\label{gs5}
\mathcal{L}_{\textrm{s-v}}^{\textrm{YM QCD}}&:= & -\frac{1}{2}\sum_{(\underline{m})}{\cal G}^{(\underline{m})a}_{\mu\bar{\nu}}{\cal G}^{(\underline{m})\mu\bar{\nu}}_{a}\nonumber\\
&=  &  \sum_{(\underline{m})}\textrm{Tr}\Big\{ \left(\mathcal{D}^{(\underline{0})}_{\mu}G^{(\underline{m})}_{\bar{\nu}'}\right)
\left(\mathcal{D}^{(\underline{0})\mu}G^{(\underline{m})}_{\bar{\nu}'}\right)
\nonumber \\
&&+2p_{\bar{\nu}}^{(\underline m)}  \mathcal{R}^{(\underline{m})}_{\bar{\nu}\bar{\nu}'}G^{(\underline{m})}_{\mu}
\mathcal{D}^{(\underline{0})\mu}G^{(\underline{m})}_{\bar{\nu}'}+m_{(\underline{m})}^{2}G^{(\underline{m})}_{\mu}
G^{(\underline{m})\mu}\nonumber\\
& & -2ig_{s}\sum_{(\underline{kr})}
\Delta'_{(\underline{krm})}\mathcal{R}^{(\underline{r})}_{\bar{\nu}\bar{\rho}'}\left[G_{\mu}^{(\underline{k})},
G^{(\underline{r})}_{\bar{\rho}'}\right]\left(\mathcal{R}^{(\underline{m})}_{\bar{\nu}\bar{\nu}'}\mathcal{D}^{(\underline{0})\mu}
G^{(\underline{m})}_{\bar{\nu}'}+p_{\bar{\nu}}^{(\underline m)} G^{(\underline{m})\mu}\right) \nonumber\\
& & -g_{s}^{2}\sum_{(\underline{rspq})}\Delta'_{(\underline{rsm})}\Delta'_{(\underline{pqm})}
\mathcal{R}^{(\underline{s})}_{\bar{\nu}\bar{\nu}'}
\mathcal{R}^{(\underline{q})}_{\bar{\nu}\bar{\rho}'}
\left[G^{(\underline{r})\mu},G^{(\underline{s})}_{\bar{\nu}'}\right]
\left[G_{\mu}^{(\underline{p})},G^{(\underline{q})}_{\bar{\rho}'}\right] \Big\} \ .
\end{eqnarray}
In this expression we identify the kinematical term of each scalar $ G^{(\underline{m})}_{\bar{\nu}'} $, the constant  $ m_{(\underline{m})}^{2} $ as the squared mass of the corresponding KK excited gauge fields $ G^{(\underline{m})}_{\mu}=G^{(\underline{m})a}_{\mu} T^{a}$, as well as various interaction terms among the SM QCD gauge field $ G^{(\underline{0})}_{\mu} $, its KK excitations $ G^{(\underline{m})}_{\mu} $, and the KK excited scalars $ G^{(\underline{m})}_{\bar{\mu}'} $. As we already commented in~\cite{NoTo,GGNNT}, it should be noted the striking resemblance of this sector with a typical Higgs kinetic term. In particular, it is remarkable that the second term can be reduced to a direct coupling between gauge excited vector fields $ G^{(\underline{m})}_{\mu} $ and massless scalar fields $ G^{(m)}_{G} $, just as it occurs in gauge theories with spontaneous symmetry breaking. Although this term could be treated as a vertex in pertubation theory, there is no need to do so in the gauge $ G^{(\underline{m})}_{G}=0 $ defined in~\cite{GGNNT}. This is what we called the unitary gauge, due to its similarity with the gauge used in the SM in which all pseudo-Goldstone bosons are set to zero. In such a gauge one has $ G^{(\underline{m})}_{\bar{\mu}} = \mathcal{R}^{(\underline{m})}_{\bar{\mu}\bar{\mu}'}G_{\bar{\mu}'}^{(\underline{m})}= \mathcal{R}^{(\underline{m})}_{\bar{\mu}\bar{n}}G_{\bar{n}}^{(\underline{m})}$.

As far as $\mathcal{L}_{\textrm{s-s}}^{\textrm{YM QCD}}$ is concerned, it can be written as follows:
\begin{eqnarray}
\label{ScaMassLPrime}
 \mathcal{L}_{\textrm{s-s}}^{\textrm{YM QCD}} &=&  -\sum_{(\underline{m})}\textrm{Tr}\Big\{ m_{(\underline{m})}^{2}G_{\bar{n}}^{(\underline{m})}G_{\bar{n}}^{(\underline{m})}\nonumber \\
 &&-2ig_{s} p_{\bar{\mu}}^{(\underline m)} \mathcal{R}^{(\underline{m})}_{\bar{\nu}\bar{\nu}'}G^{(\underline{m})}_{\bar{\nu}'}
\sum_{(\underline{rs})}\Delta'_{(\underline{rsm})}\mathcal{R}^{(\underline{r})}_{\bar{\mu}\bar{\mu}'}
\mathcal{R}^{(\underline{s})}_{\bar{\nu}\bar{\nu}'}\left[G_{\bar{\mu}'}
^{(\underline{r})},G_{\bar{\nu}'}^{(\underline{s})}\right]\nonumber\\
&&-\dfrac{1}{2}g_{s}^{2}\Big( \sum_{(\underline{rspq})}\Delta'_{(\underline{rsm})}\Delta'_{(\underline{pqm})}\mathcal{R}^{(\underline{r})}_{\bar{\mu}
\bar{\mu}'}\mathcal{R}^{(\underline{s})}_{\bar{\nu}\bar{\nu}'}\mathcal{R}^{(\underline{p})}_{\bar{\mu}
\bar{\rho}'}\mathcal{R}^{(\underline{q})}_{\bar{\nu}\bar{\sigma}'}
\left[G_{\bar{\mu}'}^{(\underline{r})},G_{\bar{\nu}'}^{(\underline{s})}\right]
\left[G_{\bar{\rho}'}^{(\underline{p})},G_{\bar{\sigma}'}^{(\underline{q})}\right]\nonumber\\
&& +\sum_{(\underline{m})}\mathcal{R}^{(\underline{p})}_{\bar{\mu}
\bar{\mu}'}\mathcal{R}^{(\underline{m})}_{\bar{\nu}\bar{\nu}'}\mathcal{R}^{(\underline{r})}_{\bar{\mu}
\bar{\rho}'}\mathcal{R}^{(\underline{p})}_{\bar{\nu}\bar{\sigma}'}
\left[G_{\bar{\mu}'}^{(\underline{m})},G_{\bar{\nu}'}^{(\underline{m})}\right]
\left[G_{\bar{\rho}'}^{(\underline{r})},G_{\bar{\sigma}'}^{(\underline{r})}\right]
\Big)\Big\}\, .
\end{eqnarray}
From this Lagrangian we immediately recognize the massive fields $ G^{(\underline{m})}_{\bar{n}} $, with mass $ m_{(\underline{m})} $, and various trilinear and quartic interaction terms among them. As shown in~\cite{GGNNT}, a proper fixing of the gauge symmetry induced by the NSGTs allows $ G^{(\underline{m})}_{G}=0 $.
\\

\subsection{The electroweak sector: Bosons}
The presence of a Higgs doublet and its minimal coupling to electroweak gauge fields, together with the occurrence of a genuine EHM at the Fermi scale, renders the mass spectrum of the excited KK scalars $ W^{(\underline{m})i}_{\bar{\mu}} $ and $ B^{(\underline{m})}_{\bar{\mu}} $ rather different from that of a pure Yang-Mills theory.

The effective Yang-Mills electroweak sector is given by
\begin{align}
\label{LYMEW}
{\mathcal{L}}^{\textrm{YM EW}}_{\textrm{eff}}  =&  -\frac{1}{4}\Big[{\mathcal{W}}^{(\underline{0})i}_{\mu \nu}{\mathcal
{W}}^{(\underline{0})\mu \nu}_i+{\mathcal{W}}^{(\underline{0})i}_{\bar{\mu} \bar{\nu}}{\mathcal{W}}^{(\underline{0})\bar{\mu}
\bar{\nu}}_i\nonumber \\
&+ \sum_{(\underline{m})}\left({\mathcal{W}}^{(\underline{m})i}_{\mu \nu}{\mathcal{W}}^{(\underline{m})\mu
\nu}_i+2{\mathcal{W}}^{(\underline{m})i}_{\mu \bar{\nu}}{\cal W}^{(\underline{m})\mu
\bar{\nu}}_i+{\mathcal{W}}^{(\underline{m})i}_{\bar{\mu} \bar{\nu}}{\mathcal{W}}^{(\underline{m})\bar{\mu}
\bar{\nu}}_i\right)\Big]\nonumber \\
& -\frac{1}{4}\Big[\mathcal{B}^{(\underline{0})}_{\mu \nu}\mathcal{B}^{(\underline{0})\,\mu \nu}\nonumber \\
&+ \sum_{(\underline{m})}\left(\mathcal{B}^{(\underline{m})}_{\mu \nu}\mathcal{B}^{(\underline{m})\,\mu \nu}+2\mathcal{B}^{(\underline{m})}_{\mu \bar{\nu}}\mathcal{B}^{(\underline{m})\,\mu \bar{\nu}}+\mathcal{B}^{(\underline{m})}_{\bar{\mu} \bar{\nu}}\mathcal{B}^{(\underline{m})\,\bar{\mu} \bar{\nu}}\right)\Big]\ .
\end{align}
%As it can be read from the general expressions in Refs.~\cite{GGNNT,OPT3}, 
The different curvature components $ \mathcal{W}^{(\underline{0})i}_{\mu\nu} $, $ \mathcal{W}^{(\underline{m})i}_{\mu\nu} $, $ \mathcal{W}^{(\underline{0})i}_{\bar{\mu}\bar{\nu}} $, $ \mathcal{W}^{(\underline{m})i}_{\bar{\mu}\bar{\nu}} $, and $ \mathcal{W}^{(\underline{m})i}_{\mu\bar{\nu}} $ are already given in~\cite{GGNNT,OPT3}, but with the appropriate $su(2)$-valued gauge-connection components ($ \mathcal{F}\mapsto\mathcal{W} $) and structure constants ($ f^{abc}\mapsto\epsilon^{ijk} $) . The abelian curvature components $ \mathcal{B}^{(\underline{0})}_{\mu\nu} $, $ \mathcal{B}^{(\underline{m})}_{\mu\nu} $, $ \mathcal{B}^{(\underline{0})}_{\bar{\mu}\bar{\nu}} $, $ \mathcal{B}^{(\underline{m})}_{\bar{\mu}\bar{\nu}} $, and $ \mathcal{B}^{(\underline{m})}_{\mu\bar{\nu}} $ are given by
\begin{subequations}
\begin{align}
\mathcal{B}^{(\underline{0})}_{\mu \nu}&={B}^{(\underline{0})}_{\mu \nu} =\partial_\mu B^{(\underline{0})}_\nu-\partial_\nu B^{(\underline{0})}_\mu \, ,\\
\mathcal{B}^{(\underline{m})}_{\mu \nu}&=\partial_\mu B^{(\underline{m})}_\nu-\partial_\nu B^{(\underline{m})}_\mu \, ,\\
\mathcal{B}^{(\underline{m})}_{\bar{\mu} \bar{\nu}}& = p_{\bar{\mu}}^{(\underline m)} B^{(\underline{m})}_{\bar{\nu}}-
p_{\bar{\nu}}^{(\underline m)}B^{(\underline{m})}_{\bar{\mu}}\, ,\\
\mathcal{B}^{(\underline{m})}_{\mu \bar{\nu}}&=\partial_\mu B^{(\underline{m})}_{\bar{\nu}}+
p_{\bar{\nu}}^{(\underline m)} B^{(\underline{m})}_{\mu}\, .
\end{align}
\end{subequations}
Notice that for any abelian gauge group we have $\mathcal{B}^{(\underline{0})}_{\bar{\mu} \bar{\nu}}=0$. This Lagrangian is invariant under the SGTs, which can be identified with the group $ SU_{L}(2,\mathcal{M}^{4})\times U_{Y}(1,\mathcal{M}^{4}) $.\\

The $(4+n)$-dimensional Lagrangian for the Higgs sector is given by
\begin{equation}\label{HSL}
{\cal L}^{\textrm{H}}_{(4+n)}=(D_M\Phi)^\dag(D^M \Phi)-V(\Phi^\dag,\Phi)\, ,
\end{equation}
where $\Phi$ is the Higgs doublet defined on the higher-dimensional spacetime. The electroweak covariant derivative $ D_{M} $, in the fundamental representation, is constructed with the connection components $ \mathcal{W}^{i}_{M} $ and $ \mathcal{B}_{M} $ as
\begin{equation}
D_M=\partial_M-ig_{4+n}\frac{\sigma^i}{2}{\cal W}^i_M-ig'_{4+n}\frac{Y}{2}{\cal B}_M \, ,
\end{equation}
while the straightforward extrapolation of the Higgs potential to higher dimensions is
\begin{equation}
V(\Phi^\dag,\Phi)=\mu^2 (\Phi^\dag \Phi)+\lambda_{(4+n)}(\Phi^\dag \Phi)^2\ ,
\end{equation}
where $ \mu $ and $ \lambda_{(4+n)} $ are real parameters with canonical dimensions 1 and $ -n $, respectively. Using the general strategy described in Ref.~\cite{GGNNT}, and in order to recover the SM Higgs sector in the small extra dimensions limit ($R_i\to 0$), we assume the Higgs doublet to be an even function under the altogether reflection of the extra spatial dimensions, that is, $ \Phi(x,-\bar{x})=\Phi(x,\bar{x}) $. Then
\begin{equation}
\Phi(x,\bar{x})= f^{(\underline{0})}_E \Phi^{(\underline{0})}(x)+\sum_{(\underline{m})}f^{(\underline{m})}_E(\bar{x}) \Phi^{(\underline{m})}(x)\, .
\end{equation}
Explicitly, $\Phi^{(\underline{0})\dag}=\left(G^{(\underline{0})-}_W, (v+H^{(\underline{0})}-iG^{(\underline{0})}_Z)/\sqrt{2}\right)$ and
$\Phi^{(\underline{m})\dag}=\left(G^{(\underline{m})-}_W,\, (H^{(\underline{m})}-iG^{(\underline{m})}_Z)/\sqrt{2}\right)$ are the SM Higgs doublet and its KK excitations, respectively. These KK excitations have the same quantum numbers as the standard Higgs doublet but we assume that they do not possess a vacuum expectation value.\\

Given the parity conditions that we established above, we conclude that the $ {\rm SO}(1,3) $-vector (-scalar) components, $D_\mu \Phi$ $ (D_{\bar{\mu}}\Phi $), of $D_M\Phi$  are even (odd) functions under the altogether reflection of the extra spatial dimensions. Hence, integrating out the extra dimensions from Eq.~\eqref{HSL} gives the effective Higgs sector
\begin{eqnarray}
\label{EffHS}
\mathcal{L}_{\textrm{eff}}^{\textrm H}&=&(D_\mu \Phi)^{(\underline{0})\dag}(D^\mu \Phi)^{(\underline{0})} +\sum_{(\underline{m})}\left[(D_\mu \Phi)^{(\underline{m})\dag}(D^\mu \Phi)^{(\underline{m})}+(D_{\bar{\mu}} \Phi)^{(\underline{m})\dag}(D^{\bar{\mu}} \Phi)^{(\underline{m})}\right]\nonumber \\
&& - V\left(\Phi^{(\underline{0})},\Phi^{(\underline{m})}\right)\, .
\end{eqnarray}
Here,
\begin{subequations}\label{DM-fourier}
\begin{align}
\label{MO}
(D_\mu \Phi)^{(\underline{0})}&= D^{(\underline{0})}_\mu \Phi^{(\underline{0})}-ig\sum_{(\underline{m})}{\cal O}^{(\underline{m})}_\mu \Phi^{(\underline{m})}\, ,\\
\label{MV}
(D_\mu \Phi)^{(\underline{m})}&=\sum_{(\underline{r})}D^{(\underline{mr})}_\mu \Phi^{(\underline{r})}-ig{\cal O}^{(\underline{m})}_\mu \Phi^{(\underline{0})}\, ,\\
\label{MS}
(D_{\bar{\mu}} \Phi)^{(\underline{m})}&=\sum_{(\underline{r})}D^{(\underline{mr})}_{\bar{\mu}} \Phi^{(\underline{r})}-ig{\cal O}^{(\underline{m})}_{\bar{\mu}}
\Phi^{(\underline{0})}\, ,
\end{align}
\end{subequations}
can be obtained by, firstly, inserting the Fourier expansion of each field into the definitions of $ D_{\mu}\Phi $ and $ D_{\bar{\mu}}\Phi $, secondly, using the parity property of each covariant derivative components and their respective Fourier expansions, and, finally, integrating out the extra dimensions by taking advantage of the orthogonality of trigonometric functions. In Eqs.~\eqref{DM-fourier}, the $SU_L(2,{\cal M}^4)\times U_Y(1,{\cal M}^4)$ covariant derivative is recovered and given by
\begin{subequations}\label{Ds}
\begin{align}
D^{(\underline{0})}_\mu & = \partial_\mu-ig{\cal O}^{(\underline{0})}_\mu \, , \ \
{\cal O}^{(\underline{0})}_\mu := \frac{\sigma^i}{2}W^{(\underline{0})i}_{\mu}+\frac{g'}{g}\frac{Y}{2}B^{(\underline{0})}_{\mu}\, ,
\end{align}
whereas
\begin{align}
D^{(\underline{mr})}_\mu &= \delta^{(\underline{mr})}D^{(\underline{0})}_\mu-ig\sum_{(\underline{s})}
\Delta_{(\underline{mrs})}{\cal O}^{(\underline{s})}_\mu \, ,  \  \
{\cal O}^{(\underline{m})}_\mu := \frac{\sigma^i}{2}W^{(\underline{m})i}_{\mu}+\frac{g'}{g}\frac{Y}{2}B^{(\underline{m})}_{\mu}\, ,\\
D^{(\underline{mr})}_{\bar{\mu}}&=-p_{\bar{\mu}}^{(\underline m)}  \delta^{(\underline{mr})}
-ig\sum_{(\underline{s})}\Delta'_{(\underline{msr})}{\cal O}^{(\underline{s})}_{\bar{\mu}}\, , \  \
{\cal O}^{(\underline{m})}_{\bar{\mu}} := \frac{\sigma^i}{2}W^{(\underline{m})i}_{\bar{\mu}}+\frac{g'}{g}\frac{Y}{2}B^{(\underline{m})}_{\bar{\mu}}\ .
\end{align}
\end{subequations}

The effective Higgs potential $ V\left(\Phi^{(\underline{0})},\Phi^{(\underline{m})}\right) $, included in Eq.~\eqref{EffHS}, is
\begin{align}\label{EffHP}
V\left(\Phi^{(\underline{0})},\Phi^{(\underline{m})}\right)  = & \ \mu^2\left(\Phi^{(\underline{0})\dag} \Phi^{(\underline{0})}\right)+\lambda \left(\Phi^{(\underline{0})\dag} \Phi^{(\underline{0})}\right)^2+ \left(\mu^2+2\lambda \Phi^{(\underline{0})\dag} \Phi^{(\underline{0})}\right)\sum_{(\underline{m})}\Phi^{(\underline{m})\dag} \Phi^{(\underline{m})}\nonumber\\
& +\lambda\sum_{(\underline{m})}\left(\Phi^{(\underline{0})\dag} \Phi^{(\underline{m})}+
\Phi^{(\underline{m})\dag} \Phi^{(\underline{0})}\right)\left(\Phi^{(\underline{0})\dag} \Phi^{(\underline{m})}+
\Phi^{(\underline{m})\dag} \Phi^{(\underline{0})}\right)\nonumber\\
& +2\lambda \sum_{(\underline{mkr})}\Delta_{(\underline{mkr})}\left(\Phi^{(\underline{k})\dag} \Phi^{(\underline{r})}\right)\left(\Phi^{(\underline{0})\dag} \Phi^{(\underline{m})}+
\Phi^{(\underline{m})\dag} \Phi^{(\underline{0})}\right)\nonumber \\
& +\lambda\sum_{(\underline{mkrs})} \Delta_{(\underline{mkrs})}\left(\Phi^{(\underline{m})\dag} \Phi^{(\underline{k})}\right)\left(\Phi^{(\underline{r})\dag} \Phi^{(\underline{s})}\right)\, .
\end{align}
The first two terms correspond to the SM Higgs potential, with the dimensionless $ \lambda $ constant being related to $ \lambda_{(4+n)} $ by  $\lambda_{(4+n)}=(R_1\cdots R_n)\lambda$.  The multi-index object $\Delta_{(\underline{mkrs})}$ is a consequence of the fourth powered self-coupling in $ \Phi $ and its definition emerges from the integral of the product of four even functions with different Fourier modes:
\begin{equation}
\int_{0}^{2\pi R_n}\ldots\int_{0}^{2\pi R_1} \mathrm{d}^{n}\overline{x} f^{(\underline{m})}_{E}(\bar{x})f^{(\underline{k})}_{E}( \bar{x})f^{(\underline{r})}_{E}(\bar{x})f^{(\underline{s})}_{E}(\bar{x})
= f^{(\underline{0})\, 2}_E\,\,\Delta_{(\underline{m} \underline{k} \underline{r} \underline{s})}\, ,
\end{equation}
where
\begin{equation}
\begin{aligned}
\Delta_{(\underline{mkrs})}  \equiv& \ \Delta_{\underline{m}_{1}\ldots \underline{m}_{n}\underline{k}_{1}\ldots \underline{k}_{n}\underline{r}_{1}\ldots \underline{r}_{n}\underline{s}_{1}\ldots \underline{s}_{n}}\\
= & \  \frac{1}{2}\left(\delta_{\underline{m}_{1}+\underline{k}_{1}\,\underline{r}_{1}+\underline{s}_{1}}\ldots\delta_{\underline{m}_{n}+\underline{k}_{n}\,\underline{r}_{n}+\underline{s}_{n}} +\delta_{\underline{m}_{1}+r_{1}\,\underline{k}_{1}+\underline{s}_{1}}\ldots\delta_{\underline{m}_{n}+\underline{r}_{n}\,\underline{k}_{n}+\underline{s}_{n}}\right.\\
&\ \ \ +\delta_{\underline{m}_{1}+\underline{s}_{1}\,\underline{k}_{1}+\underline{r}_{1}}\ldots\delta_{\underline{m}_{n}+\underline{s}_{n}\,\underline{k}_{n}+\underline{r}_{n}}
+ 2\delta_{\underline{m}_{1}\,\underline{k}_{1}+\underline{r}_{1}+\underline{s}_{1}}\ldots \delta_{\underline{m}_{n}\,\underline{k}_{n}+\underline{r}_{n}+\underline{s}_{n}}\\
& \  \ \  +\delta_{\underline{k}_{1}\,\underline{r}_{1}+\underline{s}_{1}+\underline{m}_{1}} \ldots\delta_{\underline{k}_{n}\,\underline{r}_{n}+\underline{s}_{n}+\underline{m}_{n}}+\delta_{\underline{r}_{1}\,\underline{s}_{1}+\underline{m}_{1}+\underline{k}_{1}}\ldots\delta_{\underline{r}_{n}\,\underline{s}_{n}+\underline{m}_{n}+\underline{k}_{n}}\\
& \ \ \ +\left. \delta_{\underline{s}_{1}\,\underline{m}_{1}+\underline{k}_{1}+\underline{r}_{1}}\ldots\delta_{\underline{s}_{n}\,\underline{m}_{n}+\underline{k}_{n}+\underline{r}_{n}}\right)
\end{aligned}
\end{equation}

Adding $ \mathcal{L}_{\textrm{eff}}^{\textrm{YM EW}} $ and $ \mathcal{L}_{\textrm{eff}}^{\textrm{H}} $ together gives rise to the bosonic sector of the effective electroweak Lagrangian,
\begin{equation}\label{BEWEff}
\mathcal{L}_{\textrm{eff, boson}}^{\textrm{EW}}:=\mathcal{L}_{\textrm{eff}}^{\textrm{YM EW}}+\mathcal{L}_{\textrm{eff}}^{\textrm{H}}\ .
\end{equation}
As it can be seen from the coupling between KK excited gauge fields and the Higgs doublet $ \Phi^{(\underline{0})} $ in Eq.~\eqref{MV}, there is a contribution to the mass of the former due to the vacuum expectation value $ v $. Similarly, from Eq.~\eqref{MS}, a $ v $ contribution to the mass of the scalar fields involved in $ \mathcal{O}^{(\underline{m})}_{\bar{\mu}} $ is manifest. This is the justification behind the different mass spectrum of excited KK electroweak fields with respect to that characterizing  the excited KK gluon fields.\\

In order to systematize the study of $ \mathcal{L}_{\textrm{eff, boson}}^{\textrm{EW}} $, we split it into three different terms that depend on the type of local interactions, so that modulo a constant, which does not affect the dynamics, we can write
\begin{equation}
\mathcal{L}_{\textrm{eff, boson}}^{\textrm{EW}}= \mathcal{L}_{\textrm{s-s}}^{\textrm{EW}}+\mathcal{L}_{\textrm{v-s}}^{\textrm{EW}}+\mathcal{L}_{\textrm{v-v}}^{\textrm{EW}}
\end{equation}
Below we will detail each of these Lagrangians. \\
\newline
\subsubsection{Scalar-Scalar sector $\boldsymbol{\mathcal{L}_{\textrm{s-s}}^{\textrm{EW}}}$: scalar mass spectrum and self interactions.}

The interactions among the different scalars of the theory are present in the following sector:
\begin{align}\label{LEWss}
\mathcal{L}_{\textrm{s-s, boson}}^{
\textrm{EW}}  := &-\frac{1}{4}\left[\sum_{(\underline{m})}\left({\cal W}^{(\underline{m})i}_{\bar{\mu}\bar{\nu}}{\cal W}^{(\underline{m}) \bar{\mu}\bar{\nu}}_{i}+\mathcal{B}^{(\underline{m})}_{\bar{\mu}\bar{\nu}}
\mathcal{B}^{(\underline{m})\bar{\mu}\bar{\nu}}\right)+{\cal W}^{(\underline{0})i}_{\bar{\mu}\bar{\nu}}{\cal W}^{(\underline{0}) \bar{\mu}\bar{\nu}}_{i}\right]\nonumber\\
 & +\sum_{(\underline{m})}\left(D_{\bar{\mu}}\Phi\right)^{(\underline{m})\dagger}\left(D^{\bar{\mu}}\Phi\right)^{(\underline{m})}
 -V\left(\Phi^{(\underline{0})},\Phi^{(\underline{m})}\right)\, .
\end{align}
The mass spectrum of the scalar sector, once the SM Higgs mechanism is implemented, is encoded in the quadratic part of $  \mathcal{L}_{\textrm{s-s, boson}}^{\textrm{EW}}$. In order to study the mass spectrum, it is convenient to introduce, just as in the SM, the complex combinations\footnote{These definitions, as well as those for $ A $ and $ Z $ given by Eqs.~\eqref{AandZ}, can be shown to be the Fourier-componentwise equalities implied by the corresponding higher dimensional equalities.}
 \begin{subequations}\label{Wpm}
 \begin{align}
 W^{(\underline{0})\pm}_{\mu} & :=\frac{1}{\sqrt{2}}\left(W^{(\underline{0})1}_{\mu}\mp iW^{(\underline{0})2}_{\mu}\right)\, , \\
 W^{(\underline{m})\pm}_{\mu} & :=\frac{1}{\sqrt{2}}\left(W^{(\underline{m})1}_{\mu}\mp iW^{(\underline{m})2}_{\mu}\right)\, , \\
 W^{(\underline{m})\pm}_{\bar{\mu}} & :=\frac{1}{\sqrt{2}}\left(W^{(\underline{m})1}_{\bar{\mu}}\mp iW^{(\underline{m})2}_{\bar{\mu}}\right)\, ,  \end{align}
 \end{subequations}
 and the weak angle $ \theta_{\rm W} $ through the functions $c_{\textrm{\tiny{W}}}\equiv \cos (\theta_{\rm W}) $ and  $s_{\textrm{\tiny{W}}}\equiv \sin (\theta_{\rm W}) $ in the following definitions:
\begin{subequations}\label{AandZ}
\begin{align}
Z^{(\underline{0})}_{\mu}& := W^{(\underline{0})3}_{\mu}c_{\textrm{\tiny{W}}}-B^{(\underline{0})}_{\mu}s_{\textrm{\tiny{W}}}\, ,\quad
A^{(\underline{0})}_{\mu} := W^{(\underline{0})3}_{\mu}s_{\textrm{\tiny{W}}}+B^{(\underline{0})}_{\mu}c_{\textrm{\tiny{W}}}\, ,\\
Z^{(\underline{m})}_{\mu} & := W^{(\underline{m})3}_{\mu}c_{\textrm{\tiny{W}}}-B^{(\underline{m})}_{\mu}s_{\textrm{\tiny{W}}}\, ,\quad A^{(\underline{m})}_{\mu} := W^{(\underline{m})3}_{\mu}s_{\textrm{\tiny{W}}}+B^{(\underline{m})}_{\mu}c_{\textrm{\tiny{W}}}\, ,\\
Z^{(\underline{m})}_{\bar{\mu}} &:= W^{(\underline{m})3}_{\bar{\mu}}c_{\textrm{\tiny{W}}}-B^{(\underline{m})}_{\bar{\mu}}s_{\textrm{\tiny{W}}}\, ,\quad
A^{(\underline{m})}_{\bar{\mu}} := W^{(\underline{m})3}_{\bar{\mu}}s_{\textrm{\tiny{W}}}+B^{(\underline{m})}_{\bar{\mu}}c_{\textrm{\tiny{W}}}\, .
\end{align}
 \end{subequations}

In the standard unitary gauge, using the excited KK combinations defined in Eqs.~\eqref{Wpm} and \eqref{AandZ}, and the components of $ \Phi^{(\underline{m})} $, the quadratic part of $ \mathcal{L}_{\textrm{s-s, boson}}^{\textrm{EW}} $ can conveniently be written in the following way:
\begin{eqnarray}
\label{LEWmass}
\mathcal{L}^{\textrm{EW}}_{\textrm{s-s, mass}}&=& -\frac{1}{2} m^2_{H^{(\underline{0})}}H^{(\underline{0})}H^{(\underline{0})}\nonumber \\
&& -\sum_{(\underline{m})}\Bigg\{\frac{1}{2}\left(m^2_{H^{(\underline{0})}}+m^2_{{(\underline{m})}}\right)H^{(\underline{m})}
H^{(\underline{m})}+\frac{1}{2}A^{(\underline{m})}_{\overline{\mu}}\mathfrak{M}^{(\underline{m})}_{\bar{\mu}\bar{\nu}}A^{(\underline{m})}_{\bar{\nu}}
\nonumber\\
&& +\begin{pmatrix} W^{(\underline{m})-}_{\bar{\mu}} & G^{(\underline{m})-}_{W} \end{pmatrix}\Bigg[\begin{pmatrix}
\mathfrak{M}^{(\underline{m})}_{\bar{\mu}\bar{\nu}} & 0 \\
0 & 0 \end{pmatrix}\nonumber \\
&&+\begin{pmatrix}
m^2_{W^{(\underline{0})}}\delta_{\bar{\mu}\bar{\nu}} & -im_{W^{(\underline{0})}}p_{\bar{\mu}}^{(\underline m)}  \\
im_{W^{(\underline{0})}}p_{\bar{\nu}}^{(\underline m)}  & m^2_{(\underline{m})} \end{pmatrix}\Bigg]
\begin{pmatrix} W^{(\underline{m})+}_{\bar{\nu}} \\ G^{(\underline{m})+}_{W} \end{pmatrix}
\nonumber\\
&&+\frac{1}{2}\begin{pmatrix} Z^{(\underline{m})}_{\bar{\mu}} & G^{(\underline{m})}_{Z} \end{pmatrix}\Bigg[\begin{pmatrix}
\mathfrak{M}^{(\underline{m})}_{\bar{\mu}\bar{\nu}} & 0 \\
0 & 0 \end{pmatrix}\nonumber \\
&&+\begin{pmatrix}
m^2_{Z^{(\underline{0})}}\delta_{\bar{\mu}\bar{\nu}} & -m_{Z^{(\underline{0})}}p_{\bar{\mu}}^{(\underline m)}  \\
-m_{Z^{(\underline{0})}}p_{\bar{\nu}}^{(\underline m)}  & m^2_{(\underline{m})} \end{pmatrix}\Bigg]
\begin{pmatrix} Z^{(\underline{m})}_{\overline{\nu}} \\ G^{(\underline{m})}_{Z} \end{pmatrix}\Bigg\}\ .
\end{eqnarray}
These expressions contain contributions from the Yang-Mills sector, the extra-spatial covariant derivatives, and the potential term. In Eq.~\eqref{LEWmass}, the mass $ m^{2}_{H^{(\underline{0})}}= 2\lambda v^{2}$, of the SM Higgs boson $ H^{(\underline{0})} $, is duly recovered and the squared mass of each field in the KK tower $ H^{(\underline{m})} $ is $ m^{2}_{H^{(\underline{m})}}:=m^{2}_{H^{(\underline{0})}}+m^{2}_{(\underline{m})} $. In addition, $ m^{2}_{W^{({\underline{0}})}} $ and  $ m^{2}_{Z^{({\underline{0}})}} $ are the SM squared masses of the SM fields $ W^{(\underline{0})\pm}_\mu $ and $ Z^{(\underline{0})}_\mu $. The excited scalar fields $A^{(\underline{m})}_{\bar{\mu}}$  associated to the photon field are mixed via the same structure that mixes $ G^{(\underline{m})}_{\bar{\mu}} $ in the $ \mathcal{L}_{\textrm{eff}}^{\textrm{YM QCD}} $. Here we find, as new ingredients, the balanced mixings between the charged scalars $ W^{(\underline{m})\pm}_{\bar{\mu}} $ and $ G^{(\underline{m})\pm}_{W} $, and, separately, between the real scalars $ Z^{(\underline{m})}_{\bar{\mu}} $ and $ G^{(\underline{m})}_{Z} $ as well. The above expression can be further simplified first by using some properties of the orthogonal rotation given in~\cite{GGNNT}. Once this is done, one obtains
\begin{align}\label{LEWmass2}
\mathcal{L}^{\textrm{EW}}_{\textrm{s-s, mass}}=& -\frac{1}{2} m^2_{H^{(\underline{0})}}H^{(\underline{0})}H^{(\underline{0})} -
\sum_{(\underline{m})}\left\lbrace\frac{1}{2}m^2_{H^{(\underline{m})}}H^{(\underline{m})}
H^{(\underline{m})}+\frac{1}{2}m^2_{{(\underline{m})}}A^{(\underline{m})}_{\bar{n}}A^{(\underline{m})}_{\bar{n}}
\right.\nonumber\\
&\left. +m^2_{W^{(\underline{m})}} W^{(\underline{m})-}_{\bar{n}}W^{(\underline{m})+}_{\bar{n}}+\frac{1}{2}m^2_{Z^{(\underline{m})}} Z^{(\underline{m})}_{\bar{n}}Z^{(\underline{m})}_{\bar{n}}
\right.\nonumber\\
&\left. +\begin{pmatrix} W^{(\underline{m})-} & G^{(\underline{m})-}_{W} \end{pmatrix}
\begin{pmatrix}
m^2_{W^{(\underline{0})}} & -im_{W^{(\underline{0})}}m_{(\underline{m})} \\
im_{W^{(\underline{0})}}m_{(\underline{m})} & m^2_{(\underline{m})} \end{pmatrix}
\begin{pmatrix} W^{(\underline{m})+} \\ G^{(\underline{m})+}_{W} \end{pmatrix}
\right.\nonumber\\
&\left.+\frac{1}{2}\begin{pmatrix} Z^{(\underline{m})} & G^{(\underline{m})}_{Z} \end{pmatrix}
\begin{pmatrix}
m^2_{Z^{(\underline{0})}} & -m_{Z^{(\underline{0})}}m_{(\underline{m})} \\
-m_{Z^{(\underline{0})}}m_{(\underline{m})} & m^2_{(\underline{m})} \end{pmatrix}
\begin{pmatrix} Z^{(\underline{m})} \\ G^{(\underline{m})}_{Z} \end{pmatrix}\right\rbrace\ .
\end{align}
According to this Lagrangian, the scalar excitations $G^{(\underline{m})\pm}_W$ and $G^{(\underline{m})}_Z$, which originate in the extra-dimensional Higgs doublet, mix only with $W^{(\underline{m})\pm}$ and $Z^{(\underline{m})}$, respectively, and they would correspond to massless scalars if no Higgs sector were present. So, there arise mixings between the charged KK excitations $G^{(\underline{m})\pm}_W$ and $W^{(\underline{m})\pm}$, and, similarly, between the neutral KK excitations $G^{(\underline{m})}_Z$ and $Z^{(\underline{m})}$. The $n-1$ scalar fields $W^{(\underline{m})\pm}_{\bar{n}}$ and $Z^{(\underline{m})}_{\bar{n}}$, with $\bar{n}=1,\cdots,n-1$, are not mixed  with the Higgs doublet components. These fields only receive a mass contribution given by $m_{W^{(\underline{0})}}$ in the case of $W^{(\underline{m})\pm}_{\bar{n}}$ and by $m_{Z^{(\underline{0})}}$ in the case of $Z^{(\underline{m})}_{\bar{n}}$. On the other hand, it can be seen that the mass matrix that mixes the  KK fields $G^{(\underline{m})\pm}_W$ and $W^{(\underline{m})\pm}$ and the one that mixes $G^{(\underline{m})}_Z$ and $Z^{(\underline{m})}$ have the set of eigenvalues $( m^2_{W^{(\underline{m})}}=m^2_{{(\underline{m})}}+m^2_{W^{(\underline{0})}},\, 0)$ and $( m^2_{Z^{(\underline{m})}}=m^2_{{(\underline{m})}}+m^2_{Z^{(\underline{0})}},\, 0)$, respectively. In this way, a physical scalar $W^{(\underline{m})\pm}_{n}$ ($Z^{(\underline{m})}_{n}$) and a massless scalar $W^{(\underline{m})\pm}_G$ ($Z^{(\underline{m})}_G$) associated with the gauge bosons $W^\pm$ ($Z$) emerge. The massless scalar fields correspond to the pseudo-Goldstone boson of the vectorial KK excitations of these gauge bosons. Due to the fact that the above Lagrangian also induces a mass contribution to the $n-1$ scalar fields $W^{(\underline{m})\pm}_{\bar{n}}$ ($Z^{(\underline{m})}_{\bar{n}}$) given by $m_{W^{(\underline{0})}}$ $(m_{Z^{(\underline{0})}})$, we conclude that associated to each KK tower of the massive $W^\pm$ and $Z$ gauge bosons there are a total of $n$ scalar fields with masses given by $m_{W^{(\underline{m})}}$ and $m_{Z^{(\underline{m})}}$. This in contrast with the case of gluons or the photon, which only have associated $n-1$ physical scalars with masses given by $m_{(\underline{m})}$. The above mass matrices are diagonalized through the following unitary transformations:
\begin{equation}
\label{U1}
\left(\begin{array}{ccc}
W^{(\underline{m})+} \\
\ \ \\
G^{(\underline{m})+}_W
\end{array}\right)=\mathcal{S}\left(\begin{array}{ccc}
W^{(\underline{m})+}_{n}  \\
\ \ \\
W^{(\underline{m})+}_G
\end{array}\right)\, , \, \, \, \, \,
\left(\begin{array}{ccc}
Z^{(\underline{m})} \\
\ \ \\
G^{(\underline{m})}_Z
\end{array}\right)=\mathcal{T}\left(\begin{array}{ccc}
Z^{(\underline{m})}_{n} \\
\ \ \\
Z^{(\underline{m})}_G
\end{array}\right)\, ,
\end{equation}
where
\begin{equation}
\label{U2}
\mathcal{S}=\left(\begin{array}{ccc}
\frac{m_{W^{(\underline{0})}}}{m_{W^{(\underline{m})}}} & \frac{m_{(\underline{m})}}{m_{W^{(\underline{m})}}} \\
\ \ \ & \ \ \ \\
 i \frac{m_{(\underline{m})}}{m_{W^{(\underline{m})}}} &-i\frac{m_{W{(\underline{0})}}}{m_{W^{(\underline{m})}}}
\end{array}\right)\, , \, \, \, \, \mathcal{T}=\left(\begin{array}{ccc}
 \frac{m_{Z^{(\underline{0})}}}{m_{Z^{(\underline{m})}}} & \frac{m_{(\underline{m})}}{m_{Z^{(\underline{m})}}} \\
\ \ \ & \ \ \ \\
 - \frac{m_{(\underline{m})}}{m_{Z^{(\underline{m})}}} &\frac{m_{Z{(\underline{0})}}}{m_{Z^{(\underline{m})}}}
\end{array}\right)\, .
\end{equation}

Having defined the mass spectrum of the scalar sector, it is convenient to rewrite the $ \mathcal{L}_{\textrm{s-s, boson}}^{\textrm{EW}} $ Lagrangian in a way that can be useful for phenomenological applications:
\begin{eqnarray}
\mathcal{L}_{\textrm{s-s, boson}}^{\textrm{EW}}&=&-\frac{1}{4}\left(2\,{\cal W}^{(\underline{0})-}_{\bar{\mu}\bar{\nu}} {\cal W}^{(\underline{0})+}_{\bar{\mu}\bar{\nu}}+{\cal W}^{(\underline{0})3}_{\bar{\mu}\bar{\nu}} {\cal W}^{(\underline{0})3}_{\bar{\mu}\bar{\nu}}\right)\nonumber \\
&&\sum_{(\underline{m})}\Bigg\{-\frac{1}{4}\Big[2\, {\cal W}^{(\underline{m})-}_{\bar{\mu}\bar{\nu}} {\cal W}^{(\underline{m})+}_{\bar{\mu}\bar{\nu}}+{\cal W}^{(\underline{m})3}_{\bar{\mu}\bar{\nu}} {\cal W}^{(\underline{m})3}_{\bar{\mu}\bar{\nu}}+{\cal B}^{(\underline{m})}_{\bar{\mu}\bar{\nu}}{\cal B}^{(\underline{m})}_{\bar{\mu}\bar{\nu}}\Big]\nonumber \\
&&+igp_{\bar{\mu}}^{(\underline m)}  \Big[\Phi^{(\underline{m})\dag}{\cal O}^{(\underline{m})}_{\bar{\mu}}\Phi^{(\underline{0})}-  \Phi^{(\underline{0})\dag}{\cal O}^{(\underline{m})}_{\bar{\mu}}\Phi^{(\underline{m})}\nonumber \\
&& +\sum_{(\underline{s})}\Delta'_{(\underline{msr})}\left(\Phi^{(\underline{m})\dag}{\cal O}^{(\underline{s})}_{\bar{\mu}}\Phi^{(\underline{r})}-  \Phi^{(\underline{r})\dag}{\cal O}^{(\underline{s})}_{\bar{\mu}}\Phi^{(\underline{m})}\right)\Big]\nonumber \\
&&+m^2_{(\underline{m})}\Phi^{(\underline{m})\dag}\Phi^{(\underline{m})}+g^2\Big[\Phi^{(\underline{0})\dag}{\cal O}^{(\underline{m})}_{\bar{\mu}}{\cal O}^{(\underline{m})}_{\bar{\mu}}\Phi^{(\underline{0})}\nonumber \\
&&+\sum_{(\underline{rs})}\Delta'_{(\underline{msr})}\left(\Phi^{(\underline{0})\dag}{\cal O}^{(\underline{m})}_{\bar{\mu}}{\cal O}^{(\underline{s})}_{\bar{\mu}}\Phi^{(\underline{r})}+\Phi^{(\underline{r})\dag}{\cal O}^{(\underline{s})}_{\bar{\mu}}{\cal O}^{(\underline{m})}_{\bar{\mu}}\Phi^{(\underline{0})} \right) \nonumber \\
&&+\sum_{(\underline{rspq})}\Delta'_{(\underline{mpq})}\Delta'_{(\underline{msr})}\Phi^{(\underline{r})\dag}{\cal O}^{(\underline{s})}_{\bar{\mu}}{\cal O}^{(\underline{p})}_{\bar{\mu}}\Phi^{(\underline{q})} \Big]\Bigg\}
-V(\Phi^{(\underline{0})},\Phi^{(\underline{m})}) \, .
\end{eqnarray}
To pass to mass-eigenstate fields, the following relations must be used:
\begin{eqnarray}
\label{HDT}
\Phi^{(\underline{0})}&=&\Phi^{(\underline{0})}_P+\Phi^{(\underline{0})}_G\, , \\
\label{HDT1}
\Phi^{(\underline{m})}&=&\Phi^{(\underline{m})}_P+\Phi^{(\underline{m})}_G\, ,
\end{eqnarray}
\begin{eqnarray}
\label{HDT2}
W^{(\underline{m})+}_{\bar{\mu}} & = &\mathcal{R}^{(\underline{m})}_{\bar{\mu}\bar{n}}W^{(\underline{m})+}_{\bar{n}}+
\mathcal{R}^{(\underline{m})}_{\bar{\mu}W}\left(\mathcal{S}_{11}W^{(\underline{m})+}_{n}+\mathcal{S}_{12}W^{(\underline{m})+}_{G}\right)
 \nonumber \\
 &=& \mathcal{R}^{(\underline{m})}_{\bar{\mu}\bar{n}}W^{(\underline{m})+}_{\bar{n}}+\frac{p_{\bar{\mu}}^{(\underline m)} }{m_{(\underline{m})}}
 \left(\mathcal{S}_{11}W^{(\underline{m})+}_{n}+\mathcal{S}_{12}W^{(\underline{m})+}_{G}\right)\, ,
 \end{eqnarray}
 \begin{eqnarray}
 \label{HDT3}
 Z^{(\underline{m})}_{\bar{\mu}} & = &\mathcal{R}^{(\underline{m})}_{\bar{\mu}\bar{n}}Z^{(\underline{m})}_{\bar{n}}+\,
 \mathcal{R}^{(\underline{m})}_{\bar{\mu}Z} \left(\mathcal{T}_{11}Z^{(\underline{m})}_{n}+ \mathcal{T}_{12}Z^{(\underline{m})}_{G}\right)\nonumber \\
 &=&\mathcal{R}^{(\underline{m})}_{\bar{\mu}\bar{n}}Z^{(\underline{m})}_{\bar{n}}+
 \frac{p_{\bar{\mu}}^{(\underline m)} }{m_{(\underline{m})}}\left(\mathcal{T}_{11}Z^{(\underline{m})}_{n}+ \mathcal{T}_{12}Z^{(\underline{m})}_{G}\right)\, ,
 \end{eqnarray}
 \begin{eqnarray}
 \label{HDT4}
  A^{(\underline{m})}_{\bar{\mu}} & = & \mathcal{R}^{(\underline{m})}_{\bar{\mu}\bar{n}}A^{(\underline{m})}_{\bar{n}}+\mathcal{R}^{(\underline{m})}_{\bar{\mu}A} A^{(\underline{m})}_{G}\nonumber \\
  &=&\mathcal{R}^{(\underline{m})}_{\bar{\mu}\bar{n}}A^{(\underline{m})}_{\bar{n}}+
  \frac{p_{\bar{\mu}}^{(\underline m)} }{m_{(\underline{m})}}A^{(\underline{m})}_{G}\, ,
\end{eqnarray}
with
\begin{equation}
\label{PHD}
\Phi^{(\underline{0})}_P=\left(\begin{array}{ccc}
0 \\
\ \ \\
\frac{v+H^{(\underline{0})}}{\sqrt{2}}
\end{array}\right)\, , \, \, \, \, \Phi^{(\underline{0})}_G=\left(\begin{array}{ccc}
G^{(\underline{0})+}_W \\
\ \ \\
\frac{i\, G^{(\underline{0})}_Z}{\sqrt{2}}
\end{array}\right)\, ,
\end{equation}

\begin{equation}
\label{PHD1}
\Phi^{(\underline{m})}_P=\left(\begin{array}{ccc}
\mathcal{S}_{21}W^{(\underline{m})+}_n \\
\ \ \\
\frac{H^{(\underline{m})}}{\sqrt{2}}+\frac{i}{\sqrt{2}}\mathcal{T}_{21}Z^{(\underline{m})}_n
\end{array}\right)\, , \, \, \, \, \Phi^{(\underline{m})}_G=\left(\begin{array}{ccc}
\mathcal{S}_{22}W^{(\underline{m})+}_G \\
\ \ \\
\frac{i}{\sqrt{2}}\mathcal{T}_{22}Z^{(\underline{m})}_G
\end{array}\right)\, .
\end{equation}
In addition,
\begin{eqnarray}
{\cal W}^{(\underline{m})+}_{\bar{\mu}\bar{\nu}}&=&p_{\bar{\mu}}^{(\underline m)}W^{(\underline{m})+}_{\bar{\nu}}-
p_{\bar{\nu}}^{(\underline m)} W^{(\underline{m})+}_{\bar{\mu}}+i\sum_{(\underline{rs})}\Delta'_{(\underline{rsm})}
\Big[gc_W\left(W^{(\underline{r})+}_{\bar{\mu}}Z^{(\underline{s})}_{\bar{\nu}} -
W^{(\underline{s})+}_{\bar{\nu}}Z^{(\underline{r})}_{\bar{\mu}} \right)\nonumber \\
&&+e\left(W^{(\underline{r})+}_{\bar{\mu}}A^{(\underline{s})}_{\bar{\nu}}-W^{(\underline{s})+}_{\bar{\nu}}A^{(\underline{r})}_{\bar{\mu}} \right)\Big]\, ,
\end{eqnarray}

\begin{eqnarray}
{\cal W}^{(\underline{m})3}_{\bar{\mu}\bar{\nu}}&=&c_W\left(p_{\bar{\mu}}^{(\underline m)} Z^{(\underline{m})}_{\bar{\nu}}-
p_{\bar{\nu}}^{(\underline m)} Z^{(\underline{m})}_{\bar{\mu}}\right)+s_W\left(p_{\bar{\mu}}^{(\underline m)} A^{(\underline{m})}_{\bar{\nu}}-
p_{\bar{\nu}}^{(\underline m)} A^{(\underline{m})}_{\bar{\mu}}\right) \nonumber \\
&&+ig\sum_{(\underline{rs})}\Delta'_{(\underline{rsm})}\left(W^{(\underline{r})-}_{\bar{\mu}}W^{(\underline{s})+}_{\bar{\nu}}-
W^{(\underline{r})+}_{\bar{\mu}}W^{(\underline{s})-}_{\bar{\nu}}  \right)\, ,
\end{eqnarray}
\begin{eqnarray}
{\cal W}^{(\underline{0})+}_{\bar{\mu}\bar{\nu}}&=&i\sum_{(\underline{m})}
\Big[gc_W\left(W^{(\underline{m})+}_{\bar{\mu}}Z^{(\underline{m})}_{\bar{\nu}} -
W^{(\underline{m})+}_{\bar{\nu}}Z^{(\underline{m})}_{\bar{\mu}} \right)\nonumber \\
&&+e\left(W^{(\underline{m})+}_{\bar{\mu}}A^{(\underline{m})}_{\bar{\nu}} -
W^{(\underline{m})+}_{\bar{\nu}}A^{(\underline{m})}_{\bar{\mu}} \right)\Big]\, ,
\end{eqnarray}
\begin{equation}
{\cal O}^{(\underline{m})}_{ \bar{\mu}}=\left(\begin{array}{ccc}
\frac{c_{2W}Z^{(\underline{m})}_{\bar{\mu}}+s_{2W}A^{(\underline{m})}_{\bar{\mu}}}{2c_W} & \frac{ W^{(\underline{m})+}_{\bar{\mu}}}{\sqrt{2}} \\
\ \ \ & \ \ \ \\
 \frac{W^{(\underline{m})-}_{\bar{\mu}}}{\sqrt{2}} &-\frac{Z^{(\underline{m})}_{\bar{\mu}}}{2c_W}
\end{array}\right)\,
\end{equation}
Notice that the gauge which removes the set of massless scalar KK excitations $ W^{(\underline{m})\pm}_{G} $, $ Z^{(\underline{m})}_{G} $ and $ A^{(\underline{m})}_{G} $, fixes once and for all the excited gauge parameters $ \alpha^{(\underline{m})i} $ and $ \alpha^{(\underline{m})} $ that specify the so-called NSGTs. Hence the theory remains invariant under the SGT as three of the four parameters $ \alpha^{(\underline{0})i} $ and $ \alpha^{(\underline{0})} $ are fixed by the standard unitary gauge in which $ G^{(\underline{0})\pm}_{W}=0=G^{(\underline{0})}_{Z} $. The $\mathcal{L}_{\textrm{s-s, boson}}^{\textrm{EW}}$ Lagrangian is considerably simplified when both the standard and nonstandard gauge freedom are fixed using the unitary gauge.\\

Feynman rules can be directly derived  from the above expressions. For instance, the couplings of the zero mode of the Higgs boson to pairs of physical scalars are given by the following Lagrangian:
\begin{eqnarray}
{\cal L}_{H^{(\underline{0})}S^{(\underline{m})}S^{(\underline{m})}}&=&-gH^{(\underline{0})}\sum_{(\underline{m})}\Bigg\{
m_{W^{(\underline{0})}}W^{(\underline{m})-}_{\bar{n}}W^{(\underline{m})+}_{\bar{n}}+
\frac{m_{Z^{(\underline{0})}}}{2c_W}Z^{(\underline{m})}_{\bar{n}}Z^{(\underline{m})}_{\bar{n}}\nonumber \\
&&+m_{W^{(\underline{0})}}\left(1-\frac{m^2_{H^{(\underline{0})}}}{2m^2_{W^{(\underline{0})}}}\frac{m^2_{(\underline{m})}}{m^2_{W^{(\underline{m})}}}\right)  W^{(\underline{m})-}_nW^{(\underline{m})+}_n\nonumber \\
&&+\frac{m_{Z^{(\underline{0})}}}{2c_W} \left(1-\frac{m^2_{H^{(\underline{0})}}}{2m^2_{Z^{(\underline{0})}}}\frac{m^2_{(\underline{m})}}{m^2_{Z^{(\underline{m})}}}\right) Z^{(\underline{m})}_nZ^{(\underline{m})}_n \nonumber \\
&&-\frac{3}{4}\frac{m^2_{H^{(\underline{0})}}}{m^2_{W^{(\underline{0})}}}H^{(\underline{m})}H^{(\underline{m})}
\Bigg\}\, ,
\end{eqnarray}
where an implicit sum over the index $\bar{n}$, from $1$ to $n-1$, is assumed.
Notice that, as expected, the Higgs boson distinguishes, from among all scalar fields, the excitations associated with the longitudinal components of the $W$ and $Z$ gauge bosons. It is important to stress the fact that the Higgs boson couples to pairs of charged (neutral) scalars proportionally to $m_{W^{(\underline{0})}}$ ($m_{Z^{(\underline{0})}}$), just in the same way that it couples to pairs of $W$ ($Z$) gauge bosons in the SM. This fact shows that these charged (neutral) scalars are essential parts of the gauge bosons $W$ ($Z$) in the sense that the Higgs boson does not distinguish among them.
\\

\subsubsection{Vector-scalar sector $\boldsymbol{\mathcal{L}_{\textrm{v-s}}^{\textrm{EW}}}$: Gauge-excitations mass spectrum and vector-scalar interactions.}
We now turn our attention to the vector-scalar interactions. There are two sources of vector-scalar interactions, one emerging from the Yang-Mills sector and the other coming from the Higgs kinetic sector, that is,
\begin{equation}
\mathcal{L}_{\textrm{v-s, boson}}^{\textrm{EW}}=\mathcal{L}_{\textrm{v-s,YM}}^{\textrm{EW}}+\mathcal{L}_{\textrm{v-s,HK}}^{\textrm{EW}}\, ,
\end{equation}
where
\begin{eqnarray}\label{Ls-v}
\mathcal{L}_{\textrm{v-s,YM}}^{\textrm{EW}}&=&-\frac{1}{2}\sum_{(\underline{m})} \left(\mathcal{W}^{(\underline{m})i}_{\mu\bar{\nu}}\mathcal{W}^{(\underline{m})\mu\bar{\nu}}_{i}+
\mathcal{B}^{(\underline{m})}_{\mu\bar{\nu}}\mathcal{B}^{(\underline{m})\mu\bar{\nu}}\right)\, ,\\
\mathcal{L}_{\textrm{v-s,HK}}^{\textrm{EW}}&=&
\sum_{(\underline{m})}\left(D_{{\mu}}\Phi\right)^{(\underline{m})\dagger}\left(D^{{\mu}}\Phi\right)^{(\underline{m})}+
\left(D_{{\mu}}\Phi\right)^{(\underline{0})\dagger}\left(D^{{\mu}}\Phi\right)^{(\underline{0})}\ .
\end{eqnarray}
We have found it useful to rewrite the Lagrangian $\mathcal{L}_{\textrm{v-s,YM}}^{\textrm{EW}}$ as
\begin{eqnarray}
\mathcal{L}_{\textrm{v-s,YM}}^{\textrm{EW}}&=& \sum_{(\underline{m})}\Bigg\{{\rm Tr}\Big\{ \left({\cal D}^{(\underline{0})}_\mu W^{(\underline{m})}_{\bar{n}}\right) \left({\cal D}^{(\underline{0})\mu} W^{(\underline{m})}_{\bar{n}}\right)+ \left({\cal D}^{(\underline{0})}_\mu W^{(\underline{m})}\right) \left({\cal D}^{(\underline{0})\mu} W^{(\underline{m})}\right)\nonumber \\
&&+m^2_{(\underline{m})}W^{(\underline{m})}_\mu W^{(\underline{m})\mu} +2m_{(\underline{m})}W^{(\underline{m})}_\mu {\cal D}^{(\underline{0})\mu}W^{(\underline{m})}\nonumber \\
&&-2ig  \sum_{(\underline{r} \underline{s})} \Delta'_{(\underline{rsm})}  \left( {\cal R}_{\bar \nu \bar n}^{(\underline s)} \left[ W_{\mu}^{(\underline r)} , W_{\bar n}^{(\underline s)} \right]  +{p_{\bar \nu}^{(\underline s)}  \over m_{(\underline s)}} \left[ W_{\mu}^{(\underline r)} , W^{(\underline s)} \right]   \right)   \nonumber \\
&&\times \left(  {\cal R}_{\bar \nu \bar n}^{(\underline m)}     {\cal D}^{(\underline 0) \mu}W_{\bar n}^{(\underline m)} +{p_{\bar \nu}^{(\underline m)}  \over m_{(\underline m)}} {\cal D}^{(\underline 0) \mu}W^{(\underline m)}     + p_{\bar \nu}^{(\underline m)}  W^{(\underline m) \mu}  \right)\nonumber \\
&&-g^2 \sum_{(\underline{r} \underline{s} \underline{p}  \underline{q} )} \Delta'_{(\underline{rsm})} \Delta'_{(\underline{pqm})} \left( {\cal R}_{\bar \nu \bar n}^{(\underline s)} \left[ W_{\mu}^{(\underline r)} , W_{\bar n}^{(\underline s)} \right]  +{p_{\bar \nu}^{(\underline s)}  \over m_{(\underline s)}} \left[ W_{\mu}^{(\underline r)} , W^{(\underline s)} \right]   \right)   \nonumber\\
&& \times  \left( {\cal R}_{\bar \nu \bar n}^{(\underline q)} \left[ W^{(\underline p)\mu} , W_{\bar n}^{(\underline q)} \right]  +{p_{\bar \nu}^{(\underline q)}  \over m_{(\underline q)}} \left[ W^{(\underline p) \mu} , W^{(\underline q)} \right]   \right)  \Big \}\nonumber \\
 &&+\frac{1}{2}\left(\partial_\mu B^{(\underline{m})}_{\bar{n}} \right)\left(\partial^\mu B^{(\underline{m})}_{\bar{n}} \right)+\frac{1}{2}\left(\partial_\mu B^{(\underline{m})} \right)\left(\partial^\mu B^{(\underline{m})}\right)\nonumber \\
 &&+\frac{1}{2}m^2_{(\underline{m})}B^{(\underline{m})}_\mu B^{(\underline{m})\mu}+m_{(\underline{m})}
 B^{(\underline{m})}_\mu\partial^\mu B^{(\underline{m})}
 \Bigg\}\, ,
\end{eqnarray}
where we used some results given in~\cite{GGNNT} to substitute the matrix elements $\mathcal{R}^{(\underline{m})}_{\bar{\nu}W}$ and $\mathcal{R}^{(\underline{m})}_{\bar{\nu}B}$. In this expression, ${\cal D}^{(\underline{0})}_\mu$ is the covariant derivative of $SU(2,{\cal M}^4)$ in the adjoint representation. In addition, we have introduced the following objects:
\begin{equation}
W^{(\underline{m})}_{\bar{n}}=\left(\begin{array}{ccc}
\frac{c_WZ^{(\underline{m})}_{\bar{n}}+s_WA^{(\underline{m})}_{\bar{n}}}{2} & \frac{W^{(\underline{m})+}_{\bar{n}}}{\sqrt{2}} \\
\ \ \ & \ \ \ \\
\frac{W^{(\underline{m})-}_{\bar{n}}}{\sqrt{2}}& -\frac{c_WZ^{(\underline{m})}_{\bar{n}}+s_WA^{(\underline{m})}_{\bar{n}}}{2}
\end{array}\right)\, ,
\end{equation}

\begin{equation}
W^{(\underline{m})}=\left(\begin{array}{ccc}
\frac{c_W\left(\mathcal{T}_{11}Z^{(\underline{m})}_{n}+\mathcal{T}_{12}Z^{(\underline{m})}_{G}\right)+s_WA^{(\underline{m})}_{G}}{2} & \frac{\mathcal{S}_{11}W^{(\underline{m})+}_{n}+\mathcal{S}_{12}W^{(\underline{m})+}_{G}}{\sqrt{2}} \\
\ \ \ & \ \ \ \\
\frac{\mathcal{S}_{11}W^{(\underline{m})-}_{n}+\mathcal{S}_{12}W^{(\underline{m})-}_{G}}{\sqrt{2}}& -\frac{c_W\left(\mathcal{T}_{11}Z^{(\underline{m})}_{n}+\mathcal{T}_{12}Z^{(\underline{m})}_{G}\right)+s_WA^{(\underline{m})}_{G}}{2}
\end{array}\right)\, ,
\end{equation}

\begin{equation}
W^{(\underline{m})}_{\mu}=\left(\begin{array}{ccc}
\frac{c_WZ^{(\underline{m})}_{\mu}+s_WA^{(\underline{m})}_{\mu}}{2} & \frac{W^{(\underline{m})+}_{\mu}}{\sqrt{2}} \\
\ \ \ & \ \ \ \\
\frac{W^{(\underline{m})-}_{\mu}}{\sqrt{2}}& -\frac{c_WZ^{(\underline{m})}_{\mu}+s_WA^{(\underline{m})}_{\mu}}{2}
\end{array}\right)\, ,
\end{equation}

\begin{eqnarray}
B^{(\underline{m})}_{\bar{n}}&=&-s_WZ^{(\underline{m})}_{\bar{n}}+c_WA^{(\underline{m})}_{\bar{n}}\, , \\
B^{(\underline{m})}&=&-s_W\left(\mathcal{T}_{11}Z^{(\underline{m})}_n+\mathcal{T}_{12}Z^{(\underline{m})}_G\right)+c_WA^{(\underline{m})}_{G}\, .
\end{eqnarray}

The Higgs kinetic term $\mathcal{L}_{\textrm{v-s,HK}}^{\textrm{EW}}$ can be expressed as follows:
\begin{eqnarray}
\mathcal{L}_{\textrm{v-s,HK}}^{\textrm{EW}}&=&\left(D^{(\underline{0})}_\mu \Phi^{(\underline{0})}\right)^\dag \left(D^{(\underline{0})\mu} \Phi^{(\underline{0})}\right)+\sum_{(\underline{m})}\Bigg\{\left(D^{(\underline{0})}_\mu \Phi^{(\underline{m})}\right)^\dag \left(D^{(\underline{0})\mu} \Phi^{(\underline{m})}\right)\nonumber \\
&&+ig\Bigg(\Phi^{(\underline{m})\dag} {\cal O}^{(\underline{m}) \mu}D^{(\underline{0})}_\mu \Phi^{(\underline{0})}+\Phi^{(\underline{0})\dag} {\cal O}^{(\underline{m}) \mu}D^{(\underline{0})}_\mu \Phi^{(\underline{m})}\nonumber \\
&&+\sum_{(\underline{rs})}\Delta_{(\underline{mrs})}\Phi^{(\underline{r})\dag}{\cal O}^{(\underline{s})}_\mu
 D^{(\underline{0})\mu}\Phi^{(\underline{m})}-{\rm H.\, c.}\Bigg)\nonumber \\
&&+g^2\left( \Phi^{(\underline{0})\dag}{\cal O}^{(\underline{m})}_\mu{\cal O}^{(\underline{m})\mu}\Phi^{(\underline{0})}+
\sum_{(\underline{r})}\Phi^{(\underline{m})\dag}{\cal O}^{(\underline{m})}_\mu{\cal O}^{(\underline{r})\mu}\Phi^{(\underline{r})} \right)\nonumber \\
&&+g^2\sum_{(\underline{rs})}\Delta_{(\underline{mrs})}\Bigg(\Phi^{(\underline{r})\dag}{\cal O}^{(\underline{s})}_\mu {\cal O}^{(\underline{m})\mu}\Phi^{(\underline{0})}+\Phi^{(\underline{0})\dag}{\cal O}^{(\underline{m})}_\mu {\cal O}^{(\underline{s})\mu}\Phi^{(\underline{r})}\nonumber \\
&&+\sum_{(\underline{pq})}\Delta_{(\underline{mpq})} \Phi^{(\underline{p})\dag}{\cal O}^{(\underline{q})}_\mu {\cal O}^{(\underline{s})\mu}\Phi^{(\underline{r})} \Bigg)
\Bigg\}\, ,
\end{eqnarray}
where
\begin{equation}
{\cal O}^{(\underline{m})}_\mu=\left(\begin{array}{ccc}
\frac{c_{2W}Z^{(\underline{m})}_\mu+s_{2W}A^{(\underline{m})}_\mu}{2c_W} & \frac{W^{(\underline{m})+}_\mu}{\sqrt{2}} \\
\ \ \ & \ \ \ \\
\frac{W^{(\underline{m})-}_\mu}{\sqrt{2}}&-\frac{Z^{(\underline{m})}_\mu}{2c_W}
\end{array}\right)\, .
\end{equation}
In the above expression, $D^{(\underline{0})}_\mu$ is the covariant derivative of the electroweak group in the fundamental representation. Notice that $D^{(\underline{0})}_\mu=\partial_\mu-ig {\cal O}^{(\underline{0})}_\mu$, with ${\cal O}^{(\underline{0})}_\mu$ given by the expression that defines ${\cal O}^{(\underline{m})}_\mu$ with the interchange of $(\underline{m})$ by (\underline{0}). The first term of $\mathcal{L}_{\textrm{v-s,HK}}^{\textrm{EW}}$ corresponds to the standard Higgs kinetic term.\\

Both the $\mathcal{L}_{\textrm{v-s,YM}}^{\textrm{EW}}$ and the $\mathcal{L}_{\textrm{v-s,HK}}^{\textrm{EW}}$ Lagrangians contribute to the mass spectrum of the KK excitations of the electroweak gauge bosons, the former at the compactification scale and the later at the Fermi's scale. The terms that induce the mass spectrum are
\begin{eqnarray}
\mathcal{L}_{\textrm{v-s,mass}}^{\textrm{EW}}&=&\sum_{(\underline{m})}\Big(g^2 \Phi^{(\underline{0})\dag}_0{\cal O}^{(\underline{m})}_\mu
{\cal O}^{(\underline{m})\mu}\Phi^{(\underline{0})}_0\nonumber \\
&&+ m^2_{(\underline{m})}{\rm Tr}\left(W^{(\underline{m})}_\mu W^{(\underline{m})\mu}\right)+
\frac{1}{2}m^2_{(\underline{m})}B^{(\underline{m})}_\mu B^{(\underline{m})\mu}  \Big)\, ,
\end{eqnarray}
where $  \Phi^{(\underline{0})\dag}_0=(0,\, v/\sqrt{2})$. This expression leads to squared masses $m^2_{W^{(\underline{m})}}=m^2_{(\underline{m})}+m^2_{W^{(\underline{0})}}$,  $m^2_{Z^{(\underline{m})}}=m^2_{(\underline{m})}+m^2_{Z^{(\underline{0})}}$, and $m^2_{A^{(\underline{m})}}=m^2_{(\underline{m})}$ for the KK excitations of the electroweak gauge bosons $W^{(\underline{m})\pm}_\mu$, $Z^{(\underline{m})}_\mu$, and $A^{(\underline{m})}_\mu$, respectively. Notice that the vector excitations share the mass of their scalar counterparts. In particular, it is the SM EHM the one which endows with the mass $ m_{W^{(\underline{0})}} $ (resp. $ m_{Z^{(\underline{0})} } $) to the  total mass of the vector excitations $ W^{(\underline{m})\pm}_{\mu} $ (resp. $ Z^{(\underline{m})}_{\mu} $).\\

It is worth showing the couplings of the Higgs boson to pairs of $W$ and $Z$ KK excitations. These couplings, which emerge from the $g^2 \Phi^{(\underline{0})\dag}{\cal O}^{(\underline{m})}_\mu{\cal O}^{(\underline{m})\mu}\Phi^{(\underline{0})}$ term, are given by
\begin{eqnarray}
\mathcal{L}^{\textrm{EW}}_{H^{(\underline{0})}V^{(\underline{m})}V^{(\underline{m})}}&=&g\,m_{W^{(\underline{0})}}H^{(\underline{0})}\sum_{(\underline{m})}\left(
W^{(\underline{m})-}_\mu W^{(\underline{m})+\mu}+\frac{1}{2c^2_W}Z^{(\underline{m})}_\mu Z^{(\underline{m})\mu}\right)\, ,\\
\mathcal{L}^{\textrm{EW}}_{H^{(\underline{0})}H^{(\underline{0})}V^{(\underline{m})}V^{(\underline{m})}}&=&
\frac{g^2}{4}H^{(\underline{0})}H^{(\underline{0})}\sum_{(\underline{m})}\left(
W^{(\underline{m})-}_\mu W^{(\underline{m})+\mu}+\frac{1}{2c^2_W}Z^{(\underline{m})}_\mu Z^{(\underline{m})\mu}\right)\, .
\end{eqnarray}
These expressions show us that the Higgs boson couples to $W$ and $Z$ excitations just in the same way that it couples to the corresponding standard fields. In particular, the fact that the trilinear coupling is proportional to $m_{W^{(\underline{0})}}$ guarantees the decoupling of new-physics effects. \\

The Lagrangian $ \mathcal{L}^{\textrm{EW}}_{\textrm{v-s, boson}} $ also induces a bilinear mixing of the form:
\begin{eqnarray}
 \mathcal{L}^{\textrm{EW}}_{\textrm{B-M}} &=& \sum_{(\underline{m})} \Bigg\{ p_{\bar{\nu}}^{(\underline m)}  \Big[ W^{(\underline{m})-}_{{\mu}}\partial^{\mu}W^{(\underline{m})+}_{\bar{\nu}} + W^{(\underline{m})+}_{\mu}\partial^{\mu}W^{(\underline{m})-}_{\bar{\nu}}\nonumber \\
 &&+Z^{(\underline{m})}_{\mu}\partial^{\mu}Z^{(\underline{m})}_{\bar{\nu}} +A^{(\underline{m})}_{\mu}\partial^{\mu}A^{(\underline{m})}_{\bar{\nu}}  \Big]\\
&&+ig\Big[ \Phi^{(\underline{0})\dagger}_{0}\mathcal{O}^{(\underline{m})\dagger}_{\mu}\left( \partial^{\mu}\Phi^{(\underline{m})}_{G}\right) -\left( \partial^{\mu}\Phi^{(\underline{m})}_{G}\right)^{\dagger} \mathcal{O}^{(\underline{m})}_{\mu}\Phi^{(\underline{0})}_{0}  \Big]\Bigg\} \ .
\end{eqnarray}
This term behaves as the second term in the scalar-vector sector of the QCD Yang-Mills Lagrangian, Eq.~\eqref{gs5}: only unphysical massless KK scalars couple to the $W^{(\underline{m})\pm}_\mu$, $Z^{(\underline{m})}_\mu$, and $A^{(\underline{m})}_\mu$ gauge fields.
It turns out that the whole $  \mathcal{L}^{\textrm{EW}}_{\textrm{B-M}} $ vanishes in the unitary gauge. This is analogous to what happens in the EHM. Alternatively, one can remove these mixes via a surface term by introducing a linear~\cite{LG1,LG2,LG3} or nonlinear~\cite{NLG, OPT1} $R_\xi$-gauge.

\ \\

\subsubsection{Vector-vector sector $\boldsymbol{\mathcal{L}_{\textrm{v-v}}^{\textrm{EW}}}$.} We conclude the analysis of the structure of $ \mathcal{L}_{\textrm{eff, boson}}^{\rm EW} $ with the term  $ \mathcal{L}_{\textrm{v-v}}^{\textrm{EW}} $, which comprises pure-vector interactions . This sector can be conveniently divided into two parts as follows:
\begin{equation}\label{LEWvv}
\mathcal{L}_{\textrm{v-v}}^{\textrm{EW}}:={\cal L}^{(0,{\rm KK})}_{\textrm{v-v}}+{\cal L}^{({\rm KK})}_{\textrm{v-v}}\ ,
\end{equation}
where
\begin{eqnarray}
{\cal L}^{(0,{\rm KK})}_{\textrm{v-v}}&=&-\frac{1}{4}\left({\cal W}^{(\underline{0})\,i}_{\mu \nu}{\cal W}^{(\underline{0})\,i\, \mu \nu}+
B^{(\underline{0})}_{\mu \nu}B^{(\underline{0})\, \mu \nu}\right)\,\label{L0KKEW} ,\\
{\cal L}^{({\rm KK})}_{\textrm{v-v}}&=&-\frac{1}{4}\sum_{(\underline{m})}\left({\cal W}^{(\underline{m})\,i}_{\mu \nu}{\cal W}^{(\underline{m})\,i\, \mu \nu}+B^{(\underline{m})}_{\mu \nu}B^{(\underline{m})\, \mu \nu}\right)\, \label{LKKEW}.
\end{eqnarray}
In terms of $(W^{(\underline{0})\pm}_\mu ,Z^{(\underline{0})}_\mu, A^{(\underline{0})}_\mu  ,W^{(\underline{m})\pm }_\mu, Z^{(\underline{m})}_\mu, A^{(\underline{m})}_\mu)$ (see Eqs.~\eqref{Wpm} and~\eqref{AandZ}), the Lagrangian~\eqref{L0KKEW} becomes
\begin{equation}
{\cal L}^{(0,{\rm KK})}_{VV}=-\frac{1}{2}\hat{{\cal W}}^{(\underline{0})-}_{\mu \nu}\hat{{\cal W}}^{(\underline{0})+\mu \nu}-\frac{1}{4}\left(\hat{{\cal W}}^{(\underline{0})3}_{\mu \nu}\hat{{\cal W}}^{(\underline{0})3\mu \nu}+B^{(\underline{0})}_{\mu \nu}B^{(\underline{0})\mu \nu}\right)\, ,
\end{equation}
with
\begin{eqnarray}
\hat{{\cal W}}^{(\underline{0})+}_{\mu \nu}&=&\hat{W}^{(\underline{0})+}_{\mu \nu}+ig\sum_{(\underline{m})}\left(W^{(\underline{m})+}_\mu W^{(\underline{m})3}_\nu -W^{(\underline{m})+}_\nu W^{(\underline{m})3}_\mu \right)\, ,\\
\hat{{\cal W}}^{(\underline{0})3}_{\mu \nu}&=&\hat{W}^{(\underline{0})3}_{\mu \nu}+ig\sum_{(\underline{m})}\left(W^{(\underline{m})-}_\mu W^{(\underline{m})+}_\nu -W^{(\underline{m})+}_\mu W^{(\underline{m})-}_\nu \right) \, .
\end{eqnarray}
In these expressions,
\begin{eqnarray}
\hat{W}^{(\underline{0})+}_{\mu \nu}&=&W^{(\underline{0})+}_{\mu \nu}+ig\left(W^{(\underline{0})+}_\mu W^{(\underline{0})3}_\nu -W^{(\underline{0})+}_\nu W^{(\underline{0})3}_\mu \right)\, ,\\
\hat{ W}^{(\underline{0})3}_{\mu \nu}&=&W^{(\underline{0})3}_{\mu \nu}+ig\left(W^{(\underline{0})-}_\mu W^{(\underline{0})+}_\nu -W^{(\underline{0})+}_\mu W^{(\underline{0})-}_\nu \right) \, ,
\end{eqnarray}
and
\begin{subequations}
\begin{align}
W^{(\underline{0})\pm}_{\mu \nu} & =\partial_\mu W^{(\underline{0})\pm}_\nu-\partial_\nu W^{(\underline{0})\pm}_\mu\, ,\quad W^{(\underline{m})\pm}_{\mu \nu}=\partial_\mu W^{(\underline{m})\pm}_\nu-\partial_\nu W^{(\underline{m})\pm}_\mu\, ,\\
W^{(\underline{0})3}_{\mu \nu} & =\partial_\mu W^{(\underline{0})3}_\nu-\partial_\nu W^{(\underline{0})3}_\mu\, , \quad W^{(\underline{m})3}_{\mu \nu} =\partial_\mu W^{(\underline{m})3}_\nu-\partial_\nu W^{(\underline{m})3}_\mu\, .
\end{align}
\end{subequations}

On the other hand, the ${\cal L}^{({\rm KK})}_{\textrm{v-v}}$ term can be written as follows:
\begin{equation}
{\cal L}^{({\rm KK})}_{\textrm{v-v}}=-\frac{1}{2}\hat{{\cal W}}^{(\underline{m})-}_{\mu \nu}\hat{{\cal W}}^{(\underline{m})+\mu \nu}-\frac{1}{4}\hat{{\cal W}}^{(\underline{m})3}_{\mu \nu}\hat{{\cal W}}^{(\underline{m})3\mu \nu}-\frac{1}{4}B^{(\underline{m})}_{\mu \nu}B^{(\underline{m})\mu \nu}\, ,
\end{equation}
where
\begin{eqnarray}
\hat{{\cal W}}^{(\underline{m})+}_{\mu \nu}&=&\hat{W}^{(\underline{m})+}_{\mu \nu}-ig\sum_{(\underline{rs})}
\Delta_{(\underline{mrs})}\left(W^{(\underline{r})3}_\mu W^{(\underline{s})+}_\nu- W^{(\underline{s})3}_\nu W^{(\underline{r})+}_\mu\right)\, ,\\
\hat{{\cal W}}^{(\underline{m})3}_{\mu \nu}&=&\hat{W}^{(\underline{m})3}_{\mu \nu}-ig\sum_{(\underline{rs})}
\Delta_{(\underline{mrs})}\left(W^{(\underline{r})+}_\mu W^{(\underline{s})-}_\nu- W^{(\underline{s})+}_\nu W^{(\underline{r})-}_\mu\right)\, ,
\end{eqnarray}
with
\begin{eqnarray}
\hat{W}^{(\underline{m})+}_{\mu \nu}&=&W^{(\underline{m})+}_{\mu \nu}-ig\Big(W^{(\underline{0})3}_\mu W^{(\underline{m})+}_\nu -W^{(\underline{0})3}_\nu W^{(\underline{m})+}_\mu \nonumber \\
&&+W^{(\underline{m})3}_\mu W^{(\underline{0})+}_\nu-W^{(\underline{m})3}_\nu W^{(\underline{0})+}_\mu \Big)\, ,\\
\hat{W}^{(\underline{m})3}_{\mu \nu}&=&W^{(\underline{m})3}_{\mu \nu}-ig\Big(W^{(\underline{0})+}_\mu W^{(\underline{m})-}_\nu
-W^{(\underline{0})+}_\nu W^{(\underline{m})-}_\mu\nonumber \\
&&+W^{(\underline{m})+}_\mu W^{(\underline{0})-}_\nu-W^{(\underline{m})+}_\nu W^{(\underline{0})-}_\mu \Big) .
\end{eqnarray}
From the above expressions, any Feynman rule can be obtained.

\section{The fermionic sector}
\label{FS}

Before presenting the Currents and Yukawa sectors, it is necessary to discuss the spinor representation of the Lorentz group ${\rm SO}(1,3+n)$. We assume that $n$ is even, so chirality exists in this flat spacetime. Then, there exist $4+n$ matrices $\Gamma^M$ of dimension $2^{\frac{4+n}{2}}\times 2^{\frac{4+n}{2}}$, which transform as 1-tensors under ${\rm SO}(1,3+n)$ and satisfy the Clifford's algebra
\begin{equation}
\label{ca}
\{\Gamma^M, \, \Gamma^N\}=2g^{MN}\, .
\end{equation}
The generators of this representation are defined as
\begin{equation}
S^{MN}=\frac{i}{4}\left[\Gamma^M, \, \Gamma^N \right]\, .
\end{equation}
Since the space has even dimension, there exists an additional matrix given by
\begin{equation}
\Gamma^{5+n}=i^{ {2+n}\over 2}\Gamma^0\cdots \Gamma^3 \Gamma^5\cdots \Gamma^{4+n}\, ,
\end{equation}
which transforms as a 0-tensor under ${\rm SO}(1,3+n)$. Then, one can define the left-handed and right-handed projectors $P_{\mp}=(1\mp \Gamma^{5+n})/2$.\\

Since the compactification procedure involves a canonical map from a set of fields, which are covariant objects of the ${\rm SO}(1,3+n)$ group, to another set of fields, which are covariant objects of the standard ${\rm SO}(1,3)$ group, we need to introduce a particular representation of the $\Gamma^M$ matrices that allows us to decompose the spinors $\Psi(x,\bar{x})$ of  ${\rm SO}(1,3+n)$ into spinors $\psi(x,\bar{x})$ of  ${\rm SO}(1,3)$. In other words, we need to find a representation for the $\Gamma^M$ matrices in which the six generators $S^{\mu \nu}$ of ${\rm SO}(1,3+n)$ adopt a block-diagonal form, with each block being of size $4\times 4$. For concreteness, we demand the $2^{\frac{n}{2}}$ blocks to coincide with the standard generators of ${\rm SO}(1,3)$ given by $S^{\mu \nu}_4=(i/4)[\gamma^\mu, \, \gamma^\nu]$, with $\gamma^\mu$ denoting the standard Dirac's matrices. It is easy to see that $\Gamma^\mu$ block matrices with Dirac's matrices $\gamma^\mu$ on their block-diagonal satisfy this requirement, that is, we define these four matrices as follows:
\begin{equation}
\Gamma^\mu_{ab}=\delta_{ab}\gamma^\mu \,,
\end{equation}
where the indices $a$, $b$ run from $1$ to $2^{\frac{n}{2}}$. In other words, the block-components of the $\Gamma^\mu$ matrices are the $4\times 4$ null matrix (off the diagonal) and the $4\times 4$ $\gamma^\mu$ matrices (on the diagonal). To complete the Clifford's algebra, we need to define the remainder $n$ $\Gamma^{\bar{\mu}}$ matrices, which must also be made of covariant objects of ${\rm SO}(1,3)$. Since these matrices do not have Lorentz indices, they must be constructed out of the common $\gamma^5$ Dirac matrix. These $n$ matrices not only must serve to complete the Clifford's algebra given by Eq.(\ref{ca}), but also they must satisfy the Clifford's algebra of the orthogonal group ${\rm SO}(n)\subset {\rm SO}(1,4+n)$, given by
\begin{equation}
\{i\Gamma^{\bar{\mu}}, \, i\Gamma^{\bar{\nu}}\}=2\delta^{\bar{\mu}\bar{\nu}}\, .
\end{equation}
A set of matrices that satisfy all the above requirements are of the form
\begin{equation}
\Gamma^{\bar{\mu}}_{ab}=\Lambda^{\bar{\mu}}_{ab}\gamma^5\, ,
\end{equation}
where half of the total set of matrices $\Lambda^{\bar{\mu}}$, of dimension $2^{\frac{n}{2}}\times 2^{\frac{n}{2}}$, have components that can be either zeros or $\pm 1$, whereas the other half has entries that are either zeros or $\pm i$. In fact, those matrices $ \Gamma^{\bar{\mu}} $ with $\bar{\mu}$ odd are real and antisymmetric, and those with $\bar{\mu}$ even are purely imaginary and symmetric (see \ref{AA}). The $\Gamma^{5+n}$ matrix, which is also constructed out of $\gamma^5$, is
\begin{equation}
\Gamma^{5+n}_{ab}=(-1)^{1+a}\delta_{a b}\gamma^5\, .
\end{equation}
Thus, in this representation, the $P_{\mp}$ projectors are
\begin{subequations}
\begin{align}
P_ -  &= {\rm diag}({P_L},{P_R}, \cdots ,{P_L},{P_R})\, , \\
P_ + &= {\rm diag}({P_R},{P_L}, \cdots, {P_R}, {P_L})\, ,
\end{align}
\end{subequations}
where $P_L=(1-\gamma_5)/2$ and $P_R=(1+\gamma_5)/2$. In addition, the ${\rm SO}(1,3+n)$ spinors $\Psi(x,\bar{x})$ are made of $2^{\frac{n}{2}}$ spinors $\psi(x,\bar{x})$ of the standard group ${\rm SO}(1,3)$, as follows:
\begin{equation}
\Psi_{-}(x,\bar{x})=\left(\begin{array}{ccc}
 \psi_{L(1)} \\
  \psi_{R(2)} \\
\vdots \\
\psi_{L (2^{\frac{n}{2}}-1)}\\
\psi_{R (2^{\frac{n}{2}})}
\end{array}\right)(x,\bar{x}) \, , \, \, \, \, \,
\Psi_{+}(x,\bar{x})=\left(\begin{array}{ccc}
 \psi_{R(1)} \\
  \psi_{L (2)} \\
\vdots \\
\psi_{R(2^{\frac{n}{2}}-1)}\\
\psi_{L(2^{\frac{n}{2}})}
\end{array}\right)(x,\bar{x}) \, ,
\end{equation}
where $\psi_{L(a)}$ and $\psi_{R(a)}$ stand for left-handed and right-handed spinors of ${\rm SO}(1,3)$.\\

At the level of the groups ${\rm SO}(1,3+n)$ and $G_{\rm SM}({\cal M}^d)$, right-handed singlets of $SU_L(2,{\cal M}^{d})$ will be denoted by $\hat e_{+}(x, \bar x)$ (charged lepton), $\hat d_{+}(x, \bar x)$ (quark of down type), and $\hat u_{+}(x, \bar x)$ (quark of up type). In this notation, the hat stands for a spinor of ${\rm SO}(1,3+n)$. Similarly, lepton or quark left-handed doublets of $SU_L(2,{\cal M}^{d})$ will be denoted by
\begin{equation}
 { \hat L}_{-}(x, \bar x)  = \left( {\begin{array}{*{20}{c}}
{{ \hat {\nu} }_{ -} }\\
{{\hat e}_{-} }
\end{array}} \right)(x, \bar x) \, , \, \, \, \, \, \, \, { \hat Q}_{-}(x, \bar x)  = \left( {\begin{array}{*{20}{c}}
{{ \hat {u} }_{ -} }\\
{{\hat d}_{-} }
\end{array}} \right)(x, \bar x) \, .
\end{equation}

\subsection{The Yukawa sector}
We now proceed to discuss the Yukawa sector. We propose a Yukawa Lagrangian, governed by the extended groups ${\rm ISO}(1,3+n)$ and $G_{\rm SM}({\cal M}^d)$, as follows:
\begin{eqnarray}
{\cal L}_{(4+n)}^{\rm Y} (x,\bar x) &=&  -  Y_{ {\mbox{\tiny (4+$n$)} } ij }^ { \ell   } { {\bar  {\hat L} }}_{i -} (x,\bar x) \Phi  (x,\bar x) {\hat e}_{j  +} (x,\bar x)  + {\rm H.c.} \nonumber \\
&\,& -  {Y}_{{\mbox{\tiny (4+$n$)} } ij}^{d} {\bar {\hat  Q}_{i -}} (x,\bar x) \Phi  (x,\bar x) {{ \hat d}_{j  +}} (x,\bar x)  + {\rm H.c.} \nonumber \\
&\,&-  Y_{{\mbox{\tiny (4+$n$)} }  ij}^{u} {\bar {\hat Q}_ {i  -}} (x,\bar x) {\tilde  \Phi} (x,\bar x) {{ \hat u}_{j  +}} (x,\bar x)  + {\rm H.c.}  \label{LagYukDdim} \, ,
\end{eqnarray}
where $i$ and $j$ represent flavor indices. Notice that, at this level, the Yukawa matrices have canonical dimension equal to $-\frac{n}{2}$, whereas fermion and scalar fields have dimension $\frac{n+3}{2}$ and $\frac{n+2}{2}$, respectively. Hereinafter, we will adopt the following notation: indices $a,b,\cdots$ will run from $1$ to $2^{\frac{n}{2}}$, indices labeled with a hat, $\hat{a},\hat{b}, \cdots$ will run only on even integers $2,4, \cdots, 2^{\frac{n}{2}}$, and indices labeled with a bar, $\bar{a},\bar{b},\cdots$ will run only on odd integers $1,3, \cdots, 2^{\frac{n}{2}}-1$. Also, as in the case of flavor indices, an implicit sum on repeated indices of these three types will be assumed.\\

Once the point transformation that maps spinors of ${\rm SO}(1,3+n)$ into spinors of ${\rm SO}(1,3)$ is carried out, the above Lagrangian becomes
\begin{equation}
{\cal L}_{(4+n)}^{\rm Y} (x,\bar x)={\cal L}_{ (4+n) }^{{\rm Y} \ell} (x,\bar x)+{\cal L}_{ (4+n) }^{{\rm Y} q} (x,\bar x)\, ,
\end{equation}
where
\begin{eqnarray}
{\cal L}_{ (4+n) }^{{\rm Y} \ell} (x,\bar x)&=&-  Y_{ {\mbox{\tiny (4+$n$)} } ij }^ { \ell   }    \Big [  { {\bar  { L} }}_{i L(\bar a)} (x,\bar x) \Phi  (x,\bar x) { e}_{jR(\bar a)} (x,\bar x)\nonumber \\
  &&+  { {\bar  { L} }}_{i R(\hat a)} (x,\bar x) \Phi  (x,\bar x) { e}_{j  L(\hat a)} (x,\bar x)  \Big]+ {\rm H.\ c.}\, ,
\end{eqnarray}

\begin{eqnarray}
 {\cal L}_{ (4+n) }^{{\rm Y} q} (x,\bar x)&=&  -  {Y}_{{\mbox{\tiny (4+$n$)} } ij}^{d} \Big[ { \bar {  Q}_{i L(\bar a)}} (x,\bar x) \Phi  (x,\bar x) {{ d}_{j R(\bar a)}} (x,\bar x) \nonumber \\
 &&+  {\bar {  Q}_{i R(\hat a)}} (x,\bar x) \Phi  (x,\bar x) {{  d}_{j  L(\hat a)}} (x,\bar x)\Big]  + {\rm H.\ c.} \nonumber \\
&\,&-  Y_{{\mbox{\tiny (4+$n$)} }  ij}^{u}   \Big[  {\bar { Q}_ {i  L(\bar a)}} (x,\bar x) {\tilde  \Phi} (x,\bar x) {{  u}_{j  R(\bar a)}} (x,\bar x)\nonumber \\
&& + {\bar { Q}_ {i  R(\hat a)}} (x,\bar x) {\tilde  \Phi} (x,\bar x) {{  u}_{j  L(\hat a)}} (x,\bar x) \Big]  + {\rm H.\ c.}
\end{eqnarray}
We now assume that the different left-handed and right-handed singlets, $f(x,\bar{x})$, and doublets, $F(x,\bar{x})$, of the electroweak group, are periodic functions with respect to the compact coordinates.
In order to reproduce the SM Lagrangian in the limit of a very small size of the compact manifold, a definite parity for the different spinor fields appearing in these Lagrangians must be assumed. This means that the assumed parity pattern must lead to a mass spectrum consistent with the decoupling theorem. A parity pattern that leads to an admissible mass spectrum is the following: the doublets $F_{iL(\bar{a})}$ and singlets $f_{iR(\bar{a})}$ are assumed to be even, whereas the doublets $F_{iR(\hat{a})}$ and singlets $f_{iL(\hat{a})}$ are assumed to be odd, that is,
\begin{subequations}
\begin{align}
F _{iL(1)}(x,\bar x) &= f_E^{(\underline 0)} F _{i L(1) }^{(\underline 0)}( x) + \sum\limits_{(\underline m)} f_E^{(\underline m)} (\bar{p}\cdot \bar{x})F _{i L(1) }^{(\underline m)} (x) ,  \\
 F_{iR(\hat a)}(x, \bar x) &=\sum\limits_{(\underline m)}f_O^{(\underline m)} (\bar{p}\cdot \bar{x})F _{i R(\hat a ) }^{(\underline m)} (x),    \\
 F _{iL(\bar a)}(x, \bar x) &=\sum\limits_{(\underline m)}f_E^{(\underline m)} (\bar{p}\cdot \bar{x}) F_{i L(\bar a) }^{(\underline m)} (x),   \hspace{3mm}  \bar a \ge 3 \, ,
\end{align}
\end{subequations}
\begin{subequations}
\begin{align}
f_{iR(1)}(x,\bar x) &= f_E^{(\underline 0)} f_{i R(1) }^{(\underline0)} (x)+ \sum\limits_{(\underline m)}f_E^{(\underline m)}(\bar{p}\cdot \bar{x})  f _{i R(1) }^{ (\underline m)}  (x)\\
f_{iL(\hat a)}(x, \bar x)& =\sum\limits_{(\underline m)} f_O^{(\underline m)} (\bar{p}\cdot \bar{x}) f_{i L(\hat a) }^{ (\underline m)} (x), \hspace{3mm}   \\
f_{iR(\bar a)} (x, \bar x) &=  \sum\limits_{(\underline m)} f_E^{(\underline m)}(\bar{p}\cdot \bar{x}) f_{i R(\bar a) }^{(\underline m)},  \hspace{3mm}  \bar a \ge 3 \, .
\end{align}
\end{subequations}
This map, which can be elevated into a canonical transformation at the phase-space level~\cite{OPT2}, causes, together with the integration of the compact extra dimensions at the action level, the hiding of the extended $\{{\rm ISO}(1,3+n),\, G_{\rm SM}({\cal M}^d)\}$ symmetries into the standard  $\{{\rm ISO}(1,3),\, G_{\rm SM}({\cal M}^4)\}$ groups.  Once the compact coordinates are integrated out, one obtains
\begin{equation}
{\cal L}^{\rm Y}_{\rm EDSM}={\cal L}_{\rm EDSM}^{{\rm Y} \ell}+{\cal L}_{\rm EDSM}^{{\rm Y}q} \, ,
\end{equation}
where
\begin{eqnarray}
{\cal L}_{\rm EDSM}^{{\rm Y}\ell}  (x)&=& -  Y_{ij }^ { \ell   }    \Big \{  {\bar L}_{i L(1)}^{(\underline 0)}   \Phi^{(\underline 0)}  e_{jR(1)}^{(\underline 0)} +\sum\limits_{(\underline m)}    \Big[  \Big ( {\bar L}_{i L(\bar a)}^{(\underline m)}  e_{jR(\bar a)}^{(\underline m)}+{\bar L}_{i R(\hat a)}^{(\underline m)} e_{j  L(\hat a)}^{(\underline m)}  \Big) \Phi^{(\underline 0)}  \nonumber \\
&&+  \left( {\bar L }_{i L(1)}^{(\underline 0)}  e_{jR(1)}^{(\underline m)} +{\bar L }_{i L(1)}^{(\underline m)} e_{jR(1)}^{(\underline 0)}  \right) \Phi^{(\underline m)}   \nonumber  \\
&\,& + \sum\limits_{ (\underline s \underline r)}  \Big( {\Delta }_{(\underline m \underline s \underline r)}   {\bar L}_{i L(\bar a)}^{(\underline m)}  e_{jR(\bar a)}^{(\underline s)}+{\Delta' }_{(\underline m \underline s \underline r)}  {\bar L}_{i R(\hat a)}^{(\underline m)}e_{j  L(\hat a)}^{(\underline s)} \Big )\Phi^{(\underline r)}   \Big] \Big \}\nonumber \\
&&+ {\rm H.\ c.} \, ,
\end{eqnarray}
and
\begin{eqnarray}
{\cal L}_{\rm EDSM}^{{\rm Y}q}   (x) &=&  -  Y_{ij }^ { d   }    \Big \{   {\bar Q}_{i L(1)}^{(\underline 0)}   \Phi^{(\underline 0)} d_{jR(1)}^{(\underline 0)} + \sum\limits_{(\underline m)}  \Big[ \Big ( {\bar Q}_{i L(\bar a)}^{(\underline m)}  d_{jR(\bar a)}^{(\underline m)} + {\bar Q}_{i R(\hat a)}^{(\underline m)} d_{j  L(\hat a)}^{(\underline m)}  \Big )  \Phi^{(\underline 0)} \nonumber \\
  &&+  \left( {\bar Q }_{i L(1)}^{(\underline 0)}  d_{jR(1)}^{(\underline m)} +{\bar Q }_{i L(1)}^{(\underline m)} d_{jR(1)}^{(\underline 0)}  \right) \Phi^{(\underline m)}   \nonumber  \\
&\,& + \sum\limits_{ (\underline s \underline r)}  \Big ( {\Delta }_{(\underline m  \underline s \underline r)}   {\bar Q}_{i L(\bar a)}^{(\underline m)}  d_{jR(\bar a)}^{(\underline s)}+{\Delta' }_{(\underline m \underline s \underline r)}   {\bar Q}_{i R(\hat a)}^{(\underline m)} d_{j  L(\hat a)}^{(\underline s)} \Big ) \Phi^{(\underline r)}  \Big]  \Big \}+ {\rm H.c.} \nonumber \\
&\,&-  Y_{ij }^ { u   }    \Big \{   {\bar Q}_{i L(1)}^{(\underline 0)}   {\tilde  \Phi} ^{(\underline 0)}  u_{jR(1)}^{(\underline 0)} + \sum\limits_{(\underline m)}  \Big[ \Big ( {\bar Q}_{i L(\bar a)}^{(\underline m)}  u_{jR(\bar a)}^{(\underline m)} +  {\bar Q}_{i R(\hat a)}^{(\underline m)} u_{j  L(\hat a)}^{(\underline m)} \Big )  {\tilde  \Phi} ^{(\underline 0)}\nonumber \\
  &&+ \left( {\bar Q }_{i L(1)}^{(\underline 0)}  u_{jR(1)}^{(\underline m)} +{\bar Q }_{i L(1)}^{(\underline m)} u_{jR(1)}^{(\underline 0)}  \right) {\tilde  \Phi} ^{(\underline m)} \nonumber  \\
&\,& + \sum\limits_{ (\underline s  \underline r)}  \Big( {\Delta }_{(\underline m   \underline s  \underline r)}   {\bar Q}_{i L(\bar a)}^{(\underline m)}  u_{jR(\bar a)}^{(\underline s)}+ {\Delta' }_{(\underline m  \underline s  \underline r)}{\bar Q}_{i R(\hat a)}^{(\underline m)} u_{j  L(\hat a)}^{(\underline s)} \Big ){\tilde  \Phi} ^{(\underline r)}  \Big ] \Big \}\nonumber \\
&& + {\rm H.\ c.}
\end{eqnarray}
In the above expressions,
\begin{equation}
Y^{\ell, d,u}=    {\textstyle{{Y_{ {\mbox{\tiny (4+$n$)} } }^{\ell, d,u}  } \over  {\sqrt {\prod\limits_{\alpha= 1}^n {  (2\pi {R_\alpha} ) } } }   }} \, .
\end{equation}

\subsection{The Currents sector}
The Lorentz and gauge structure of the Currents sector $ \mathcal{L}_{4+n}^{\rm C}(x,\bar{x}) $ is dictated by the extended groups ${\rm SO}(1,3+n)$ and $G_{\rm SM}({\cal M}^d)$, so that the $(4+n)$-dimensional counterpart of the SM version is assumed to be
\begin{eqnarray}
{\cal L}_{(4+n)}^{\rm C}  (x,\bar x)&=&   i   {{\bar  {\hat L}}_{i-}}{\Gamma ^M}{D_M}{{ \hat L}_{i-}}   +  i { {\bar {\hat e}}_{i+}}{\Gamma ^M} {{D_M} { \hat e}_{i +}}       \nonumber \\
&\,& +\, i  {{\bar  {\hat Q}}_{i-}}{\Gamma ^M}{D_M}{ {\hat Q}_{i-}} +  i{{\bar{ \hat u}}_{i+}}{\Gamma ^M}{D_M} { {\hat u}_{i+}}    + i {{\bar  {\hat d}}_{i+}}{\Gamma ^M}{D_M}{{\hat d}_{i+}}      \label{LagCorr4masN}\, ,
\end{eqnarray}
where $D_M$ is the covariant derivative associated with the extended gauge group; it is given by
\begin{equation}
 {D_M } = {\partial _M }-ig_{s\mbox{\tiny (4+$n$)}}\frac{\lambda^a}{2}{\cal G}^a_M - ig_{\mbox{\tiny (4+$n$)}} {\textstyle{{{\sigma ^i}} \over 2}}{\cal W}_M ^i - i{g'_{\mbox{\tiny (4+$n$)}}} {\textstyle{Y \over 2}} {\cal B}_M \, .
\end{equation}
Once performed the canonical map from ${\rm SO}(1,3+n)$ to ${\rm SO}(1,3)$, one obtains
\begin{equation}
{\cal L}_{(4+n)}^{\rm C}   (x,\bar x)=  {\cal L}_{(4+n)}^ {\rm VC}(x,\bar x) + {\cal L}_{(4+n)}^ {\rm SC}(x,\bar x)\, ,
\end{equation}
where
\begin{eqnarray}
{\cal L}_{(4+n)}^ {\rm VC}   (x,\bar x) &=&   i  {{\bar  L}_{iL(\bar a)}} {\gamma ^\mu }   {D_\mu } {{ L}_{iL(\bar a)}}   +   i  { {\bar {e}}_{iR(\bar a)}} {\gamma ^\mu }   {D_\mu }  { e}_{i R(\bar a)}   \nonumber \\
&\,&   +\,  i  {{\bar  L}_{iR(\hat a)}} {\gamma ^\mu }   {D_\mu } {{ L}_{iR(\hat a)}} +   i { {\bar {e}}_{iL(\hat a)}} {\gamma ^\mu }   {D_\mu }  { e}_{i L(\hat a)}     \nonumber \\
&\,& +\, i  {{\bar  Q}_{iL(\bar a)}} {\gamma ^\mu }   {D_\mu } { Q_{iL(\bar a)}}\nonumber \\
&&+  i  {{\bar  d}_{iR(\bar a)}} {\gamma ^\mu }   {D_\mu } {d_{iR(\bar a)}} + i  {{\bar u}_{iR(\bar a)}} {\gamma ^\mu }   {D_\mu } { u_{iR(\bar a)}}   \nonumber\\
&\,& +\,   i {{\bar  Q}_{iR(\hat a)}} {\gamma ^\mu }   {D_\mu } { Q_{iR(\hat a)}}\nonumber \\
&&+ i {{\bar  d}_{iL(\hat a)}} {\gamma ^\mu }   {D_\mu } {d_{iL(\hat a)}}  +  i {{\bar u}_{iL(\hat a)}} {\gamma ^\mu }   {D_\mu }  { u_{iL(\hat a)}}\, ,
\end{eqnarray}
and
\begin{eqnarray}
{\cal L}_{(4+n)}^ {\rm SC}   (x,\bar x)&=&   i \Lambda_{\bar a\, \hat b}^{\bar \mu}\Big( {{\bar  L}_{iL(\bar a)}}{D_{\bar \mu}}{{ L}_{iR(\hat b)}}   -   { {\bar {e}}_{iR(\bar a)}} {{D_{\bar \mu}} { e}_{i L(\hat b )}}  \Big)  \nonumber \\
&\,&  -\, i \Lambda_{\hat b\, \bar a}^{\bar \mu}\Big( {{\bar  L}_{iR(\hat b)}}{D_{\bar \mu}}{{ L}_{iL(\bar a)}}         - { {\bar {e}}_{iL(\hat b)}} {{D_{\bar \mu}} { e}_{i R(\bar a)}} \Big) \nonumber \\
&\,&  +\,     i\Lambda_{\bar a\, \hat b}^{\bar \mu} \Big( {{\bar  Q}_{iL(\bar a)}}{D_{\bar \mu}}{ Q_{iR(\hat b)}}\nonumber \\
&&- {{\bar  d}_{iR(\bar a)}}{D_{\bar \mu}}{d_{iL(\hat b)}}  - {{\bar u}_{iR(\bar a)}}{D_{\bar \mu}} { u_{iL(\hat b)}}  \Big) \nonumber \\
&\,&-\, i \Lambda_{\hat b\, \bar a}^{\bar \mu}\Big(  {{\bar  Q}_{iR(\hat b)}}{D_{\bar \mu}}{ Q_{iL(\bar a)}}\nonumber \\
&&- {{\bar  d}_{iL(\hat b)}}{D_{\bar \mu}}{d_{iR(\bar a)}}      -{{\bar u}_{iL(\hat b)}}{D_{\bar \mu}} { u_{iR(\bar a)}}     \Big)  .
\end{eqnarray}
Then, after integrating out the compact coordinates, one obtains
\begin{equation}
{\cal L}^{\rm C}_{\rm EDSM} ={\cal L}^{\rm  VC }_{\rm EDSM} +{\cal L}^{\rm SC }_{\rm EDSM}\, ,
\end{equation}
where
\begin{eqnarray}
\label{FVC}
{\cal L}^{\rm VC}_{\rm EDSM}  (x)&=&i \sum\limits_{F=L,Q}  \Big \{  {\bar F}_{i L(1) }^{(\underline 0)} {\gamma ^\mu } {\left( {{D_\mu }   F_{i L(1) }}    \right)^{(\underline 0)}}\nonumber \\
&&+   \sum\limits_{(\underline m)}  \Big[    {{\bar  F}_{iL(\bar a)}} ^{(\underline m)}  {\gamma ^\mu }    \left( {{D_\mu } { F}_{iL(\bar a)} } \right)^{(\underline m)} +    {{\bar  F}_{iR(\hat a )}}^{(\underline m)} {\gamma ^\mu }  \left( {D_\mu } {{ F}_{iR(\hat a)}} \right)^{(\underline m)}   \Big ] \Big \} \nonumber \\
&\,&+\,i \sum\limits_{f=e,u,d} \Big \{  f_{i R(1) }^{(\underline 0)} {\gamma ^\mu } {\left( {{D_\mu }   f_{i R(1) } }    \right)^{(\underline 0)}}\nonumber  \\
&&+   \sum\limits_{(\underline m)}  \Big[  { {\bar {f}}_{iR(\bar a)}^{(\underline m)} } {\gamma ^\mu } \left( {D_\mu }  { f}_{i R(\bar a)} \right)^{(\underline m)}        +    {\bar {f}}_{iL(\hat a)}^{(\underline m)} {\gamma ^\mu }  \left(  {D_\mu }  { f}_{i L(\hat a)}  \right)^{(\underline m)} \Big ]  \Big \}\, ,   \nonumber \\
\end{eqnarray}
and
\begin{eqnarray}
\label{FSC}
{\cal L}^{\rm SC}_{\rm EDSM}  (x)&=& i  \sum\limits_{F=L,Q}  \Big \{ \Lambda_{1\, \hat b}^{\bar \mu}{\bar  F}_{iL(1)}^{(\underline 0)} \big({ {D_{\bar \mu}}{ F}_{iR(\hat b)} }\big) ^{(\underline 0)}\nonumber \\
&&+   \sum\limits_{(\underline m)}  \Big[ \Lambda_{\bar a\, \hat b}^{\bar \mu}    {\bar  F}_{iL(\bar a) }^{(\underline m)} \big( {D_{\bar \mu}}{{ F}_{iR(\hat b)}} \big)^{(\underline m)}    -   \Lambda_{\hat b\, \bar a}^{\bar \mu}  {\bar  F}_{iR(\hat b)}^{(\underline m)} \left({D_{\bar \mu}}{{ F}_{iL(\bar a)}}   \right)^{(\underline m)}  \Big ] \Big \}   \nonumber \\
&\,&-\, i \sum\limits_{f=e,u,d} \Big \{     \Lambda_{1\, \hat b}^{\bar \mu}   {\bar {f}}_{iR(1)}^{(\underline 0)} \big ( {{D_{\bar \mu}} { f}_{i L(\hat b )}} \big )^{(\underline 0)}\nonumber \\
&&+  \sum\limits_{(\underline m)}  \Big[  \Lambda_{\bar a\, \hat b}^{\bar \mu}   {\bar {f}}_{iR(\bar a)}^{(\underline m)} \big( {{D_{\bar \mu}} { f}_{i L(\hat b )}} \big)^{(\underline m)}   -  \Lambda_{\hat b\, \bar a}^{\bar \mu}{\bar {f}}_{iL(\hat b)}^{(\underline m)}  \left( {{D_{\bar \mu}} { f}_{i R(\bar a)}} \right )^{(\underline m)}  \Big ] \Big \}\, . \nonumber \\
\end{eqnarray}
The different covariant objects appearing in these expressions are given in~\ref{AB}.

\subsection{Mass spectrum}
The mass spectrum of the fermionic sector emerges from both the Yukawa and the Currents sectors. After spontaneous symmetry breaking, the zero modes and the KK excitations of quarks and leptons are endowed with mass terms that are proportional to the Fermi scale. The diagonalization of the standard sector allows us to pass from the gauge basis to the flavor basis, which is achieved through the well-known unitary transformations
\begin{eqnarray}
N^{(\underline{0})}_L&=&U^l_LN'^{(\underline{0})}_L \, , \\
E^{(\underline{0})}_L&=&U^l_LE'^{(\underline{0})}_L \, ,  \ \ \ \ \ E^{(\underline{0})}_R=U^l_RE'^{(\underline{0})}_R \, , \\
U^{(\underline{0})}_L&=&U^u_LU'^{(\underline{0})}_L \, ,  \ \ \ \ \ U^{(\underline{0})}_R=U^u_RU'^{(\underline{0})}_R \, , \\
D^{(\underline{0})}_L&=&U^d_LD'^{(\underline{0})}_L \, ,  \ \ \ \ \ D^{(\underline{0})}_R=U^d_RD'^{(\underline{0})}_R \, ,
\end{eqnarray}
where $N_L$, $E_{L,R}$, $U_{L,R}$ and $D_{L,R}$ are vectors in the flavor space. The corresponding transformation laws for the KK excitations arise as a consequence of the fact that this flavor diagonalization commutes with the compactification, that is, one can first carry out the flavor space diagonalization and then implement compactification, or the other way around. In other words, we assume the flavor space transformation to be $ \bar{x}$-independent. This means that the $\hat{N}_{-}$, $\hat{E}_{-,+}$, $\hat{U}_{-,+}$ and $\hat{D}_{-,+}$ flavor vectors must satisfy the same transformation relations given above for the zero modes. From this fact we have, for the KK excitations of quarks, that
\begin{eqnarray}
U^{(\underline{m})}_{\bar{L}}&=&U^u_L U'^{(\underline{m})}_{\bar{L}}\, , \ \ \ \ \ U^{(\underline{m})}_{\bar{R}}=U^u_R U'^{(\underline{m})}_{\bar{R}}\, ,\\
U^{(\underline{m})}_{\hat{R}}&=&U^u_L U'^{(\underline{m})}_{\hat{R}}\, , \ \ \ \ \ U^{(\underline{m})}_{\hat{L}}=U^u_R U'^{(\underline{m})}_{\hat{L}}\, ,
\end{eqnarray}
\begin{eqnarray}
D^{(\underline{m})}_{\bar{L}}&=&U^d_L D'^{(\underline{m})}_{\bar{L}}\, , \ \ \ \ \ D^{(\underline{m})}_{\bar{R}}=U^d_R D'^{(\underline{m})}_{\bar{R}}\, ,\\
D^{(\underline{m})}_{\hat{R}}&=&U^d_L D'^{(\underline{m})}_{\hat{R}}\, , \ \ \ \ \ D^{(\underline{m})}_{\hat{L}}=U^d_R D'^{(\underline{m})}_{\hat{L}}\, ,
\end{eqnarray}
where we have introduced the notation
\begin{equation}
 U^{(\underline{m})}_{\bar{L}}= \left( {\begin{array}{*{20}{c}}
U^{(\underline{m})}_{L(1)}\\
\vdots \\
U^{(\underline{m})}_{L (2^{\frac{n}{2}}-1)}
\end{array}} \right)\, , \, \, \, \, \, \, \, U^{(\underline{m})}_{\hat{L}}= \left( {\begin{array}{*{20}{c}}
U^{(\underline{m})}_{L(2)}\\
\vdots \\
U^{(\underline{m})}_{L(2^{\frac{n}{2}})}
\end{array}} \right) \, .
\end{equation}
Similar expressions are used for all the other quantities involved. In the lepton sector, the corresponding unitary transformations are given by
\begin{eqnarray}
N^{(\underline{m})}_{\bar{L}}&=&U^l_L N'^{(\underline{m})}_{\bar{L}}\, ,\\
N^{(\underline{m})}_{\hat{R}}&=&U^l_L N'^{(\underline{m})}_{\hat{R}}\, ,
\end{eqnarray}
\begin{eqnarray}
E^{(\underline{m})}_{\bar{L}}&=&U^l_L E'^{(\underline{m})}_{\bar{L}}\, , \ \ \ \ \ E^{(\underline{m})}_{\bar{R}}=U^l_R E'^{(\underline{m})}_{\bar{R}}\, ,\\
E^{(\underline{m})}_{\hat{R}}&=&U^l_L E'^{(\underline{m})}_{\hat{R}}\, , \ \ \ \ \ E^{(\underline{m})}_{\hat{L}}=U^l_R E'^{(\underline{m})}_{\hat{L}}\, .
\end{eqnarray}
The especial way in which neutrinos are transformed arises as consequence of the fact that there are no right-handed neutrinos $\hat{\nu}_+(x,\bar{x})$. Notice that we can define vectors of dimension $2^{\frac{n}{2}}$ as follows:
\begin{equation}
 F^{(\underline{m})}_{VL,VR}= \left( {\begin{array}{*{20}{c}}
F^{(\underline{m})}_{\hat{L},\hat{R}}\\
\ \\
F^{(\underline{m})}_{\bar{L},\bar{R}}
\end{array}} \right)\, ,
\end{equation}
where $F=U,D,E$. Using this notation, we can write the mass term for the KK excitations as
\begin{equation}
\label{MMF}
{\cal L}^{\rm KK}_{{\rm Mass}}=-\sum_{F=U,D,E,N}\sum_{(\underline{m})}\bar{F}'^{(\underline{m})}_{VL}\mathrm{M}_FF'^{(\underline{m})}_{VR}+{\rm H.\, c.\,} ,
\end{equation}
where
\begin{equation}
 M_F=\left(\begin{array}{ccc}
M_{F^{(\underline{0})}} & i\Lambda_{(\underline{m})} \\
\ \ \ & \ \ \ \\
 i \Lambda^\dag_{(\underline{m})} & M_{F^{(\underline{0})}}
\end{array}\right)\, , \ \ \ \ \ F=U,D,E\, ,
\end{equation}
and
\begin{equation}
 M_N=\left(\begin{array}{ccc}
0 & 0 \\
\ \ \ & \ \ \ \\
 i \Lambda_{(\underline m)}^\dag & 0
\end{array}\right)\, .
\end{equation}
In these expressions $\Lambda_{(\underline{m})}=p_{\bar{\mu}}^{(\underline m)} \Theta^{\bar{\mu}}$ is a matrix of dimension $(2^{\frac{n}{2}}/2) \times (2^{\frac{n}{2}}/2)$, $\Theta^{\bar{\mu}}$ is defined in \ref{AA}, and the $M_{F^{(\underline{0})}}$ terms are the standard diagonal flavor matrices of dimension $3\times 3$. For instance, $M_{U^{(\underline{0})}}={\rm diag}(m_{u^{(\underline{0})}},m_{c^{(\underline{0})}},m_{t^{(\underline{0})}})$. Here, it is understood that $M_{F^{(\underline{0})}}$ and $\Lambda_{(\underline{m})}=p_{\bar{\mu}}^{(\underline m)} \Theta^{\bar{\mu}}$ are multiplied by the identity matrices of dimension $(2^{\frac{n}{2}}/2) \times (2^{\frac{n}{2}}/2)$, denoted by $\tilde{\mathbf{1}}$, and $3\times 3$, respectively. Notice that the contributions to (\ref{MMF}) that are proportional to $M_{F^{(\underline{0})}}$ arise from the Yukawa sector, whereas those contributions which are proportional to $\Lambda_{(\underline m)}$ emerge from the Currents sector. It should be noted that while all massive fermions are characterized by $2^{\frac{n}{2}}$ fermionic towers of KK excitations, a neutrino is described only by $2^{\frac{n}{2}}/2$ towers. This is due to the absence of right-handed spinors $\hat{\nu}_+$ in the theory governed by the extended groups $\{{\rm ISO}(1,3+n),\, G_{SM}({\cal M}^d)\}$.

The diagonalization of the mass matrices in Eq.~(\ref{MMF}) is simple. In fact, notice that
\begin{equation}
M_FM^\dag_F=\left(\begin{array}{ccc}
M^2_{F^{(\underline{m})}} & 0\\
\ \ \ & \ \ \ \\
0 & M^2_{F^{(\underline{m})}}
\end{array}\right)\, , \ \ \ \ \ F=U,D,E\, ,
\end{equation}
where $M^2_{F^{(\underline{m})}}=M^2_{F^{(\underline{0})}}+m^2_{(\underline{m})}$. In the case of the neutrino mass matrix, we have
\begin{equation}
M_NM^\dag_N=\left(\begin{array}{ccc}
0 & 0\\
\ \ \ & \ \ \ \\
0 & M^2_{N^{(\underline{m})}}
\end{array}\right)\, ,
\end{equation}
where $M^2_{N^{(\underline{m})}}=m^2_{(\underline{m})}$. In the above expressions we have used the Clifford's algebra satisfied by the $\Gamma^{\bar{\mu}}$ matrices to show that $\Lambda_{(\underline{m})} \Lambda^\dag_{(\underline{m})}=\Lambda^\dag_{(\underline{m})} \Lambda_{(\underline{m})}=m^2_{(\underline{m})}$. From these results, it is immediate to conclude that the mass matrices $M_F$ ($F=U,D,E$) are diagonalized by the matrices
 \begin{eqnarray}
V_F^{(\underline{m})}&=&\frac{1}{M_{F^{(\underline{m})}}} M^\dag_F \nonumber \\
 &=&\frac{M_{F^{(\underline{0})}}}{M_{F^{(\underline{m})}}}-\frac{i\Omega^{(\underline m)}}{M_{F^{(\underline{m})}}}\, ,
 \end{eqnarray}
where
\begin{equation}
\Omega^{(\underline m)}=\left(\begin{array}{ccc}
0 & \Lambda_{(\underline{m})} \\
\ \ \ & \ \ \ \\
\Lambda^\dag_{(\underline{m})} & 0
\end{array}\right)\, .
\end{equation}
As far as the mass matrix of neutrinos is concerned, it is diagonalized by the matrix
\begin{equation}
V_N^{(\underline{m})}=-\frac{i\Omega^{(\underline m)}}{M_{N^{(\underline{m})}}}\, .
\end{equation}
There are two possible paths to go from the flavor basis to the KK-mass basis. One way consists in performing the transformations
\begin{eqnarray}
F'^{(\underline{m})}_{V\,L}&=&F''^{(\underline{m})}_{V\,L}\, , \ \ \ \ \  F'^{(\underline{m})}_{V\,R}=V_F^{(\underline{m})} F''^{(\underline{m})}_{V\,R}\, , \\
N'^{(\underline{m})}_{V\,L}&=&N''^{(\underline{m})}_{V\,L}\, , \ \ \ \ \  N'^{(\underline{m})}_{V\,R}=V_N^{(\underline{m})} N''^{(\underline{m})}_{V\,R}\, ,
\end{eqnarray}
where the subindex $V$ stands for a vector in the space of dimension $2^{\frac{n}{2}}$ in which the $V_F^{(\underline{m})}$ and $V_N^{(\underline{m})}$ unitary matrices are defined. Each component of these vectors corresponds to a vector of the flavor space. The other possible way consists in making the transformation
\begin{eqnarray}
F'^{(\underline{m})}_{V\,L}&=&V_F^{(\underline{m})\, \dag}F''^{(\underline{m})}_{V\,L}\, , \ \ \ \ \  F'^{(\underline{m})}_{V\,R}=F''^{(\underline{m})}_{V\,R}\, , \\
N'^{(\underline{m})}_{V\,L}&=&V_N^{(\underline{m})\, \dag}N''^{(\underline{m})}_{V\,L}\, , \ \ \ \ \  N'^{(\underline{m})}_{V\,R}=N''^{(\underline{m})}_{V\,R}\, .
\end{eqnarray}
However, these two types of transformations are equivalent because one can go from one to another through an unitary transformation given by the matrices $V_F^{(\underline{m})}$ and $V_N^{(\underline{m})}$. We will adopt the first type of transformation. In the KK-mass basis,
\begin{equation}
F''^{(\underline{m})}_{V\,L,R}= \left( {\begin{array}{*{20}{c}}
F''^{(\underline{m})}_{L,R (1)}\\
F''^{(\underline{m})}_{L,R (2)} \\
\vdots \\
F''^{(\underline{m})}_{L,R (2^{\frac{n}{2}})}
\end{array}} \right)\, , \ \ \ \ \ F=U,D,E \, ,
\end{equation}
so that each flavor has associated $2^{\frac{n}{2}}$ towers of KK spinor excitations. Concerning neutrinos, there is a difference due to the absence of right-handed zero modes. In this case,
\begin{equation}
N''^{(\underline{m})}_{V\, L,R}= \left( {\begin{array}{*{20}{c}}
0\\
\vdots\\
0\\
\ \\
N''^{(\underline{m})}_{L,R (1)}\\
N''^{(\underline{m})}_{L,R  (2)} \\
\vdots \\
N''^{(\underline{m})}_{L,R (\frac{2^{\frac{n}{2}}}{2}) }\, ,
\end{array}} \right)
\end{equation}
so that each neutrino flavor has associated $2^{\frac{n}{2}}/2$ towers of KK spinor excitations. To write, in a compact way, the diverse interactions that emerge from this sector, it will be convenient to use the following notation for the standard fermions:
\begin{equation}
 F''^{(\underline{0})}_{V\, L,R}=   \left( {\begin{array}{*{20}{c}}
0\\ \vdots \\ 0\\ F^{(\underline{0})}_{L,R}\\ \vdots \\ 0\end{array}} \right)  \, , \ \ \ \ \ N''^{(\underline{0})}_{V\, L}=   \left( {\begin{array}{*{20}{c}}
0\\ \vdots \\ 0\\ N^{(\underline{0})}_{L}\\ \vdots \\ 0\end{array}} \right)\, ,
\end{equation}
where $ F^{(\underline{0})}_{L,R}$ and $N^{(\underline{0})}_{L}$ appear as the ${2^{n\over 2}\over 2}+1$ components of the $ F''^{(\underline{0})}_{V\, L,R}$ and $N''^{(\underline{0})}_{V\, L}$ vectors, respectively.

It is important to stress that, for a given flavor, there is no way of distinguishing one tower from another, as they have the same mass, which is given by
\begin{equation}
m^2_{f^{(\underline{m})}}=m^2_{(\underline{m})}+m^2_{f^{(\underline{0})}}\, , \ \ \ \ \ f=q,l,\nu_l\, .
\end{equation}
\subsection{Weak currents}
The weak interactions in this sector of the model are characterized by charged and neutral vector and scalar currents mediated by the families of fields $\{W^{(\underline{0})\pm }_\mu, W^{(\underline{m})\pm }_\mu, W^{(\underline{m})\pm }_{\bar{n}}, W^{(\underline{m})\pm }_n\}$ and $\{Z^{(\underline{0})}_\mu, Z^{(\underline{m})}_\mu, Z^{(\underline{m})}_{\bar{n}}, Z^{(\underline{m})}_n\}$. From now on, the double-prime symbol on the fermion fields will be omitted.

The vector charged currents can be written as
\begin{eqnarray}
{\cal L}_{\rm VCC} &=&W_\mu ^{  (\underline 0) +}  \Big[ J_W^{(\underline{0})\, \mu} + \sum\limits_{(\underline m)} \Big ( J_{WL}^{\mbox{\tiny{$(\underline m)(\underline m)$}}\, \mu } + J_{W R}^{\mbox{\tiny{$(\underline m)(\underline m)$}}\, \mu}  \Big )\Big ] \nonumber \\
&\,&+ \sum\limits_{(\underline m)}  W_\mu ^{ (\underline m)  + } \Big [  J_{W}^{\mbox{\tiny{$(\underline 0)(\underline m)$}}\, \mu}   +  \sum\limits_{(\underline r)(\underline s)}  J_{W}^{\mbox{\tiny{$(\underline r)(\underline s) (\underline m)$}}\, \mu} \Big ]+ {\rm H. \, c.} \, ,
\end{eqnarray}
where
\begin{subequations}
\begin{equation}
 J_W^{(\underline{0})\, \mu}=\frac{g}{{\sqrt 2 }}  \left( \bar N^ {(\underline 0)}{\gamma ^\mu }P_L E^{(\underline 0)} + \bar U^ {(\underline 0)}  K {\gamma ^\mu }P_L D^{(\underline 0)} \right),
\end{equation}
is the standard charged current. The gauge boson $W_\mu ^{  (\underline 0) +}$ also couples to currents involving only KK excitations. These currents are given by
\begin{eqnarray}
J_{W L}^{\mbox{\tiny{$(\underline m)(\underline m)$}}\, \mu }&=&\frac{g}{{\sqrt 2 }}  \left(  \bar N_{V}^{(\underline m)} {\gamma ^\mu }{P_L}   E_{V}^{(\underline m)}   +{1\over 2}\bar U_{V}^{(\underline m)} K {\gamma ^\mu } {P_L}   D_{V}^{(\underline m)}\right),   \\
J_{W R}^{\mbox{\tiny{$(\underline m)(\underline m)$}}\, \mu}&=&\frac{g}{{\sqrt 2 }}   \Big( \bar N_{V}^{(\underline m)} {\gamma ^\mu } {P_R} V_N^{(\underline{m}) \dag} V_{E}^{(\underline{m})}   E_{ V}^{(\underline m)}\nonumber \\
&&+ \bar U_{ V}^{(\underline m)}  {\gamma ^\mu }  {P_R} {\hat V}_U^{(\underline{m}) \dag } K {\hat V}_{D }^{(\underline{m})} D_{ V}^{(\underline m)} \Big)\, ,
\end{eqnarray}
\end{subequations}
where we have divided the transformation matrix $V_{F}^{(\underline{m})}$ into two parts: $V_{F}^{(\underline{m})}={\hat V}_{F}^{(\underline{m})}+{\bar V}_{F}^{(\underline{m})}$ ($F=E,D,U$), with
\begin{equation}
{\hat V}_{F}^{(\underline{m})}={1\over M_{F^{(\underline m)}}}\left( {\begin{array}{*{20}{c}}
M_{F^{(\underline 0)}} &{-i \Lambda_{(\underline{m})}} \\
0 &0
\end{array}} \right),  \hspace{5mm}  {\bar V}_{F}^{(\underline{m})}={1\over M_{F^{(\underline m)}}}\left( {\begin{array}{*{20}{c}}
{0} &0 \\
{ -i \Lambda^{\dag}_{(\underline{m})} } &M_{F^{(\underline 0)}}
\end{array}} \right)\, .
\end{equation}
The KK excitations $W_\mu ^{ (\underline {m})+} $ couple to the following currents:
\begin{subequations}
\begin{eqnarray}
 J_{W}^{\mbox{\tiny{$(\underline 0)(\underline m)$}}\, \mu}  &=& {g \over {\sqrt 2 } }        \Big [\bar N_{ V}^{(\underline 0)}  {\gamma ^\mu } {P_L} E_{ V}^{(\underline m)} + \bar N_{ V}^{(\underline m)} {\gamma ^\mu }{P_L} E_{ V}^{(\underline 0)} \nonumber \\
&\,&+\,  \bar U_{ V}^{(\underline 0)} K {\gamma ^\mu }  {P_L} D_{ V}^{(\underline m)} + \bar U_{ V}^{(\underline m)}K {\gamma ^\mu }{P_L}    D_{V}^{(\underline 0)} \Big ]\, .
\end{eqnarray}
Notice the absence of the $V_F^{(\underline{m})}$ matrix in this case. The currents involving only combinations of KK excitations are given by
\begin{eqnarray}
J_{W}^{\mbox{\tiny{$(\underline r)(\underline s) (\underline m)$}}\, \mu} &=&  {g \over {\sqrt 2 }} \Big\{    \bar  N_{V}^{(\underline r)} {\gamma ^\mu } \Big [ \Delta {'_{(\underline {rsm})}}       V_N^{(\underline {r}) \dag}   V_{E }^{(\underline{s})}  {P_R}  +{\Delta _{(\underline{rsm})}}  {P_L} \Big ]  E_{V}^{(\underline s)}  \nonumber \\
&\,& +\,  \bar U_{ V}^{(\underline r)}{\gamma ^\mu } \Big [      \Delta {'_{(\underline{rsm})}}    {\hat V}_U^{(\underline{r})  \dag}  K {\hat V}_{D  }^{(\underline{s})}   {P_R}    + {1\over 2}{\Delta _{(\underline{rsm})}}   K {P_L}   \Big  ] D_{V}^{(\underline s)}   \Big \}\, .
\end{eqnarray}
\end{subequations}

We now turn to describe the scalar currents. In all cases we  have implemented the unitary transformations (and their associated useful relations) that allow us to pass from gauge fields to mass eigenstate fields (see Ref.~\cite{GGNNT} and Eqs. \eqref{U1},\eqref{U2},\eqref{HDT}, \eqref{HDT1}, \eqref{HDT2},\eqref{HDT3},\eqref{HDT4}, \eqref{PHD},\eqref{PHD1}).
The scalar charged currents can be organized as follows:
\begin{equation}
{\cal L}_{\rm SCC}={\cal L}_{{ G_W}} +{\cal L}_{{ W_G}} +{\cal L}_{{ W_{\bar{n}}}} +{\cal L}_{{ W_n}} \, .
\end{equation}
The currents mediated by the pseudo-Goldstone bosons  $G^{(\underline{0})\pm}_W$ and $W^{(\underline{m})\pm}_G$ are given by
\begin{eqnarray}
\label{csc}
{\cal L}_{{ G_W}} &=&G_W ^{ (\underline 0)  +}  \Big[ J^{(\underline{0})} _{G_W} + \sum\limits_{(\underline m)} J_{G_W}^{ \mbox{\tiny{$(\underline m)(\underline m)$}} } \Big] + {\rm H.\, c.\,} ,\\
\label{cnsc}
{\cal L}_{{ W_G}} &=& \sum\limits_{(\underline m)}  W_G ^{ (\underline m)  + } \Big [  J_{W_G}^{ \mbox{\tiny{$(\underline 0)(\underline m)$}}\, +}   +  \sum\limits_{(\underline {rs})}  J_{W_G}^{ \mbox{\tiny{$(\underline r)(\underline s) (\underline m)$}} \, +} \Big ]  + {\rm H.\, c.}\, .
\end{eqnarray}
The charged currents mediated by the physical scalars $W^{(\underline{m})\pm}_{\bar{n}}$ ($\bar{n}=1,\cdots, n-1$) and $W^{(\underline{m})\pm}_{n}$ are given by
\begin{eqnarray}
\label{cwbnc}
{\cal L}_{W_{\bar n}} &=& \sum\limits_{\bar n =1}^{n-1}  \sum\limits_{(\underline m)}  W_{\bar n} ^{ (\underline m)  + } \Big [  J_{W_{\bar n}}^{ \mbox{\tiny{$(\underline 0)(\underline m)$}} }   +  \sum\limits_{(\underline {rs})}  J_{W_{\bar n } }^{\mbox{\tiny{$(\underline r)(\underline s) (\underline m)$}}} \Big ]+ {\rm H.\, c.} \, , \\
\label{cwnc}
{\cal L}_{{ W_n}} &=& \sum\limits_{(\underline m)}  W_n ^{ (\underline m)  + } \Big [  J_{W_n}^{ \mbox{\tiny{$(\underline 0)(\underline m)$}}}  +  \sum\limits_{(\underline {rs})}  J_{W_n}^{ \mbox{\tiny{$(\underline r)(\underline s) (\underline m)$}} } \Big ] + {\rm H.\, c.}
\end{eqnarray}
Notice that the currents mediated by the $W^{(\underline{m})\pm}_{n}$ scalar differ from those associated with the $n-1$ scalars $W^{(\underline{m})\pm}_{\bar{n}}$, which is a consequence of the fact that the couplings involving the former field come from both the currents and Yukawa sectors, whereas the couplings of the latter ones emerge only from the currents sector. The various currents appearing in the above expressions are given in \ref{AC}.

We now proceed to describe the neutral currents mediated by the family of fields $\{Z^{(\underline{0})}_\mu , Z^{(\underline{m})}_\mu, Z^{(\underline{m})}_{\bar{n}}, Z^{(\underline{m})}_n\}$, associated with the SM $Z$ gauge boson. The vector currents are given by
\begin{eqnarray}
{\cal L}_{\rm VNC} = Z_\mu ^{(\underline 0)}   \Big [ J_{Z }^{(\underline{0})\, \mu  } +\sum\limits_{(\underline m)} J_{Z}^{\mbox{\tiny{$(\underline m)(\underline m)$}}\, \mu   }   \Big ]   + \sum\limits_{(\underline m)}   Z_\mu ^{(\underline m)}   \Big [  J_{Z}^{\mbox{\tiny{$(\underline 0)(\underline m)$}}\, \mu   } +\sum\limits_{(\underline {rs})} J_{Z}^{\mbox{\tiny{$(\underline r)(\underline s) (\underline m)$}}\, \mu} \Big ].
\end{eqnarray}
In this expression, $J_{Z }^{(\underline{0})\, \mu  }$ is the standard current, which is given by
\begin{eqnarray}
J_{Z }^{(\underline{0})\, \mu  } &=&\frac{g}{{2{c_W}}} \Big [  \bar E^{{(\underline 0)}}{\gamma ^\mu }\left( { \mbox{g}_V^{E} -\mbox{g}_A^{E}{\gamma ^5}} \right) E^{(\underline 0)} + \bar N^{(\underline 0)}{\gamma ^\mu }P_L N^{(\underline 0)}   \nonumber \\
&\,& + \, \bar D^{(\underline 0)}{\gamma ^\mu }\left(  \mbox{g}_V^{D} - \mbox{g}_A^{D}{\gamma ^5} \right) D^{(\underline 0)} +\bar U^{(\underline 0)}{\gamma ^\mu }\left(  \mbox{g}_V^{U} - \mbox{g}_A^{U}{\gamma ^5} \right) U^{(\underline 0)}  \Big]\, ,
\end{eqnarray}
where we have introduced the following definitions:
\begin{equation}
\mbox{g}_V^{F}=t_{3\,L(F)} - 2{s_W^2}{Q_F},      \hspace{4mm} \mbox{g}_A^{F}= t_{3\,L(F)}  , \hspace{3mm} \, \, \, \,  \hspace{4mm} t_{3\,L(F)}  = \left\{ {\begin{array}{*{20}{l}}{\textstyle{1 \over 2}} , \hspace{5mm} F=U, \\-{\textstyle{ 1 \over 2}} , \hspace{2mm} F=E,D,  \end{array}} \right.
\end{equation}
with $Q_F$ the charge of the fermions $F=E,D,U$ in units of the positron charge. It is convenient to express the currents $J_{Z}^{\mbox{\tiny{$(\underline m)(\underline m)$}}\, \mu   }$, in terms of its left-handed and right-handed components, as follows:
\begin{equation}
J_{Z}^{\mbox{\tiny{$(\underline m)(\underline m)$}}\, \mu   }=J_{ZL}^{\mbox{\tiny{$(\underline m)(\underline m)$}}\, \mu   }+J_{ZR}^{\mbox{\tiny{$(\underline m)(\underline m)$}}\, \mu   }\, ,
\end{equation}
where
\begin{subequations}
\begin{align}
 J_{ZL}^{\mbox{\tiny{$(\underline m)(\underline m)$}}\, \mu   }=&\frac{g}{{2{c_W}}} \Big [  \sum\limits_{F=E,D,U}   \mbox{g}_{V}^{F}  \bar F_{ V}^{(\underline m)}{\gamma ^\mu }  P_L F_{V}^{(\underline m)}  +\bar N_{V}^{(\underline m)}{\gamma ^\mu }   P_L N_{V}^{(\underline m)} \Big ], \\
J_{ZR}^{\mbox{\tiny{$(\underline m)(\underline m)$}}\, \mu   }=&\frac{g}{{2{c_W}}}  \Big \{ \sum\limits_{F=E,D,U}   \bar F_{ V}^{(\underline m)}{\gamma ^\mu } P_R \Big [   \mbox{g}_{+}^{F}   {\hat V}_F^{(\underline{m}) \dag} {\hat V}_{F}^{(\underline{m})}   + \mbox{g}_{-}^{F}   {\bar V}_F^{(\underline{m}) \dag} {\bar V}_{F }^{(\underline{m})} \Big ]  F_{V}^{(\underline m)}  \nonumber \\
&+ \, \bar N_{V}^{(\underline m)}{\gamma ^\mu }  P_R N_{V}^{(\underline m)} \Big \}\, ,
\end{align}
with $\mbox{g}_{\pm}^{F}=\mbox{g}_V^{F} \pm \mbox{g}_A^{F}$. Finally,
\begin{eqnarray}
 J_{Z}^{\mbox{\tiny{$(\underline 0)(\underline m)$}}\, \mu   }&=&{g \over {2{c_W}}} \Big \{ \sum\limits_{ F=E,D,U}     \bar  F_{V}^{(\underline 0)}  {\gamma ^\mu }  \Big [ \mbox{g}_{-}^{F}   V_{F }^{(\underline{m})} {P_R} + \mbox{g}_{+}^{F}  {P_L} \Big ] F_{ V}^{(\underline m)}    + \bar  N_{ V}^{(\underline 0)}{\gamma ^\mu }  {P_L}   N_{ V}^{(\underline m)} \Big \}\nonumber \\
 && +{\rm H.\, c.} \, ,
\end{eqnarray}
\begin{eqnarray}
 J_{Z}^{\mbox{\tiny{$(\underline r)(\underline s) (\underline m)$}}\, \mu}&=&{{g \over {2{c_W}}}}   \Big \{\sum\limits_{F=E,D,U}  \bar  F_{ V}^{(\underline r)} {\gamma ^\mu }\Big [      \Big (   \Delta {'_{(\underline{rsm})}}  \mbox{g}_{+}^{F}    {\hat V}_F^{(\underline{r}) \dag} {\hat V}_{F }^{(\underline{s})} +{\Delta _{(\underline{rsm})}}  \mbox{g}_{-}^{F}    {\bar V}_F^{(\underline{r}) \dag}  {\bar V}_{F }^{(\underline{s})}   \Big )   {P_R}    \nonumber \\
&\,& +\,{1\over 2} \big(\Delta {'_{(\underline{rsm})}}   \mbox{g}_{-}^{F}  +{\Delta _{(\underline{rsm})}} \mbox{g}_{+}^{F}  \big )  {P_L}     \Big ] F_{V}^{(\underline s)}  \nonumber \\
&\,&+\,  \bar  N_{ V }^{(\underline r)}{\gamma ^\mu }  \Big[  \Delta {'_{(\underline{rsm})}}     V_N^{(\underline{r}) \dag}  V_{N }^{(\underline{s})}     {P_R}   +  {\Delta _{(\underline{rsm})}}   {P_L}    \Big ]N_{V }^{(\underline s)}  \Big \} \, .
\end{eqnarray}
\end{subequations}

The scalar currents can be organized as
\begin{equation}
{\cal L}_{\rm SNC}={\cal L}_{G_Z}+{\cal L}_{Z_G}+{\cal L}_{Z_{\bar{n}}}+{\cal L}_{Z_n}\, ,
\end{equation}
where
\begin{eqnarray}
{\cal L}_{G_Z } &=& G_Z^{(\underline 0)}   \Big [ J^{(\underline 0)}_{G_Z} +\sum\limits_{(\underline m)} J_{G_Z}^{\mbox{\tiny{$(\underline m)(\underline m)$}}}    \Big ] \, , \\
{\cal L}_{Z_G}&=&   \sum\limits_{(\underline m)}   Z_G^{(\underline m)}   \Big [  J_{Z_G} ^{\mbox{\tiny{$(\underline 0)(\underline m)$}}} +\sum\limits_{(\underline {rs})} J_{Z_G}^{\mbox{\tiny{$(\underline r)(\underline s) (\underline m)$}}} \Big ]\, ,\\
{\cal L}_{Z_{\bar n}}&=&  \sum\limits_{\bar n =1}^{n-1}  \sum\limits_{(\underline m)}  Z_{\bar n} ^{ (\underline m)  } \Big [  J_{Z_{\bar n}}^{ \mbox{\tiny{$(\underline 0)(\underline m)$}} }   +  \sum\limits_{(\underline {rs})}  J_{Z_{\bar n } }^{\mbox{\tiny{$(\underline r)(\underline s) (\underline m)$}}} \Big ] \, , \\
{\cal L}_{Z_n }  &=&   \sum\limits_{(\underline m)}  Z_n^{(\underline m)}   \Big [ J_{Z_n} ^{\mbox{\tiny{$(\underline 0)(\underline m)$}}} +\sum\limits_{(\underline{rs})} J_{Z_n}^{\mbox{\tiny{$(\underline r)(\underline s) (\underline m)$}}} \Big ].
\end{eqnarray}
The various currents appearing in these expressions are given in~\ref{AD}.

\subsection{QED currents}
This section is devoted to discuss the currents involving the family of fields $\{A^{(\underline{0})}_\mu, A^{(\underline{m})}_\mu, A^{(\underline{m})}_{\bar{n}},  A^{(\underline{m})}_G \}$, associated with the photon. The vector currents are given by
\begin{eqnarray}
{\cal L}_{\rm VQEDC} &=& A_\mu ^{(\underline 0)}  \Big [ J_A^{{(\underline 0)}\,\mu   }\nonumber \\
&&+  \sum\limits_{(\underline m)} J_{A }^{\mbox{\tiny{$(\underline m)(\underline m)$}}\, \mu   } \Big ] +\sum\limits_{(\underline m)}  A_\mu ^{(\underline m)}  \Big [ J_{A}^{\mbox{\tiny{$(\underline 0)(\underline m)$}}\, \mu   } +\sum\limits_{(\underline {rs})}J_{A}^{\mbox{\tiny{$(\underline r)(\underline s) (\underline m)$}}\, \mu}  \Big ] \, ,
\end{eqnarray}
where $J_A^{{(\underline 0)}\,\mu   }$ is the standard electromagnetic current, which is given by
\begin{subequations}
\begin{equation}
  J_A^{{(\underline 0)}\,\mu   }=e    \Big [ Q_E\bar E^{(\underline 0)}{\gamma ^\mu }E^{(\underline 0)}  + {Q_D} \bar D^{ (\underline 0)}{\gamma ^\mu }D^{ (\underline 0)}+ {Q_U} \bar U^{ (\underline 0)}{\gamma ^\mu }U^{ (\underline 0)} \Big ]\, .
\end{equation}
The currents which contain KK excitations are given by
\begin{equation}
 J_{A }^{\mbox{\tiny{$(\underline m)(\underline m)$}}\, \mu   }=e\sum\limits_{F=E,D,U} {Q_F}  \bar F_{V}^{(\underline m)}{\gamma ^\mu } F_{V}^{(\underline m)}\, ,
\end{equation}
\begin{equation}
J_{A }^{\mbox{\tiny{$(\underline 0)(\underline m)$}}\, \mu   }=e  \sum\limits_{F=E,D,U}   {Q_F}  \bar F_{ V}^{(\underline 0)}{\gamma ^\mu }  \big [ V_{F }^{(\underline{m})} {P_R} +{P_L} \big ] F_{ V}^{(\underline m)}     +{\rm H.\, c.}\, ,
\end{equation}
\begin{eqnarray}
J_{A}^{\mbox{\tiny{$(\underline r)(\underline s) (\underline m)$}}\, \mu} &=& e     \sum\limits_{F=E,D,U}  {Q_F}  \bar  F_{ V}^{(\underline r)}{\gamma ^\mu }   \Big [   \Big (   \Delta {'_{(\underline{rsm})}}  {\hat V}_F^{(\underline{r}) \dag } {\hat V}_{F}^{(\underline{s})}   +  {\Delta _{(\underline {rsm})}} {\bar V}_F^{(\underline{r}) \dag} {\bar V}_{F}^{(\underline{s})}  \Big )     {P_R}  \nonumber  \\
&\,&+ \,  {1\over 2}\big( \Delta {'_{(\underline{rsm})}}  + {\Delta _{(\underline {rsm})}} \big ) {P_L}   \Big ] F_{V}^{(\underline s)}\, .
\end{eqnarray}
\end{subequations}

The scalar currents mediated by the pseudo-Goldstone boson  $A_G^{(\underline m)}$ and the $n-1$ physical scalars $A_{\bar n}^{(\underline m)}$, are given by
\begin{eqnarray}
{\cal L}_{\rm SQEDC} &=& \sum\limits_{(\underline m)} \Big \{  A_{G} ^{ (\underline m)  } \Big [  J_{A_G}^{ \mbox{\tiny{$(\underline 0)(\underline m)$}} }   +  \sum\limits_{(\underline {rs})}  J_{A_{G} }^{\mbox{\tiny{$(\underline r)(\underline s) (\underline m)$}}} \Big ]   \\
&\,&+\, \sum\limits_{\bar n =1}^{n-1}    A_{\bar n} ^{ (\underline m)  } \Big [  J_{A_{\bar n}}^{ \mbox{\tiny{$(\underline 0)(\underline m)$}} }   +  \sum\limits_{(\underline {rs})}  J_{A_{\bar n } }^{\mbox{\tiny{$(\underline r)(\underline s) (\underline m)$}}} \Big ]  \Big \}\, ,
\end{eqnarray}
where the KK excited currents are
\begin{subequations}
\begin{equation}
J_{A_{G}}^{ \mbox{\tiny{$(\underline 0)(\underline m)$}} }=-e \sum\limits_{F=E,D,U}    {Q_F}   \bar F_{V}^{(\underline 0)}   \Pi ^{(\underline m)}  \left[ V_{F}^{(\underline{m})} {P_R} - {P_L} \right] F_{V}^{(\underline m)}+ {\rm H.\, c.} \, ,
\end{equation}
\begin{equation}
J_{A_{\bar n}}^{ \mbox{\tiny{$(\underline 0)(\underline m)$}} }=-e \sum\limits_{F=E,D,U}    {Q_F}   \bar F_{V}^{(\underline 0)}  \mathcal{R}^{(\underline{m})}_{\bar{\mu}\bar{n}} \Pi^{\bar \mu} \left[ V_{F}^{(\underline{m})} {P_R} - {P_L} \right] F_{V}^{(\underline m)}+ {\rm H.\, c.}\, ,
\end{equation}
\begin{eqnarray}
J_{A_{G} }^{\mbox{\tiny{$(\underline r)(\underline s) (\underline m)$}}}&=& -e \sum\limits_{F=E,D,U}   {Q_F} \bar  F_{V}^{(\underline r)} \Big [ \big( \Delta {'_{(\underline {msr})}}    \Pi ^{(\underline m)} -  \Delta {'_{(\underline {mrs})}}     \Pi ^{(\underline m) \dag}  \big ) V_{F}^{(\underline{s})} {P_R}  \nonumber\\
&\,&+\,  V _F^{(\underline{r}) \dag} \big ( \Delta {'_{(\underline {mrs})}} \Pi ^{(\underline m) \dag} - \Delta {'_{(\underline {msr})}}   \Pi ^{(\underline m)} \big)  {P_L}    \Big ] F_{V}^{(\underline s)} \, ,
\end{eqnarray}
\begin{eqnarray}
J_{A_{\bar n } }^{\mbox{\tiny{$(\underline r)(\underline s) (\underline m)$}}}&=& -e \sum\limits_{F=E,D,U}  {Q_F} \bar  F_{V}^{(\underline r)}    \mathcal{R}^{(\underline{m})}_{\bar{\mu}\bar{n}}   \Big [ \big( \Delta {'_{(\underline {msr})}}     \Pi^{\bar \mu}  -  \Delta {'_{(\underline{mrs})}}     \Pi^{\bar \mu  \dag }   \big ) V_{F }^{(\underline{s})} {P_R}  \nonumber\\
&\,&+ \, V _F^{(\underline{r}) \dag} \big ( \Delta {'_{(\underline {mrs})}}  \Pi^{\bar \mu \dag }  - \Delta {'_{(\underline {msr})}}    \Pi^{\bar \mu} \big)  {P_L}    \Big ] F_{V}^{(\underline s)} \, .
\end{eqnarray}
\end{subequations}
with
\begin{equation}
\Pi^{\bar \mu}= \left( {\begin{array}{*{20}{c}} 0&0\\ { {{\Theta} ^{\bar \mu \, \dag}}}&0 \end{array}} \right) \, , \hspace{3mm}  \Pi ^{(\underline m)}= \Pi^{\bar{\mu}}\mathcal{R}_{\bar{\mu} n} ={1\over m_{(\underline{m})} }\left( {\begin{array}{*{20}{c}} 0&0\\ {\Lambda}^{\dag}_{(\underline{m})}&0 \end{array}} \right)    .
\end{equation}
which arises after using some properties of the orthogonal rotation given in~\cite{GGNNT}.

\subsection{QCD currents}
In this part, we describe the strong currents associated with the family of gluon fields $\{ G_{\mu}^{ (\underline 0) \, a}, G_{\mu}^{ (\underline m) \, a}, G_{\bar{n}}^{ (\underline m) \, a}, G_{G}^{ (\underline m) \, a}\}$. The vector interactions are given by

\begin{eqnarray}
{\cal L}_{\rm VQCDC} &=& G_\mu ^{(\underline 0) \, a}  \Big [ J_G^{(\underline 0)\, a \, \mu } \nonumber \\
 &&+  \sum\limits_{(\underline m)} J_{G }^{\mbox{\tiny{$(\underline m)(\underline m)$}}\, a \, \mu } \Big ] +\sum\limits_{(\underline m)}  G_\mu ^{(\underline m) \, a}  \Big [ J_{G }^{\mbox{\tiny{$(\underline 0)(\underline m)$}} \, a   \, \mu} +\sum\limits_{(\underline {rs})}J_{G }^{\mbox{\tiny{$(\underline r)(\underline s) (\underline m)$}}\, a \, \mu}  \Big ] \, ,
\end{eqnarray}
where
\begin{subequations}
\begin{equation}
  J_G^{\mu   \, a}  = g_s \Big [\bar D^{ (\underline 0)}{\gamma ^\mu } {\lambda^{a} \over 2 } D^{ (\underline 0)}+  \bar U^{ (\underline 0)}{\gamma ^\mu }{\lambda^{a} \over 2 } U^{ (\underline 0)} \Big ]\, ,
\end{equation}
\begin{equation}
 J_{G }^{ \mbox{\tiny{$(\underline m)(\underline m)$}}\, a \, \mu  }=g _s \sum\limits_{F=D,U} \bar F_{V}^{(\underline m)}{\gamma ^\mu }{\lambda^{a} \over 2 } F_{V}^{(\underline m)}\, ,
\end{equation}
\begin{equation}
J_{G }^{\mbox{\tiny{$(\underline 0)(\underline m)$}}\, a  \, \mu}= g_s \sum\limits_{F=D,U}  \bar F_{ V}^{(\underline 0)}{\gamma ^\mu }  {\lambda^{a} \over 2 } \big [ V_{F }^{(\underline{m})}{P_R} +{P_L} \big ] F_{ V}^{(\underline m)}     +{\rm H.\, c.} \, ,
\end{equation}
\begin{eqnarray}
J_{G }^{\mbox{\tiny{$(\underline r)(\underline s) (\underline m)$}}\, a \, \mu} &=&    g_s  \sum\limits_{F=D,U}  \bar  F_{V}^{(\underline r)}{\gamma ^\mu }  {\lambda^{a} \over 2 }   \Big \{     \Big [   \Delta {'_{(\underline {rsm})}}  {\hat V}_F^{(\underline{r}) \dag } {\hat V}_{F }^{(\underline{s})}   +  {\Delta _{(\underline {rsm})}} {\bar V}_F^{(\underline{r}) \dag} {\bar V}_{F }^{(\underline{s})}  \Big ] P_R \nonumber  \\
&\,&+  {1 \over 2}\big( \Delta {'_{(\underline {rsm})}}    + {\Delta _{(\underline{rsm})}} \big ) {P_L}    \Big \}  F_{V}^{(\underline s)}.
\end{eqnarray}
\end{subequations}
The scalar interactions are given by
\begin{eqnarray}
{\cal L}_{\rm SQCDC} &=&  \sum\limits_{(\underline m)} \Big \{   G_{G} ^{ (\underline m) \, a  } \Big [  J_{G_{G}}^{ \mbox{\tiny{$(\underline 0)(\underline m)$}} \, a}   +  \sum\limits_{(\underline {rs})}  J_{G_{G} }^{\mbox{\tiny{$(\underline r)(\underline s) (\underline m)$}}  \, a}  \Big ] \nonumber \\
 &&+\sum\limits_{\bar n=1}^{n-1}    G_{\bar n} ^{ (\underline m)  \, a} \Big [  J_{G_{\bar n}}^{ \mbox{\tiny{$(\underline 0)(\underline m)$}} \, a}   +  \sum\limits_{(\underline {rs})}  J_{G_{\bar n } }^{\mbox{\tiny{$(\underline r)(\underline s) (\underline m)$}} \, a} \Big ] \Big \}\, ,
\end{eqnarray}
where
\begin{subequations}
\begin{equation}
J_{G_{G}}^{ \mbox{\tiny{$(\underline 0)(\underline m)$}} \, a}=-g_s \sum\limits_{F=D,U}      \bar F_{V}^{(\underline 0)} {\lambda^{a}\over 2}   \Pi ^{(\underline m)}  \left[ V_{F}^{(\underline{m})}{P_R} - {P_L} \right] F_{V}^{(\underline m)}+ {\rm H.\, c.} \, ,
\end{equation}
\begin{equation}
J_{G_{\bar n}}^{ \mbox{\tiny{$(\underline 0)(\underline m)$}} \,}=-g_s \sum\limits_{F=D,U}      \bar F_{V}^{(\underline 0)} {\lambda^{a}\over 2} \mathcal{R}^{(\underline{m})}_{\bar{\mu}\bar{n}}   \Pi^{\bar \mu}\left[ V_{F}^{(\underline{m})}{P_R} - {P_L} \right] F_{V}^{(\underline m)}+ {\rm H.\, c.} \, ,
\end{equation}
\begin{eqnarray}
J_{G_{G} }^{\mbox{\tiny{$(\underline r)(\underline s) (\underline m)$}}  \, a} &=& - g_s\sum\limits_{F=D,U}  \bar  F_{V}^{(\underline r)} {\lambda^{a}\over 2}   \Big [ \big( \Delta {'_{(\underline {msr})}}    \Pi ^{(\underline m)}  -  \Delta {'_{(\underline {mrs})}}   \Pi ^{(\underline m) \dag}     \big ) V_{F  }^{(\underline{s})}  {P_R}\nonumber\\
&\,&+ \, V_F^{(\underline{r}) \dag}  \big ( \Delta {'_{(\underline{mrs})}}  \Pi ^{(\underline m) \dag}  - \Delta {'_{(\underline{msr})}}  \Pi ^{(\underline m)}  \big ) {P_L}    \Big ]  F_{ V}^{(\underline s)} \, ,
\end{eqnarray}
\begin{eqnarray}
J_{G_{\bar n } }^{\mbox{\tiny{$(\underline r)(\underline s) (\underline m)$}}  \, a} &=& - g_s\sum\limits_{F=D,U}  \bar  F_{V}^{(\underline r)} {\lambda^{a}\over 2} \mathcal{R}^{(\underline{m})}_{\bar{\mu}\bar{n}}   \Big [ \big( \Delta {'_{(\underline{msr})}}   \Pi^{\bar \mu} -  \Delta {'_{(\underline{mrs})}}    \Pi^{\bar \mu \dag}    \big ) V_{F  }^{(\underline{s})}  {P_R}\nonumber\\
&\,&+ \, V_F^{(\underline{r}) \dag}  \big (\Delta {'_{(\underline{mrs})}} \Pi^{\bar \mu \dag}  - \Delta {'_{(\underline{msr})}}     \Pi^{\bar \mu} \big ) {P_L}    \Big ]  F_{ V}^{(\underline s)} \, .
\end{eqnarray}
\end{subequations}

\subsection{Higgs-fermion interactions}
The couplings of the family of fields $\{H ^{(\underline 0)}, H ^{(\underline m)}\}$, associated with the Higgs boson, can be written as follows:
\begin{equation}
{\cal L}_{ H ff} = H ^{(\underline 0)}   \Big [ J^{(\underline 0)}_H +\sum\limits_{(\underline m)} J_H^{\mbox{\tiny{$(\underline m)(\underline m)$}}}    \Big ]   + \sum\limits_{(\underline m)}   H^{(\underline m)}   \Big [  J_H ^{\mbox{\tiny{$(\underline 0)(\underline m)$}}} +\sum\limits_{(\underline {rs})} J_H^{\mbox{\tiny{$(\underline r)(\underline s) (\underline m)$}}} \Big ]\, ,
\end{equation}
where
\begin{subequations}
\begin{equation}
J_H^{(\underline{0})}=  -  {g \over {2 m_{W^{(\underline 0)}}} }  \Big [  \bar E _{V}^{(\underline 0)} M_{E^{(\underline 0)}}E_{V}^{(\underline 0)}+ \bar D_{V}^{(\underline 0)} M_{D^{(\underline 0)}}  D_{V}^{(\underline 0)}+\bar U_{V}^{(\underline 0)} M_{U^{(\underline 0)}}  U_{V}^{(\underline 0)}\Big ] \, ,
\end{equation}
\begin{equation}
 J_H^{\mbox{\tiny{$(\underline m)(\underline m)$}}}=  -  {g \over {2 m_{W^{(\underline 0)}}} }   \sum\limits_{F=E,D,U}    \bar F_{ V}^{(\underline m)}  M_{F^{(\underline 0)}} {\tilde V}_{F}^{(\underline{m})} F_{ V}^{(\underline m)}   \, ,
\end{equation}
\begin{eqnarray}
 J_H ^{\mbox{\tiny{$(\underline 0)(\underline m)$}}} =- {g  \over {2 m_{W^{(\underline 0)}} } }     \sum\limits_{F=E,D,U}        \bar F_{ V}^{(\underline 0)}  M_{F^{(\underline 0)}} \big [ V_{F}^{(\underline{m})} {P_R}+ {P_L}  \big ] F_{ V}^{(\underline m)}   +{\rm H.\, c .}\, ,
\end{eqnarray}
\begin{eqnarray}
 J_H^{\mbox{\tiny{$(\underline r)(\underline s) (\underline m)$}}} &=&- {g\over {2 m_{W^{(\underline 0)}} } }   \sum\limits_{F=E,D,U}   \bar F_{V}^{(\underline r)} M_{F^{(\underline 0)}}  \Big \{    \Big [  {\Delta _{(\underline{rsm})}}      {\bar V}_{F }^{(\underline{s})} +\Delta {'_{(\underline {rsm})}} {\hat V}_{F }^{(\underline{s})}  \Big]  {P_R}   \nonumber \\
&\,&  +\,  \Big [   {\Delta _{(\underline{rsm})}} {\bar V}_F^{(\underline{r}) \dag}  + \Delta {'_{(\underline{rsm})}} {\hat V}_F^{(\underline{r}) \dag}   \Big] {P_L}  \Big \} F_{V}^{(\underline s)} \, .
\end{eqnarray}
\end{subequations}
In the above expressions,
\begin{equation}
 {\tilde V}_{F }^{(\underline{m})}={1\over M_{F^{(\underline m)}}}\left( {\begin{array}{*{20}{c}}
M_{F^{(\underline 0)}} & -i \Lambda_{(\underline{m})} \gamma^5 \\
{-i \Lambda^{\dag}_{(\underline{m})}}\gamma^5 &M_{F^{(\underline 0)}}
\end{array}} \right)\, .
\end{equation}

\section{Concluding remarks and summary}
\label{C}

In this paper, we have proposed an effective theory for the SM that incorporates $n$ flat compact dimensions. Our starting point has been a field theory that is valid at energies far above the compactification scale $R^{-1}$, which respects the extended  $\{ {\rm ISO}(1,3+n),\, G({\cal M}^d)\}$ symmetries. It is assumed that the size $R$ of the extra dimensions are so large compared with the distance scales to which this theory is valid that such extra dimensions can be practically considered as infinite.\\

To describe the physical phenomena at energies of order of the compactification scale $R^{-1}$, we resort to the notion of hidden symmetry and to a mass-generating mechanism, in this case the Kaluza-Klein mass-generating mechanism. In order to hide the $\{ {\rm ISO}(1,3+n),\, G_{\rm SM}({\cal M}^d)\}$ symmetries into the $\{ {\rm ISO}(1,3),\, G_{\rm SM}({\cal M}^4)\}$ ones, two canonical transformations were implemented. First, a canonical map was implemented in order to accommodate ${\rm SO}(1,3+n)$ representations into ${\rm SO}(1,3)$ representations. This map allows one to hide the ${\rm SO}(1,3+n)$ symmetry into the ${\rm SO}(1,3)$ one. Next, a second nontrivial canonical map was implemented in order to remove any manifest dynamical role of the inhomogeneous ${\rm ISO}(n)$ group. Crucial to this map is to assume the existence of a set of orthonormal functions $\{ f^{(\underline{0})}, \, f^{(\underline{m})}(\bar x)\}$ defined on the compact manifold. The presence of the constant function $f^{(\underline{0})}$, which may be common to any compactification scheme, plays a central role in defining the connections and gauge parameters of the standard gauge group $G_{\rm SM}({\cal M}^4)$. The main ideas behind the approach to extra flat dimensions followed here, which have already been presented by some of us in the context of pure Yang-Mills theories in~\cite{NoTo,GGNNT}, were extended to include the Engler-Higgs mechanism.\\

 %For instance, in a pure Yang-Mills theory, the components of ${\cal A}^a_\mu(x,\bar x)$ and $\alpha^a(x,\bar x)$ along $f^{(\underline{0})}$ can be identified, respectively, as the gauge fields and gauge parameters of the $SU(N,{\cal M}^4)$ group; while their components along the $f^{(\underline{m})}(\bar x)$ directions emerge in the adjoint representation of this group. This map also allows one to identify the KK fields $A^{(\underline{m})a}_\mu(x)$ as genuine gauge fields because it turns out that there is a one-to-one relation with the gauge parameters $\alpha^{(\underline{m})a}(x)$. As far as the scalar fields  ${\cal A}^a_{\bar \mu}(x,\bar x)$ are concerned, their components, either along  $f^{(\underline{0})}$ or along $f^{(\underline{m})}(\bar x)$, transform in the adjoint representation of the $SU(N,{\cal M}^4)$ group. The components of both ${\cal A}^a_\mu(x,\bar x)$ and ${\cal A}^a_{\bar \mu}(x,\bar x)$ fields along the $f^{(\underline{m})}(\bar x)$ directions can be endowed with mass, since they appear as matter fields from the $SU(N,{\cal M}^4)$ group perspective. In general, this is enough to correctly identify the $\{ {\rm ISO}(1,3),\, G_{SM}({\cal M}^4)\}$ covariant structure of the new basic fields. This means that to hide the extended symmetries into the standard ones, it is not necessary to specify the geometry of the compact manifold.\\

The fact that the $f^{(\underline{0})}$ direction allows us to identify the connections of the standard gauge group, means that any class of field with component along this direction does not receive mass at the $R^{-1}$ scale. Only those fields along the $f^{(\underline{m})}(\bar x)$ directions are endowed with mass by the Kaluza-Klein mechanism. In order to recover the known particle spectrum of the standard theory, the basis $\{ f^{(\underline{0})}, \, f^{(\underline{m})}(\bar x)\}$ was divided into two sub-bases: the basis of even functions, $\{ f^{(\underline{0})}, \, f^{(\underline{m})}_E(\bar x)\}$, and the basis of odd functions, $\{ f^{(\underline{m})}_O(\bar x)\}$. It was postulated that fields with standard counterpart are necessarily even, while fields without standard counterpart are odd. This classification obeys the physical requirement that new-physics effects decouple at $R^{-1}\gg v$.\\

With results given in Refs.~\cite{NoTo,GGNNT} at hand, a comprehensive study of an extension of the SM to more than four dimensions was presented. Starting from a straightforward extension of all the SM sectors to $4+n$ dimensions, an effective theory preserving the standard $\{ {\rm ISO}(1,3),\, G_{\rm SM}({\cal M}^4)\}$ symmetries was derived. In the end, one has an effective field theory in which each SM particle is described by a family of fields into which the original field is unfolded. Our main results can be summarized as follows:\\

\noindent $\bullet$ In the effective theory, the Higgs boson is described by the family of fields $\{ H^{(\underline{0})}, \, H^{(\underline{m})}\}$, where $ H^{(\underline{0})}$ is the SM field and  $H^{(\underline{m})}$ are its KK excitations, whose masses are given by $m^2_{H^{(\underline{m})}}=m^2_{(\underline{m})}+m^2_{H^{(\underline{0})}}$. Here $m_{H^{(\underline{0})}}$ is the physical mass of the Higgs boson and $m^2_{(\underline{m})}=p^{(\underline{m})}_{\bar \mu} p^{(\underline{m})}_{\bar \mu}$. Here, $p^{(\underline{m})}_{\bar \mu}$ is the eigenvalue of the $P_{\bar \mu}$ generator of the translations group $T(n)$ to which the eigenfunctions $f^{(\underline{m})}(\bar x)$ are associated. The $m_{(\underline{m})}$ mass is a gauge mass in the sense that it appears in terms which are invariant under SGT, as it is required by the decoupling theorem.

\noindent $\bullet$ Massless gauge bosons, as gluons or the photon, are described by vector and scalar KK excitations. So, a gluon is described by the family of fields $\{G^{(\underline{0})a}_\mu,\, G^{(\underline{m})a}_\mu, \, G^{(\underline{m})a}_{\bar{n}} \}$, with $G^{(\underline{m})a}_{\bar{n}}$ representing $n-1$ scalar fields. The photon is described by a similar family of fields, $\{A^{(\underline{0})}_\mu,\, A^{(\underline{m})}_\mu, \, A^{(\underline{m})}_{\bar{n}} \}$. The masses of all KK excitations, vectorial or scalar, are given by the $p^{(\underline{m})}_{\bar \mu}$ scale. On the other hand, it should be recalled that the vector field ${\cal A}^a_M$ of ${\rm SO}(1,3+n)$ is mapped into the vector field $A_\mu$ and $n$ scalar fields of ${\rm SO}(1,3)$. However, it was shown that one of these scalar fields, which is massless, can be removed from the theory through a specific NSGT, so it can be recognized as a pseudo-Goldstone boson in this sense. In this way, the NSGT allow us not only to show that the vectorial KK excitations $G^{(\underline{m})a}_\mu$ or $A^{(\underline{m})}_\mu$ are gauge fields, but to define unitary propagators for them as well.

\noindent $\bullet$ In contrast with the case of massless gauge bosons, the $W$ and $Z$ gauge bosons have associated, besides their vectorial KK excitations, $n$ scalar KK excitations, namely, $\{ W^{(\underline{0})\pm}_\mu,\, W^{(\underline{m})\pm}_\mu, \, W^{(\underline{m})\pm}_{\bar{n}}, \, W^{(\underline{m})\pm}_n\}$ and $\{ Z^{(\underline{0})}_\mu,\, Z^{(\underline{m})}_\mu, \, Z^{(\underline{m})}_{\bar{n}}, \, Z^{(\underline{m})}_n\}$. The masses of all KK excitations are  given by $m^2_{W^{(\underline{m})}}=m^2_{(\underline{m})}+m^2_{W^{(\underline{0})}}$ and $m^2_{Z^{(\underline{m})}}=m^2_{(\underline{m})}+m^2_{Z^{(\underline{0})}}$, respectively. In this case, the presence of the scalar excitations $W^{(\underline{m})\pm}_n$ and $Z^{(\underline{m})}_n$ is due to the longitudinal components of these gauge bosons. In this case, there is also a massless scalar excitation for each gauge field (denoted by $W^{(\underline{m})\pm}_G$ and $Z^{(\underline{m})}_G$), which can be identified as a pseudo-Goldstone boson, as it can be removed from the theory via a NSGT. The pairs of scalar excitations $(W^{(\underline{m})\pm}_n, W^{(\underline{m})\pm}_G)$ and $(Z^{(\underline{m})}_n, Z^{(\underline{m})}_G)$ emerge from a mixing among the massless scalars induced by the Yang-Mills sector and the KK excitations of the Higgs doublet. Due to this, the couplings of $H^{(\underline{0})}$ to pairs of $W^{(\underline{m})\pm}_{\bar{n}}$ ($Z^{(\underline{m})}_{\bar{n}}$) scalars differ from those to pairs of $W^{(\underline{m})\pm}_n$ ($Z^{(\underline{m})}_n$) scalars. It is important to stress that the Higgs boson distinguishes the families of fields $\{ W^{(\underline{0})\pm}_\mu,\, W^{(\underline{m})\pm}_\mu, \, W^{(\underline{m})\pm}_{\bar{n}}, \, W^{(\underline{m})\pm}_n\}$ and $\{ Z^{(\underline{0})}_\mu,\, Z^{(\underline{m})}_\mu, \, Z^{(\underline{m})}_{\bar{n}}, \, Z^{(\underline{m})}_n\}$ as collectively describing the gauge bosons $W$ and $Z$, for it couples to them proportionally to $g\,m_{W^{(\underline{0})}}$ and $g\,m_{Z^{(\underline{0})}}$, respectively. Finally, let us remark that the zero mode fields $W^{(\underline{0})\pm}_\mu$ and  $Z^{(\underline{0})}_\mu$ (SM fields) have associated the well-known pseudo-Goldstone bosons $G^{(\underline{0})\pm}_W$ and  $G^{(\underline{0})}_Z$, which can be removed from the theory via a specific SGT (unitary gauge), whereas the KK excitations of these gauge fields, $W^{(\underline{m})\pm}_\mu$ and  $Z^{(\underline{m})}_\mu$, have associated the pseudo Goldstone bosons mentioned above, namely, $W^{(\underline{m})\pm}_G$ and $Z^{(\underline{m})}_G$, which can be removed from the theory through a specific NSGT. These facts show in turn that both the SGT and NSGT can be fixed independently of one another.

 \noindent $\bullet$ In the fermionic sector, each charged lepton or quark is described by a family of $2^{\frac{n}{2}}+1$ KK towers of spinor fields, $\{ f^{(\underline{0})}, \, f^{(\underline{m})}_1, \cdots, f^{(\underline{m})}_{2^{\frac{n}{2}}}\}$. Neutrinos, on the other hand, are described by a set of $2^{\frac{n}{2}}/2+1$ KK towers of spinor fields, $\{ f^{(\underline{0})}, \, f^{(\underline{m})}_1, \cdots, f^{(\underline{m})}_{2^{\frac{n}{2}}/2}\}$. The masses of all these KK excitations are given by $m^2_{f^{(\underline{m})}}=m^2_{(\underline{m})}+m^2_{f^{(\underline{0})}}$. The Higgs boson $H^{(\underline{0})}$ couples to each family of fields in the same way. In other words, the Higgs boson identifies each family  $\{ f^{(\underline{0})}, \, f^{(\underline{m})}_1, \cdots, f^{(\underline{m})}_{2^{\frac{n}{2}}}\}$ as belonging to the same flavor $f^{(\underline{0})}$.

 For practical purposes the explicit expressions for the Lagrangians of all sectors of the model have been presented, and any Feynman rule can be extracted from them. This turns the model into a phenomenological laboratory in order to study the impact of the presence of extra dimensions on the four-dimensional SM realm.\\

The SM extension to extra dimensions  derived in this work is characterized by three energy scales, namely, the unknown scale $\Lambda$, the compactification scale $R^{-1}$, and the Fermi scale $v$. At distance scales much smaller than the size of the extra dimensions, the theory is described by the extended groups $\{{\rm ISO}(1,3+n),\, G_{\rm SM}({\cal M}^d)\}$; but at much larger distance scales, of order of $R$, the effective theory is governed by the standard groups $\{{\rm ISO}(1,3),\, G_{\rm SM}({\cal M}^4)\}$. Implicit to this is the hierarchy $v < R^{-1} \ll \Lambda$. This means that in practice, effects proportional to the  $\Lambda$ scale can be ignored. As it was stressed throughout the work, all nonrenormalizable interactions in the Dyson's sense are proportional to this scale. The theory that emerges from deleting all interactions proportional to the $\Lambda$ scale only contains interactions that are renormalizable in the Dyson's sense. This theory, which is characterized by the $v$ and $R^{-1}$ scales, resembles a conventional renormalizable extension of the SM, except because it contains an infinite number of particles.

\section*{Acknowledgments}
The authors acknowledge financial support from CONACYT (M\' exico), and M.A.L-O., E.M-P., H.N-S., and J.J.T. also acknowledge financial support from SNI (M\' exico). J.J.T. also acknowledges financial support from VIEP-BUAP (M\' exico).

\appendix

\section{ Dirac's matrices of dimension higher than four}
\label{AA} This appendix is devoted to present explicit expressions for the $\Lambda^{\bar \mu}$, $\Lambda_{(\underline{m})}$, $\Theta^{\bar \mu}$ matrices given in the fermion sector of the EDSM. In a spacetime of even dimension $4+n$, there exists a set of $4+n$ Dirac's matrices
 $\Gamma^{M}=(\Gamma^{\mu},\Gamma^{\bar \mu})$ ($\mu=0,1,2,3;\, \bar{\mu}=5,\cdots, 4+n$), which can be written in terms of covariant objects of the standard ${\rm SO}(1,3)$ group as follows
\begin{equation}
\Gamma_{ab}^{\mu}=\delta_{ab}\gamma^{\mu} \hspace{3mm}\, , \, \, \, \,   \hspace{3mm} \Gamma_{ab}^{\bar \mu}=\Lambda_{ab}^{\bar \mu} \gamma^{5}\, ,
\end{equation}
where the $\Lambda^{\bar \mu}$ matrices of dimension $2^{\frac{n}{2}}\times2^{\frac{n}{2}}$ are given by the following direct products of Pauli's matrices:
\begin{equation}
\Lambda^{\bar \mu}= i  ( {\bf 1}\otimes)^{S_{\bar \mu}}(\sigma_{2}\otimes)^{\Delta_{2}^{\bar \mu}-\Delta_{4}^{\bar \mu}}(-\sigma_{3}\otimes)^{\Delta_{1}^{\bar \mu}-\Delta_{3} ^{\bar \mu}}(-\sigma_{1}\otimes)^ {S_{n}^{ \bar \mu}} (\sigma_{2}\otimes)^{1-\Delta_{4} ^{\bar \mu}}\, . \label{LambdaBM}
\end{equation}
In this expression,
\begin{subequations}
\begin{align}
 \Delta_1^{\bar \mu}=    \left\{ {\begin{array}{*{20}{l}} 1, \, \mbox{for} \, \bar \mu \, \, \mbox{odd},\\ 0, \, \mbox{otherwise,} \end{array}} \right.   &     \, \, \, \, \,    \Delta_2^{\bar \mu} = \left\{ {\begin{array}{*{20}{l}} 1, \,\mbox{for} \, \bar \mu \, \, \mbox{even}, \\ 0, \, \mbox{otherwise,}  \end{array}} \right.   \, \, \, \, \,  \Delta_3^{\bar \mu}=\delta_{5}^{\bar \mu}, \; \, \, \, \,  \Delta_4^{\bar \mu}=\delta_{n+4}^{\bar \mu} \, ,
 \\ %%
S_{\bar \mu} = \textstyle{1\over 2} (\bar \mu -\Delta_1^{\bar \mu})- & 3+\Delta_{3}^{\bar \mu}, \; \, \, \, \, S_{n}^{ \bar \mu}= \textstyle{1\over 2} ( n- \bar \mu +2+\Delta_1^{\bar \mu}) +2\Delta_4^{\bar \mu} \, .
\end{align}
\end{subequations}
In addition, $\bf 1$ is the $2 \times 2$ identity matrix, $(\sigma_{i}\otimes)^0=1$, $(\sigma_{i}\otimes)^1=\sigma_{i}\otimes$, $(\sigma_{i}\otimes)^2=\sigma_{i}\otimes \sigma_{i}\otimes$, etc. From (\ref{LambdaBM}), one obtains
\begin{equation}
\Lambda_{ab}^{\bar \mu}=  ( - 1)^{V_{an}^{\bar \mu}}    i^{ \Delta _2^{\bar \mu} }   \delta _{S_{an}^{\bar \mu}  - a}^{b}\, , \label{LambdaComp}
\end{equation}
where
\begin{equation}
V_{an}^{\bar \mu}=S_{n}^{ \bar \mu} +(\Delta _1^{\bar \mu}  - \Delta _3^{\bar \mu}) K_{a}^{ \bar \mu}+ (\Delta _2^{\bar \mu} - \Delta _4^{\bar \mu} ){X_a^{\bar \mu}} +(1 - \Delta _4^{\bar \mu} )(a+1)\, .
\end{equation}
For a given $\bar \mu$, the quantities appearing in this expression are given by
\begin{subequations}  \label{DefSSDaK}
\begin{align}
S_{an}^{\bar \mu}=  2^{S} (2{K_{a}^{ \bar \mu}} -& 1) + 1 \, , \; \, \, \, \,  S=\textstyle{1\over 2}(n - \bar \mu  -  {\Delta _1^{\bar \mu}}) +3, \; \, \, \,   K= {1\over 2}(\bar \mu  + \Delta _1^{\bar \mu}) - 3 \, ,  \\
K_{a}^{ \bar \mu} =& \sum\limits_{k = 1}^{2^{ S}} {\sum\limits_{r = 1}^{{2^{K }}} {r\delta _{2^{S} (r - 1) + k}^a} }\, ,   \; \, \, \, \,  {X_a^{\bar \mu}}= \sum\limits_{k = 1}^{2^{S-1}} \sum\limits_{r=1}^{2^{K } } {\delta _{2^{S}(r - 1) + k}^a} \, .
\end{align}
\end{subequations}
Notice that $K_{a}^{ \bar \mu}$ and ${X_a^{\bar \mu}}$ depend on the subindex $a$ of $\Lambda_{ab}^{\bar \mu}$.

A careful analysis of the above expressions allows us to determine the symmetry properties of the $\Lambda^{\bar \mu}$ matrices mentioned in subsection \ref{FS}. These properties imply that
\begin{equation}
\Gamma^{\bar \mu \dag}=-\Gamma^{\bar \mu}\, .
\end{equation}
Due to the fact that $S_{an}^{\bar \mu}$ is odd, the no vanishing entries of $\Lambda^{\bar \mu}$  are of the form $\Lambda_{\bar a \hat a}^{\bar \mu}$ and $\Lambda_{\hat a \bar a}^{\bar \mu}$, where $\hat a$ and $\bar a$ stand for even and odd numbers, respectively.

From (\ref{LambdaBM}), it can be shown that the $\Gamma^{5+n}$ matrix can be written as
\begin{equation}
\Gamma^{5+n}=i^{ {2+n}\over 2}\Gamma^0\cdots \Gamma^3 \Gamma^5\cdots \Gamma^{4+n}= {\left( { {\bf 1} \otimes } \right)^{{n-2}\over 2 }}{\sigma _3} \otimes {\gamma ^5}  \, .
\end{equation}

 On the other hand, the ${\Theta}^{\bar \mu}$  matrices of dimension ${2^{n\over2} / 2} \times {2^{n\over2} / 2} $, are given by
\begin{equation}
 {\Theta}^{\bar \mu} \equiv \Lambda_{\hat a \bar a}^{\bar \mu}  =   \left( {\begin{array}{*{20}{c}}
{\Lambda _{21}^{\bar \mu}}&{\Lambda _{23}^{\bar \mu}}& \cdots \\
{\Lambda _{41}^{\bar \mu}}&{\Lambda _{43}^{\bar \mu}}& \cdots \\
 \vdots & \vdots & \ddots
\end{array}} \right),    \label{MatricesT}
\end{equation}
where the entries of the $\Lambda^{\bar \mu}$ matrices are obtained from the expression (\ref{LambdaComp}).
As far the $\Lambda_{(\underline{m})}$ matrix is concerned, it is given by
\begin{equation}
\Lambda_{(\underline{m})} \equiv  p_{\bar \mu}^{(\underline m)} \Lambda_{\hat a \bar a}^{\bar \mu} = \sum\limits_{\alpha=1}^{n} {\textstyle{m_{\alpha} \over R_{\alpha}} } {\Theta}^{4+\alpha}  \hspace{3mm}\, , \hspace{3mm}   \Lambda^{\dag}_{(\underline{m})} \equiv  -p_{\bar \mu}^{(\underline m)} \Lambda_{\bar a \hat a}^{\bar \mu \,} = \sum\limits_{\alpha=1}^{n} {\textstyle{m_{\alpha} \over R_{\alpha}} } {\Theta}^{4+\alpha \, \dag} \, ,
\end{equation}
where the matrices ${\Theta}^{4+\alpha} $ are given by (\ref{MatricesT}). Finally, using Clifford's algebra, it can be shown that
\begin{equation}
\Lambda_{(\underline{m})} \Lambda^{\dag}_{(\underline{m})}=\Lambda^{\dag}_{(\underline{m})}\Lambda_{(\underline{m})}=m_{(\underline m )}^2\, .
\end{equation}

\section{Covariant terms of currents sector}
\label{AB}
The covariant objets appearing in the expressions given by Eqs. (\ref{FVC}) and (\ref{FSC}) are given by
\begin{subequations}
\begin{align}
{\left( {D_\mu }  { F_{i L(1)} }   \right)^{(\underline 0)}} =&\;  D_\mu ^{(\underline 0)} F_{i L(1)}^{ (\underline 0)} -  \sum\limits_{(\underline m)} \left(ig{\cal O}_\mu ^{(\underline m)}+ig_s\mathbf{ O}_\mu ^{(\underline m)} \right) F_{i L(1)}^{(\underline m)}  ,\\
\left( {D_\mu }F_{i L(1)}  \right)^{(\underline m)} =&\;  \sum\limits_{(\underline s)} D_\mu ^{(\underline m \underline s)}F_{i L(1)}^{ (\underline s)} - \left(ig{\cal O}_\mu ^{(\underline m)}+ig_s\mathbf{ O}_\mu ^{(\underline m)} \right)F_{i L(1)}^{ (\underline 0)}  \\
{\left( D_\mu  F_{i L(\bar a)}  \right)^{(\underline m)}} =&\;  \sum\limits_{(\underline s)} D_\mu ^{(\underline m \underline s)}F_{i L(\bar a)}^{ (\underline s)}   , \hspace{3mm} \bar a\ge3, \\
{\left( {{D_\mu }  F_{i R(\hat a)} } \right)^{(\underline m)}} =&\; \sum\limits_{(\underline s)} {D'}_\mu ^{(\underline m \underline s)}F_{i R(\hat a)}^{ (\underline s)},   \\
{\left( {{D_{\bar \mu }}  F_{i L(1)}   } \right)^{(\underline m)}} =&\;  \sum\limits_{(\underline s)} D_{\bar \mu} ^{(\underline m \underline s)}F_{i L(1)}^{( \underline s)} - \left(ig{\cal O}_{\bar{\mu}} ^{(\underline m)}+ig_s\mathbf{ O}_{\bar{\mu}} ^{(\underline m)} \right) F_{i L(1)}^{(\underline 0)} ,  \\
{\left( {{D_{\bar \mu }}F_{i L(\bar a)} } \right)^{(\underline m)}} =&\; \sum\limits_{(\underline s)} D_{\bar \mu} ^{(\underline m \underline s)}F_{i L(\bar a)}^{ (\underline s)} , \hspace{3mm} \bar a \ge 3 \\
{\left( {{D_{\bar \mu }} F_{i R(\hat a)} } \right)^{(\underline 0)}} =&\;   - \sum\limits_{(\underline m)}  \left(ig{\cal O}_{\bar{\mu}} ^{(\underline m)}+ig_s\mathbf{ O}_{\bar{\mu}} ^{(\underline m)} \right)   F_{i R(\hat a)}^{ (\underline m)} ,  \hspace{3mm}  \\
{\left( {{D_{\bar \mu }} F_{i R(\hat a)}  } \right)^{(\underline m)}} =&\; \sum\limits_{(\underline s)} {D'}_{\bar \mu} ^{(\underline m \underline s)} F_{i R(\hat a)}^{ (\underline s)} \, ,
\end{align}
\end{subequations}
\begin{subequations}
\begin{align}
{\left( {D_\mu }  f_{i R(1) }    \right)^{(\underline 0)}} =&\;  D_\mu ^{(\underline 0)} f_{i R(1) }^{ (\underline 0)} - \sum\limits_{(\underline m)} \left( ig \hat{\cal O}_\mu ^{(\underline m)}+ ig_s\mathbf{ O}_\mu ^{(\underline m)} \right)f_{i R(1) }^{ (\underline m)}  , \\
{\left( {{D_\mu } f_{i R(1) }} \right)^{(\underline m)}} =&\; \sum\limits_{(\underline s)} D_\mu ^{(\underline m \underline s)} f_{i R(1) }^{  (\underline s)} - \left( ig \hat{\cal O}_\mu ^{(\underline m)}+ ig_s\mathbf{ O}_\mu ^{(\underline m)} \right)f_{i R(1) }^{  (\underline 0)}    ,\\
{\left( {{D_\mu } f_{i R(\bar a) } } \right)^{(\underline m)}} =&\;  \sum\limits_{(\underline s)} D_\mu ^{(\underline m \underline s)}  f_{i R(\bar a) }^{  (\underline s)} , \hspace{3mm} \bar a\ge 3 \\
{\left( {{D_\mu } f_{i L(\hat a) } } \right)^{(\underline m)}} =&\; \sum\limits_{(\underline s)} {D'}_\mu ^{(\underline m \underline s)}  f_{i L(\hat a) }^{ (\underline s)}   ,  \\
{\left( {{D_{\bar \mu }}f_{i R(1) } } \right)^{(\underline m)}} =&\;  \sum\limits_{(\underline s)} D_{\bar \mu} ^{(\underline m \underline s)}f_{i R(1) }^{ (\underline s)} - \left( ig \hat{\cal O}_{\bar{\mu}} ^{(\underline m)}+ ig_s\mathbf{ O}_{\bar{\mu}} ^{(\underline m)} \right)f_{i R(1) }^{ (\underline 0)}  , \\
{\left( {{D_{\bar \mu }}f_{i R(\bar a) }  } \right)^{(\underline m)}} =&\; \sum\limits_{(\underline s)} D_{\bar \mu} ^{(\underline m \underline s)}f_{i R(\bar a) }^{(\underline s)}  ,  \hspace{3mm} \bar a\ge 3 \\
{\left( {{D_{\bar \mu }} f_{i L(\hat a) } } \right)^{(\underline 0)}} =&\;  - \sum\limits_{(\underline m)} \left( ig \hat{\cal O}_{\bar{\mu}} ^{(\underline m)}+ ig_s\mathbf{ O}_{\bar{\mu}} ^{(\underline m)} \right)f_{i L(\hat a) }^{ (\underline m)} ,   \\
{\left( {{D_{\bar \mu }} f_{i L(\hat a) } } \right)^{(\underline m)}} =&\; \sum\limits_{(\underline s)} {D'}_{\bar \mu} ^{(\underline m \underline s)} f_{i L(\hat a) }^{ (\underline s)} \, ,
\end{align}
\end{subequations}
where
\begin{subequations}
\begin{align}
D_\mu ^{(\underline 0)} =& \; {\partial _\mu } - \left(ig{\cal O}_\mu ^{(\underline m)}+ig_s\mathbf{ O}_\mu ^{(\underline m)} \right)\, , \\
{D}_{\mu}^{( \underline m) ( \underline s)} =& \delta^{( \underline {ms})}  D_{\mu}^{( \underline 0)}-  \sum\limits_{(\underline r)} {\Delta _{(\underline m \underline r \underline s)}}\left(ig{\cal O}_\mu ^{(\underline r)}+ig_s\mathbf{ O}_\mu ^{(\underline r)} \right)\, ,\\
{D'}_{\mu}^{( \underline m) ( \underline s)} =&\; \delta^{( \underline m \underline s)}  D_{\mu}^{( \underline 0)}- i g \sum\limits_{(\underline r)} {\Delta' _{(\underline m  \underline s  \underline r)}}\left(ig{\cal O}_\mu ^{(\underline r)}+ig_s\mathbf{ O}_\mu ^{(\underline r)} \right)\, ,\\
{D}_{\bar \mu}^{( \underline m) ( \underline s)} =&\; -p_{\bar{\mu}}^{(\underline m)} \delta^{( \underline m  \underline s)} - \sum\limits_{(\underline r)} {\Delta' _{(\underline m \underline r  \underline s)}}\left(ig{\cal O}_{\bar{\mu}} ^{(\underline r)}+ig_s\mathbf{ O}_{\bar{\mu}} ^{(\underline r)} \right)\, ,\\
{D'}_{\bar \mu}^{( \underline m) ( \underline s)} =&\; p_{\bar{\mu}}^{(\underline m)} \delta^{( \underline m \underline s)} - \sum\limits_{(\underline r)} {\Delta' _{(\underline s \underline r \underline m)}}\left(ig{\cal O}_{\bar{\mu}} ^{(\underline r)}+ig_s\mathbf{ O}_{\bar{\mu}} ^{(\underline r)} \right)\, .
\end{align}
\end{subequations}
In these expressions, the quantities of the type ${\cal O}^{(\underline m)}$ were introduced in the Higgs sector, whereas
\begin{subequations}
\begin{align}
\hat{{\cal O}}^{(\underline m)}=&\; \frac{g'}{g}\frac{Y}{2}B^{(\underline m)} \, ,\\
\mathbf{O}^{(\underline m)}=& \; \frac{\lambda^a}{2}G^{a(\underline m)}\, .
\end{align}
\end{subequations}
Notice that when the above covariant derivatives act on singlets, the operators ${\cal O}^{(\underline m)}$ must be replaced by the operators $\hat{{\cal O}}^{(\underline m)}$. Of course, the operator $\mathbf{O}^{(\underline m)}$ is not present in the case of leptons.

\section{Scalar charged currents associated with the $W$ gauge boson}
\label{AC} In this Appendix we present the charged scalar currents mediated by scalar fields. The currents mediated by the standard pseudo Goldstone bosons $G^{(\underline{0})+}_W$, Eq. (\ref{csc}), are given by
\begin{equation}
 J^{(\underline{0})} _{G_W}=- {g \over {\sqrt 2 {m_{{W^{(0)}}}}} } \Big \{ \bar N ^{(\underline 0)} M_{E^{(\underline 0)}} P_R E ^{(\underline 0)}  +\bar U^{(\underline 0)}  \big[ KM_{D^{(\underline 0)}}{P_R}  +M_{U^{(\underline 0)}} K{P_L}  \big ]   D^{(\underline 0)} \Big \}\, .
\end{equation}
Here, $M_{F^{(\underline 0)}}$ ($F=E,D,U$) is the standard diagonal mass matrix. For instance, $M_{E^{(\underline 0)}}=diag(m_{e^{(\underline 0)}},m_{{\mu}^{(\underline 0)}},m_{{\tau}^{(\underline 0)}})$. In addition,
\begin{eqnarray}
 J_{G_W}^{ \mbox{\tiny{$(\underline m)(\underline m)$}} } &=& - {g \over { {\sqrt 2} m_{W^{(\underline 0)} } } }   \Big \{  \bar N_{V }^{(\underline m)}   \big [ V_{E }^{(\underline{m})}  {P_R}       +   V_N^{(\underline{m}) \dag} {P_L}    \big ]  M_{E^{(\underline 0)}} E_{ V}^{(\underline m)}  \nonumber \\
&\,&+\,  \bar U_{ V}^{(\underline m)}  \Big [ \big ( KM_{D^{(\underline 0)}}{\bar V}_{D }^{(\underline{m})} -  M_{U^{(\underline 0)}}K  {\hat V}_D^{(\underline{m})}\big ) {P_R}  \nonumber \\
&\,&   +\,  \big(   {\hat V}_U^{(\underline{m}) \dag}  K M_{D^{(\underline 0)}} -      {\bar V}_U^{(\underline{m}) \dag} M_{U^{(\underline 0)}}K  \big)  P_L  \Big]D_{ V}^{(\underline m)}  \Big \} \, .
\end{eqnarray}
On the other hand, the currents mediated by the KK pseudo Goldstone bosons $W^{(\underline{m})+}_G$, Eq. (\ref{cnsc}), can be expressed as
\begin{eqnarray}
J_{W_G}^{ \mbox{\tiny{$(\underline 0)(\underline m)$}}} &=& J_{W_G\, \ell }^{ \mbox{\tiny{$(\underline 0)(\underline m)$}}}  + J_{W_G \, q}^{ \mbox{\tiny{$(\underline 0)(\underline m)$}}}\, ,  \\
 J_{W_G}^{ \mbox{\tiny{$(\underline r)(\underline s) (\underline m)$}} } &=&J_{W_G \, \ell }^{ \mbox{\tiny{$(\underline r)(\underline s) (\underline m)$}}} +J_{W_G \, q}^{ \mbox{\tiny{$(\underline r)(\underline s) (\underline m)$}}}\, ,
\end{eqnarray}
where the lepton and quark currents are given by
\begin{eqnarray}
 J_{W_G \, \ell}^{ \mbox{\tiny{$(\underline 0)(\underline m)$}}} &=&  {i g \over{ {\sqrt 2 } m_{W^{(\underline m)}}  } }  \Big \{  \bar N_{V}^{(\underline 0)} \big [    M_{E^{(\underline 0)}} + i m_{(\underline m)} \Pi ^{(\underline m)}  \big ] V_{E}^{(\underline{m})}  {P_R} E_{V}^{(\underline m)} \nonumber \\
&\,& + \,      \bar N_{V}^{(\underline m)} \big [ M_{E^{(\underline 0)}}  {P_R} +  i  m_{(\underline m)} V_N ^{(\underline{m}) \dag}     \Pi ^{(\underline m) \dag }  {P_L}  \big]   E_{V}^{(\underline 0)} \Big\} \, ,
\end{eqnarray}
\begin{eqnarray}
J_{W_G \, q}^{ \mbox{\tiny{$(\underline 0)(\underline m)$}}} &=&  {i g  \over{ {\sqrt 2} m_{W^{(\underline m)}} } }  \Big \{    \bar U_{V}^{(\underline 0)}  \big [ K\big (   M_{D^{(\underline 0)}}+  i m_{(\underline m)}\Pi ^{(\underline m)}  \big) V_{D}^{(\underline{m})}{P_R}   -  M_{U^{(\underline 0)}} K{P_L}  \big ]   D_{V}^{(\underline m)}  \nonumber  \\
&\,&+ \,  \bar U_{ V}^{(\underline m)}  \big [     KM_{D^{(\underline 0)}} {P_R} -  V_U^{(\underline{m}) \dag} \big (  M_{U^{(\underline 0)}} -im_{(\underline m)} \Pi ^{(\underline m) \dag }   \big )K{P_L}   \big ]  D_{ V}^{(\underline 0)} \Big \}\, ,
\end{eqnarray}
\begin{eqnarray}
 J_{W_G \, \ell}^{ \mbox{\tiny{$(\underline r)(\underline s) (\underline m)$}} } &=&   {i g \over{ {\sqrt 2 } m_{W^{(\underline m)}} } }  \bar N_{V}^{(\underline r)}  \Big \{   \big [ {\Delta _{(\underline r  \underline s  \underline m)}}       M_{E^{(\underline 0)}}  + i \Delta {'_{(\underline m  \underline s  \underline r)}}  m_{(\underline m)}    \Pi ^{(\underline m)}     \big ] V_{E}^{(\underline{s})}  {P_R}          \nonumber \\
&\,&+ \,    V_N^{(\underline{r}) \dag}\big [  \Delta {'_{(\underline r  \underline s  \underline m)}}    M_{E^{(\underline 0)}} +i \Delta {'_{(\underline m  \underline r  \underline s)}}    m_{(\underline m)} \Pi ^{(\underline m) \dag }    \big ] {P_L}    \Big \}E_{V}^{(\underline s)}\, ,
\end{eqnarray}
\begin{eqnarray}
J_{W_G \, q}^{ \mbox{\tiny{$(\underline r)(\underline s) (\underline m)$}}}&=&   {i g  \over{ {\sqrt 2} m_{W^{(\underline m)}} } }   \bar U_{ V}^{(\underline r)} \Big \{   \Big [  K  \big ( {\Delta _{(\underline r  \underline s \underline m)}}    M_{D^{(\underline 0)}} {\bar V}_{D }^{(\underline{s})} + i\Delta {'_{(\underline m  \underline s  \underline r)}}   m_{(\underline m)} \Pi ^{(\underline m)}   V_{D }^{(\underline{s})}  \big ) \nonumber \\
&\,&- \, \Delta' _{(\underline r  \underline s  \underline m)}    M_{U^{(\underline 0)}} K {\hat V}_{D }^{(\underline{s})} \Big ]    {P_R}      +  \Big [  \Delta' _{(\underline r \underline s  \underline m)}  {\hat V}_U ^{(\underline{r}) \dag}   K M_{D^{(\underline 0)}} \nonumber \\
&\,&-  \,  \big ({\Delta _{(\underline r \underline s  \underline m)}}       {\bar V}_U ^{(\underline{r}) \dag}  M_{U^{(\underline 0)}} -i\Delta {'_{(\underline m  \underline r  \underline s)}}  m_{(\underline m)} V_U ^{(\underline{r}) \dag} \Pi ^{(\underline m) \dag }  \big ) K  \Big ] {P_L}   \Big \}D_{V}^{(\underline s)}\,   .
\end{eqnarray}

The charged currents in Eq. (\ref{cwbnc}) mediated by physical scalar fields  $W^{(\underline{m})+}_{\bar{n}}$ ($\bar{n}=1,\cdots,n-1$) fields are

\begin{eqnarray}
J_{W_{\bar n}}^{ \mbox{\tiny{$(\underline 0)(\underline m)$}}}&=&  -{g \over {\sqrt 2 } }      \mathcal{R}^{(\underline{m})}_{\bar{\mu}\bar{n}}   \Big [  \bar N_{V}^{(\underline 0)} \Pi^{\bar \mu}  V_{E }^{(\underline{m})} {P_R} E_{V}^{(\underline m)} +\bar N_{V}^{(\underline m)} V_N^{(\underline{m}) \dag}\Pi ^{\bar \mu \dag} {P_L} E_{V}^{(\underline 0)} \nonumber  \\
&\,&+ \,\bar U_{V}^{(\underline 0)} \Pi ^{\bar \mu} KV_{D}^{(\underline{m})} {P_R} D_{V}^{(\underline m)} + \bar U_{V }^{(\underline m)} V_U ^{(\underline{m}) \dag }K \Pi ^{\bar \mu \dag} {P_L} D_{V}^{(\underline 0)} \Big] \, ,
\end{eqnarray}
\begin{eqnarray}
 J_{W_{\bar n } }^{\mbox{\tiny{$(\underline r)(\underline s) (\underline m)$}}} &=&- {g \over {\sqrt 2 }}   \mathcal{R}^{(\underline{m})}_{\bar{\mu}\bar{n}}   \Big \{   \bar N_{ V}^{(\underline r)} \big [  \Delta {'_{(\underline m \underline s \underline r)}}   \Pi ^{\bar \mu} V_{E}^{(\underline{s})} {P_R} +  \Delta {'_{(\underline m  \underline r  \underline s)}}  V _N^{(\underline{r}) \dag} \Pi ^{\bar \mu \dag} {P_L} \big ]  E_{ V}^{(\underline s)}\nonumber   \\
&\,& +\, \bar U_{  V}^{(\underline r)}   \big [ \Delta {'_{(\underline m \underline s  \underline r)}}   K \Pi ^{\bar \mu} V_{D}^{(\underline{s})} {P_R}  +  \Delta {'_{(\underline m  \underline r  \underline s)}}   V_U ^{(\underline{r}) \dag}\Pi ^{\bar \mu \dag} K{P_L} \big ]  D_{V}^{(\underline s)}  \Big \}\, .
\end{eqnarray}
The currents in Eq. (\ref{cwnc}) are conveniently divided in lepton and quark currents as follows:
\begin{eqnarray}
J_{W_n}^{ \mbox{\tiny{$(\underline 0)(\underline m)$}}}&=&  J_{W_n \, \ell}^{ \mbox{\tiny{$(\underline 0)(\underline m)$}}}+J_{W_n \, q}^{ \mbox{\tiny{$(\underline 0)(\underline m)$}}}\, , \\
J_{W_n}^{ \mbox{\tiny{$(\underline r)(\underline s) (\underline m)$}}}&=& J_{W_n \, \ell}^{ \mbox{\tiny{$(\underline r)(\underline s) (\underline m)$}}}+J_{W_n \, q}^{ \mbox{\tiny{$(\underline r)(\underline s) (\underline m)$}}}\, ,
\end{eqnarray}
where
\begin{eqnarray}
 J_{W_n \, \ell}^{ \mbox{\tiny{$(\underline 0)(\underline m)$}}} &=&- {i g m_{W^{(\underline 0) }} \over{ {\sqrt 2 } m_{W^{(\underline m)}}   } }  \Big \{  \bar N_{V}^{(\underline 0)} \big [ \textstyle {m_{(\underline m)} \over m_{W^{(\underline 0)}}^2 }   M_{E^{(\underline 0)}} - i   \Pi ^{(\underline m)}   \big ] V_{E }^{(\underline{m})}  {P_R} E_{V}^{(\underline m)} \nonumber \\
&\,& +  \,    \bar N_{V}^{(\underline m)} \big [\textstyle {m_{(\underline m)} \over m_{W^{(\underline 0)}}^2 } M_{E^{(\underline 0)}}  {P_R} -  i V_N ^{(\underline{m}) \dag}    \Pi ^{(\underline m)\dag }   {P_L}  \big]   E_{V}^{(\underline 0)} \Big\}\, ,
\end{eqnarray}
\begin{eqnarray}
J_{W_n \, q}^{ \mbox{\tiny{$(\underline 0)(\underline m)$}}} &=& - {i g m_{W^{(\underline 0)}} \over{ {\sqrt 2} m_{W^{(\underline m)}}   } }  \Big \{    \bar U_{V}^{(\underline 0)}  \big [ K\big (  \textstyle {m_{(\underline m)} \over m_{W^{(\underline 0)}}^2 }  M_{D^{(\underline 0)}}-  i   \Pi ^{(\underline m)}    \big) V_{D }^{(\underline{m})} {P_R}   - \textstyle {m_{(\underline m)} \over m_{W^{(\underline 0)}}^2 } M_{U^{(\underline 0)}} K{P_L}  \big ]   D_{V}^{(\underline m)}  \nonumber  \\
&\,&+ \, \bar U_{ V}^{(\underline m)}  \Big [    \textstyle {m_{(\underline m)} \over m_{W^{(\underline 0)}}^2 }  KM_{D^{(\underline 0)}} {P_R} -  V_U^{(\underline{m}) \dag} \big ( \textstyle {m_{(\underline m)} \over m_{W^{(\underline 0)}}^2 }   M_{U^{(\underline 0)}} +i  \Pi ^{(\underline m)\dag }   \big )K{P_L}   \Big ]  D_{ V}^{(\underline 0)} \Big \}\, ,
\end{eqnarray}
\begin{eqnarray}
 J_{W_n \, \ell}^{ \mbox{\tiny{$(\underline r)(\underline s) (\underline m)$}}} &=&  - {i g m_{W^{(\underline 0) }} \over{ {\sqrt 2 } m_{W^{(\underline m)}}   } }  \bar N_{V}^{(\underline r)}  \Big \{   \big [ {\Delta _{(\underline r  \underline s  \underline m)}}  \textstyle {m_{(\underline m)} \over m_{W^{(\underline 0)}}^2 }  M_{E^{(\underline 0)}}  - i\Delta {'_{(\underline m  \underline s  \underline r)}}     \Pi ^{(\underline m)}    \big ] V_{E }^{(\underline{s})}  {P_R}          \nonumber \\
&\,&+ \,    V^{N \dag}\big [  \Delta {'_{(\underline r  \underline s  \underline m)}}   \textstyle {m_{(\underline m)} \over m_{W^{(\underline 0)}}^2 }  M_{E^{(\underline 0)}} -i\Delta {'_{(\underline m  \underline r  \underline s)}}    \Pi ^{(\underline m) \dag}  \big ] {P_L}    \Big \}E_{V}^{(\underline s)}\, ,
\end{eqnarray}

\begin{eqnarray}
J_{W_n \, q}^{ \mbox{\tiny{$(\underline r)(\underline s) (\underline m)$}}}&=& - {i  g m_{W^{(\underline 0)}}  \over{ {\sqrt 2} m_{W^{(\underline m)}}  } }   \bar U_{ V}^{(\underline r)} \Big \{   \Big [  K  \big ( {\Delta _{(\underline r  \underline s \underline m)}}    \textstyle {m_{(\underline m)} \over m_{W^{(\underline 0)}}^2 }  M_{D^{(\underline 0)}} {\bar V}_{D }^{(\underline{s})}  - i\Delta {'_{(\underline m  \underline s  \underline r)}}  \Pi ^{(\underline m)}     V_{D }^{(\underline{s})}  \big ) \nonumber \\
&\,&- \, \Delta' _{(\underline r  \underline s  \underline m)}   \textstyle {m_{(\underline m)} \over m_{W^{(\underline 0)}}^2 }  M_{U^{(\underline 0)}} K {\hat V}_{D }^{(\underline{s})}  \Big ]   {P_R}      +  \Big [  \Delta' _{(\underline r  \underline s  \underline m)}  \textstyle {m_{(\underline m)} \over m_{W^{(\underline 0)}}^2 }    {\hat V}_U ^{(\underline{r}) \dag} K M_{D^{(\underline 0)}} \nonumber \\
&\,&-  \,  \big ({\Delta _{(\underline r  \underline s  \underline m)}}   \textstyle {m_{(\underline m)} \over m_{W^{(\underline 0)}}^2 }     {\bar V}_U ^{(\underline{r}) \dag} M_{U^{(\underline 0)}} +i\Delta {'_{(\underline m  \underline r  \underline s)}}  V_U ^{(\underline{r}) \dag}    \Pi ^{(\underline m) \dag}   \big ) K  \Big ] {P_L}   \Big \}D_{V}^{(\underline s)}\, .
\end{eqnarray}

\section{Scalar neutral currents associated with the $Z$ gauge boson}
\label{AD}
In this Appendix we present the neutral scalar currents mediated by the different scalar fields associated with the $Z$ gauge boson. The current mediated by the standard, $G^{(\underline{0})}_Z$, and nonstandard, $Z^{(\underline{m})}_G$, pseudo Goldstone bosons are given by
\begin{eqnarray}
 J^{(\underline{0})}_{G_Z}&=& - { ig \over {2 m_{W^{(\underline 0)}} } } \Big [  \bar E^{(\underline 0)}M_{E^{(\underline 0)}}  {\gamma ^5}E^{(\underline 0)}\nonumber \\
 &&+\bar D^{(\underline 0)} M_{D^{(\underline 0)}} \gamma^{5}D^{(\underline 0)} -  \bar U^{(\underline 0)} M_{U^{(\underline 0)}} \gamma^{5}U^{(\underline 0)}\Big ] \, , \\
J_{G_Z}^{\mbox{\tiny{$(\underline m)(\underline m)$}}}   &=&  { ig \over {2 m_{W^{(\underline 0)}} } }   \sum\limits_{F=E,D,U}  \epsilon_F      \bar F _{ V}^{(\underline m)} M_{F^{(\underline 0)}}   V_{F -}^{(\underline{m})}  \gamma^5 F_{ V}^{(\underline m)}\, ,
\end{eqnarray}
\begin{eqnarray}
J_{Z_G} ^{\mbox{\tiny{$(\underline 0)(\underline m)$}}} &=&- { ig  \over {2 c_W  m_{Z^{(\underline m)}}  } }  \Big \{   \sum\limits_{F=E,D,U}     \bar F_{V}^{(\underline 0)} \Big [  \big ( \epsilon_F      M_{F^{(\underline 0)}}  -i \mbox{g}_{+}^{F} m_{(\underline m)} \Pi ^{(\underline m)}   \big ) V_{F}^{(\underline{m})} {P_R} \nonumber \\
&\,& -\, \big(  \epsilon_F      M_{F^{(\underline 0)}}   -i\mbox{g}_{-}^{F} m_{(\underline m)} \Pi ^{(\underline m)}  \big ){P_L}\Big ] F_{V}^{(\underline m)}         -i  \bar N_{V}^{(\underline 0)}  m_{(\underline m)} \Pi ^{(\underline m)}V_{N }^{(\underline m)}  {P_R}   N _{V}^{(\underline m)}    \Big\}\nonumber \\
&& +H.\, c.\,
\end{eqnarray}
\begin{eqnarray}
J_{Z_G}^{\mbox{\tiny{$(\underline r)(\underline s) (\underline m)$}}} &=& - { ig m_{(\underline m)}  \over {2 c_W  m_{Z^{(\underline m)}}  } }  \Big \{   \sum\limits_{F=E,D,U}  \bar F _{V}^{(\underline r)}\Big [ \mbox{g}_{Z_G }^{F_{ R }^{rsm}} {P_R}+  \mbox{g}_{Z_G }^{F_{ L }^{rsm}} {P_L}    \Big ]  F_{V}^{(\underline s)} \nonumber \\
&\,&-\, i\bar N_{ V}^{(\underline r)} \big [  \Delta {'_{(\underline m  \underline s  \underline r)}}   \Pi ^{(\underline m)}    V_N^{(\underline{s})} {P_R}  +\Delta {'_{(\underline m  \underline r  \underline s)}}   V_N^{(\underline{r}) \dag}  \Pi ^{(\underline m) \dag}   {P_L} \big ] N_{ V}^{(\underline s)}     \Big \}\, ,
\end{eqnarray}
where $V_{F -}^{(\underline m)}={\hat V}_{F}^{(\underline m)} - {\bar V}_{F}^{(\underline m)}$,
\begin{eqnarray}
\mbox{g}_{Z_ G}^{F_{ R }^{rsm}} &=&  \epsilon_F   \big( {\Delta _{(\underline r  \underline s  \underline m)}}     {\bar V}_{F}^{(\underline{s})}  -  \Delta {'_{(\underline r  \underline s  \underline m)}}  {\hat V}_{F }^{(\underline{s})}   \big)  \textstyle{M_{F^{(\underline 0)}}  \over m_{(\underline m )} }. \nonumber \\
  &&+   i\big(  \Delta {'_{(\underline m  \underline r \underline s)}}  \mbox{g}_{-}^{F}   \Pi ^{(\underline m) \dag }   - \Delta {'_{(\underline m  \underline s  \underline r)}}   \mbox{g}_{+}^{F}  \Pi ^{(\underline m)}       \big ){ V}_{F }^{(\underline{s})}  ,  \\
\mbox{g}_{Z_G}^{F_{ L }^{rsm}} &=&  \epsilon_F  \textstyle{ M_{F^{(\underline 0)}} \over m_{(\underline m )}}\big ( \Delta {'_{(\underline r \underline s  \underline m)}} {\hat V}_F^{(\underline{r}) \dag }    -  {\Delta _{(\underline r  \underline s  \underline m)}}  {\bar V}_F^{(\underline{r}) \dag }  \big)\nonumber \\
  &&+ i { V}_F^{(\underline{r}) \dag } \big ( \Delta {'_{(\underline m  \underline s  \underline r)}} \mbox{g}_{-}^{F}       \Pi ^{(\underline m)}    - \Delta {'_{(\underline m \underline r  \underline s)}} \mbox{g}_{+}^{F}    \Pi ^{(\underline m) \dag}  \big) \,  .
\end{eqnarray}
The currents mediated by the physical scalars $Z^{(\underline{m})}_{\bar{n}}$ and $Z^{(\underline{m})}_{n}$ are
\begin{eqnarray}
 J_{Z_{\bar n}}^{ \mbox{\tiny{$(\underline 0)(\underline m)$}} } &=&  - {g \over {2 {c_W}}}    \mathcal{R}^{(\underline{m})}_{\bar{\mu}\bar{n}} \Big \{ \sum\limits_{F=E,D,U}       \bar F_{V}^{(\underline 0)}   \Pi^{\bar \mu} \big [  \mbox{g}_{+}^{F} V_{F}^{(\underline{m})} {P_R} - \mbox{g}_{-}^{F}
 {P_L}  \big ]  F_{ V}^{(\underline m)} \nonumber \\
   &&+      \bar N_{V}^{(\underline 0)}   \Pi^{\bar \mu} V_{N}^{(\underline{m})} {P_R}   N _{V }^{(\underline m)}    \Big \} +H.\, c. \, ,
\end{eqnarray}
\begin{eqnarray}
 J_{Z_{\bar n } }^{\mbox{\tiny{$(\underline r)(\underline s) (\underline m)$}}}&=&- {g \over {2 c_W} }      \mathcal{R}^{(\underline{m})}_{\bar{\mu}\bar{n}} \Big \{    \sum\limits_{F=E,D,U}  \bar F _V^{(\underline r)}       \Big [ \Big ( \Delta {'_{(\underline m \underline s  \underline r)}}   \mbox{g}_{+}^{F}    \Pi^{\bar \mu} - \Delta {'_{(\underline m  \underline r  \underline s)}}  \mbox{g}_{-}^{F}   \Pi^{\bar \mu \dag} \Big ) V_{F }^{(\underline{s})} {P_R}  \nonumber\\
&\,& + \, V_F^{(\underline{r}) \dag} \Big(  \Delta {'_{(\underline m  \underline r  \underline s)}} \mbox{g}_{+}^{F}  \Pi^{\bar \mu \dag}    - \Delta {'_{(\underline m  \underline s  \underline r)}} \mbox{g}_{-}^{F}  \Pi^{\bar \mu}    \Big ){P_L} \Big ]F_V^{(\underline s)}  \nonumber \\
&\,&+ \, \bar N_{ V}^{(\underline r)}  \Big [  \Delta {'_{(\underline m  \underline s  \underline r)}}   \Pi^{\bar \mu}V_{N }^{(\underline{s})} {P_R}  +\Delta {'_{(\underline m  \underline r  \underline s)}}   V_N^{(\underline{r}) \dag}\Pi^{\bar \mu \dag}{P_L} \Big ] N_{ V}^{(\underline s)}     \Big \}\, ,
\end{eqnarray}
\begin{eqnarray}
J_{Z_n} ^{\mbox{\tiny{$(\underline 0)(\underline m)$}}} &=& { ig m_{Z^{(\underline 0)}} \over {2 c_W  m_{Z^{(\underline m)}}   } }  \Big \{   \sum\limits_{F=E,D,U}     \bar F_{V}^{(\underline 0)} \Big [  \Big ( \epsilon_F     \textstyle {m_{(\underline m)} \over m_{Z^{(\underline 0)}}^2 }  M_{F^{(\underline 0)}}  + i \mbox{g}_{+}^{F}   \Pi ^{(\underline m)}  \Big ) V_{F}^{(\underline{m})} {P_R} \nonumber \\
&\,& -\, \Big(  \epsilon_F     \textstyle {m_{(\underline m)} \over m_{Z^{(\underline 0)}}^2 }   M_{F^{(\underline 0)}}   +i\mbox{g}_{-}^{F}  \Pi ^{(\underline m)}  \Big ){P_L}\Big ] F_{V}^{(\underline m)}         +i  \bar N_{V}^{(\underline 0)}   \Pi ^{(\underline m)} V_{N }^{(\underline{m})} {P_R}   N _{V}^{(\underline m)}    \Big\}\nonumber \\
&&  +H.\, c.\, ,
\end{eqnarray}
\begin{eqnarray}
J_{Z_n}^{\mbox{\tiny{$(\underline r)(\underline s) (\underline m)$}}} &=&  { ig m_{Z^{(\underline 0)}} \over {2 c_W  m_{Z^{(\underline m)}}   } }  \Big \{   \sum\limits_{F=E,D,U}  \bar F _{V}^{(\underline r)}\Big [ \mbox{g}_{Z_n }^{F_{ R }^{rsm}} {P_R}+  \mbox{g}_{Z_n }^{F_{ L }^{rsm}} {P_L}    \Big ]  F_{V}^{(\underline s)} \nonumber \\
&\,&+\, i\bar N_{ V}^{(\underline r)} \big [  \Delta {'_{(\underline m  \underline s  \underline r)}}   \Pi ^{(\underline m)}   V_{N}^{(\underline{s})} {P_R}  +\Delta {'_{(\underline m  \underline r  \underline s)}}   V_N^{(\underline{r}) \dag }  \Pi ^{(\underline m)\dag}  {P_L} \big ] N_{V}^{(\underline s)}     \Big \}\, ,
\end{eqnarray}
where
\begin{eqnarray}
\mbox{g}_{Z_n }^{F_{ R }^{rsm}} &=&  \epsilon_F  \textstyle {m_{(\underline m)} \over m_{Z^{(\underline 0)}}^2 }  \big( {\Delta _{(\underline r  \underline s  \underline m)}}  {\bar V}_{F}^{(\underline{s})}  -   \Delta {'_{(\underline r  \underline s  \underline m)}}  {\hat V}_{F }^{(\underline{s})}  \big) M_{F^{(\underline 0)}}\nonumber \\
 &&+  i \big( \Delta {'_{(\underline m  \underline s  \underline r)}}   \mbox{g}_{+}^{F}   \Pi ^{(\underline m)}     - \Delta {'_{(\underline m  \underline r  \underline s)}}  \mbox{g}_{-}^{F}   \Pi ^{(\underline m)\dag}    \big) { V}_{F }^{(\underline{s})}  \, ,  \\
\mbox{g}_{Z_n}^{F_{ L }^{rsm}} &=& \epsilon_F  \textstyle {m_{(\underline m)} \over m_{Z^{(\underline 0)}}^2 }    \big (\Delta {'_{(\underline r  \underline s  \underline m)}} {\hat V}_F^{(\underline{r}) \dag}  -  {\Delta _{(\underline r  \underline s  \underline m)}} {\bar V}_F^{(\underline{r}) \dag}     \big)    M_{F^{(\underline 0)}}\nonumber \\
    &&+  i { V}_F ^{(\underline{r}) \dag} \big( \Delta {'_{(\underline m  \underline r  \underline s)}} \mbox{g}_{+}^{F}   \Pi ^{(\underline m)\dag} -\Delta {'_{(\underline m \underline s  \underline r)}} \mbox{g}_{-}^{F}    \Pi ^{(\underline m)}     \big) \,  .
\end{eqnarray}
In the above expressions,
\begin{equation}
\epsilon_F = \left\{ {\begin{array}{*{20}{l}}
{1,  \hspace{3.5mm} F =E,D, }\\
{ - 1, \, F =U .}
\end{array}} \right.
\end{equation}

\section{Feynman rules for the fermionic sector}
\label{FR}
In this appendix, we present the Feynman rules for the fermionic sector of the EDSM. In order to write our expressions more concisely, it is convenient to introduce some slight changes in the notation introduced in Sec.~\ref{FS}. The unitary matrices of dimension $2^{n\over 2} \times 2^{n\over 2}$ that transform right-handed KK excitations are rewritten as:
\begin{equation}
V_f^{(\underline{m})}={1\over m_{f^{(\underline m)}}}\left(\begin{array}{ccc}
m_{f^{(\underline{0})}}\tilde{\mathbf{1}} \,  & -i\Lambda_{(\underline m)} \\
 -i \Lambda_{(\underline m)}^\dag \, & m_{f^{(\underline{0})}}\tilde{\mathbf{1}}
\end{array}\right) \, ,  \hspace{3mm}  V_{\nu_\ell}^{(\underline{m})}=-\frac{i\Omega^{(\underline m)}}{m_{(\underline{m})}} \, , \hspace{3mm} \Omega^{(\underline m)}=\left(\begin{array}{ccc}
0 & \Lambda_{(\underline m)} \\
\Lambda_{(\underline m)}^\dag & 0
\end{array}\right)\, ,   \label{DVfqe}
\end{equation}
where $f$ denotes a quark or charged lepton, whereas $\nu_{\ell}$ represents one of the neutrinos of the three known species. $f^{(\underline{0})}$ and $\nu_{\ell}^{(\underline{0})}$ represent the SM particles, whereas $f^{(\underline{m})}$ and $\nu_{\ell}^{(\underline{m})}$ are their respective KK excitations. The label $(a)$ denote components of the vectors
\begin{equation}\label{VSSi}
f^{(\underline{m})}= \left( {\begin{array}{*{20}{c}}
f^{(\underline{m})}_{(1)}\\
\vdots \\
f^{(\underline{m})}_{(a)} \\
\vdots \\
f^{(\underline{m})}_{(2^{\frac{n}{2}})}
\end{array}} \right)\, , \hspace{4mm}     \nu_\ell^{(\underline{m})}= \left( {\begin{array}{*{20}{c}}
0\\
\vdots\\
0\\
\nu^{(\underline{m})}_{\ell\, (1)}\\
\vdots \\
\nu^{(\underline{m})}_{\ell\,  (a)} \\
\vdots \\
\nu^{(\underline{m})}_{\ell\, (\frac{2^{\frac{n}{2}}}{2}) }
\end{array}} \right)  \, , \hspace{3mm} \mbox{with}  \, f=\ell, q \,   .
\end{equation}
Notice that the first $2^{\frac{n}{2}}/2$ entries of $\nu_\ell^{(\underline{m})}$ are zero. Due to this, it is convenient to introduce the following matrix:
\begin{equation}\label{bOne}
{\bf \bar 1 }=\left( {\begin{array}{*{20}{c}}
0 \, &0\\
0 \, &{\bf \tilde 1}
\end{array}} \right) \, .
\end{equation}
Remember that ${\bf \tilde 1}$ is the unity matrix of dimension $2^{n\over 2}/2 \times 2^{n\over 2}/2$. \\

For practical purposes, it is convenient to express the matrix $V_f^{(\underline{m})}$, with $f=q, \ell$, as the sum of two submatrices, namely,  $V_f^{(\underline{m})}={\hat V}_f^{(\underline{m})}+{\bar V}_f^{(\underline{m})}$, where
\begin{equation}\label{MRFCi}
{\hat V}_{f}^{(\underline{m})}={1\over m_{f^{(\underline m)}}}\left( {\begin{array}{*{20}{c}}
m_{f^{(\underline 0)}} &{-i \Lambda_{(\underline m)}} \\
0 &0
\end{array}} \right),  \hspace{5mm}  {\bar V}_{f}^{(\underline{m})}={1\over m_{f^{(\underline m)}}}\left( {\begin{array}{*{20}{c}}
{0} &0 \\
{ -i \Lambda_{(\underline m)}^{\dag} } &m_{f^{(\underline 0)}}
\end{array}} \right)\, .
\end{equation}
Because SM fermions have been placed in the $2^{n\over 2}+1$ entry of the vector $\bar f^{(\underline 0)}=(0, \cdots,0, \bar f_{(2^{n\over 2}+1)}^{(\underline 0)},0, \cdots, 0)$, the label $(a)$ in $f_{(a)}^{(\underline 0)}$ only can take the value ${2^{n\over 2}\over 2}+1$.\\

We now proceed to list complete set of Feynman rules for vector and scalar currents.\\

\noindent \textbf{W-mediated charged currents}\\

\noindent \textit{Vector charged currents}

\begin{equation*}
\begin{minipage}[r]{0.3\linewidth}
\includegraphics[width=2.0in]{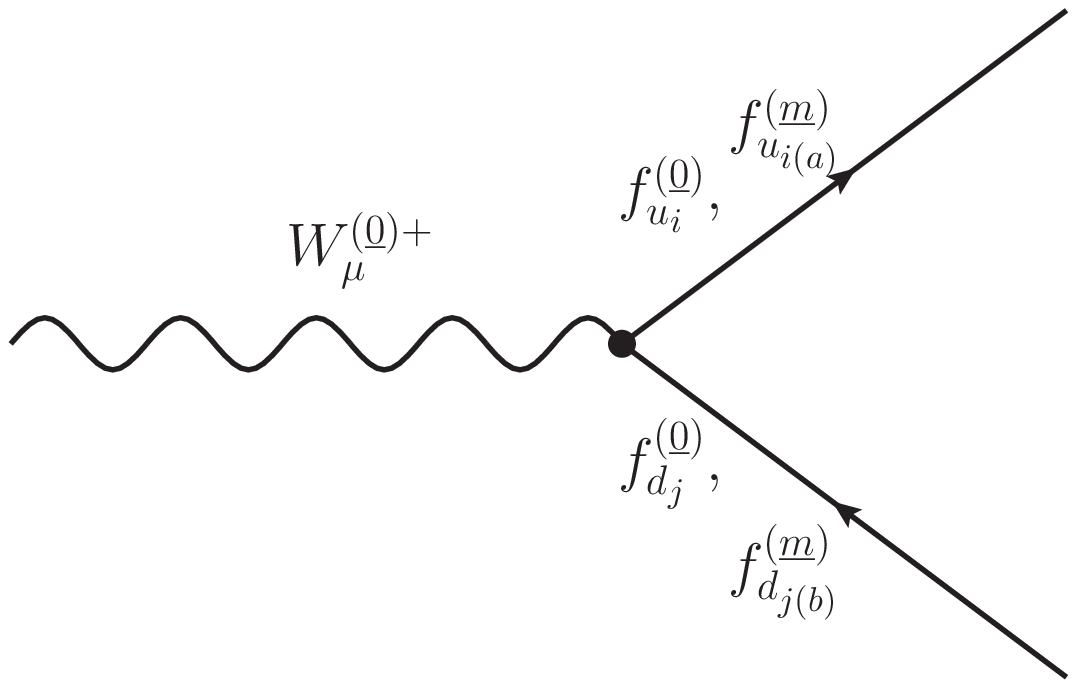}
\end{minipage}
\begin{minipage}[l]{0.55 \textwidth}
\begin{equation*}
:= \frac{ig}{\sqrt{2}}K_{ij}\gamma^\mu P_L \, , \ \  i {\rm W}_{(ab)}^{(\underline {m}) \, \mu} \, ,
\end{equation*}
\end{minipage}
\end{equation*}
where ${\rm W}_{(ab)}^{(\underline {m}) \, \mu}$  comprises quarks or leptons interactions, which are respectively given by:
\begin{align}
{\rm W}_{ij(ab)}^{(\underline {m}) \, \mu } =&\frac{g}{{\sqrt 2 }} K_{ij} {\gamma ^\mu } \Big[    {\hat V}_{u_i}^{(\underline{m}) \dag }  {\hat V}_{d_j }^{(\underline{m})} {P_R} +{1\over 2}  {P_L}     \Big]_{ab} \, ,  \\
{\rm W}_{\ell (ab)}^{ (\underline {m}) \, \mu} = &\frac{g}{{\sqrt 2 }}{\gamma ^\mu } \Big[ {\bf \bar 1}\Big(  V_{\nu_\ell}^{(\underline{m}) \dag} V_{\ell}^{(\underline{m})}{P_R}    + {P_L}    \Big)\Big]_{ab} \, .
\end{align}

\begin{equation*}
\begin{minipage}[r]{0.3\linewidth}
\includegraphics[width=2.0in]{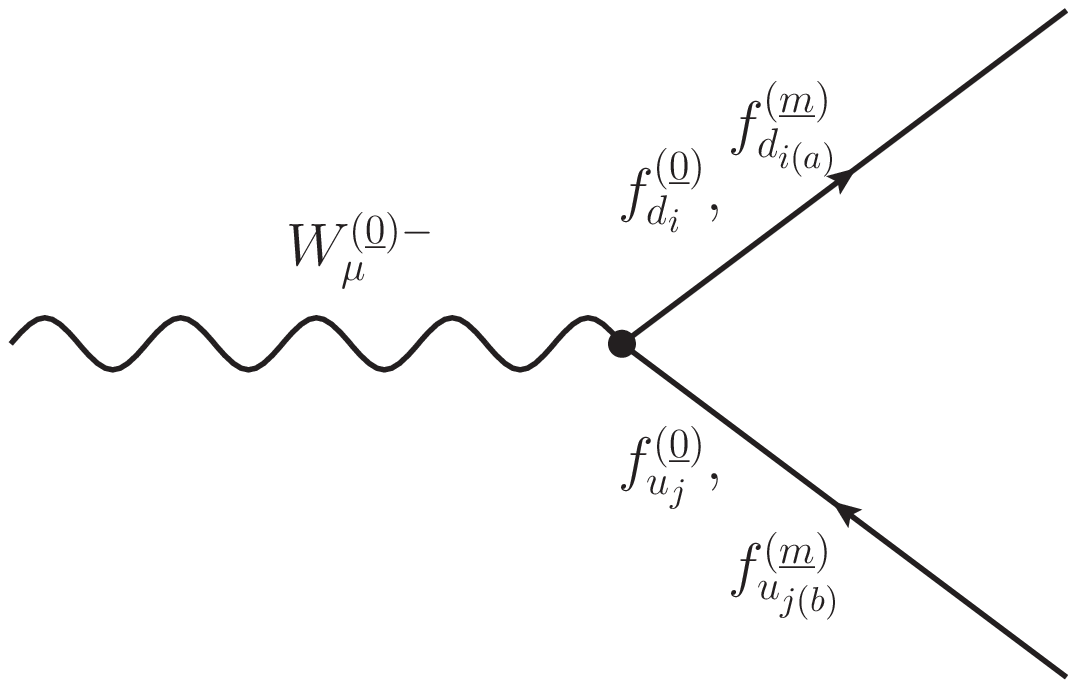}
\end{minipage}
\begin{minipage}[l]{0.55 \textwidth}
\begin{equation*}
:= \frac{ig}{\sqrt{2}}K^\dag_{ij}\gamma^\mu P_L \, , \   i {\rm W}_{(ab)}^{(\underline {m}) \, \mu \, \dag} \, ,
\end{equation*}
\end{minipage}		
\end{equation*}
where
\begin{align}
{\rm W}_{ij(ab)}^{(\underline {m}) \, \mu \dag} =&\frac{g}{{\sqrt 2 }} K^\dag_{ij} {\gamma ^\mu } \Big[    {\hat V}_{d_i}^{(\underline{m}) \dag }  {\hat V}_{u_j }^{(\underline{m})} {P_R} +{1\over 2}  {P_L}     \Big]_{ab} \, ,  \\
{\rm W}_{\ell (ab)}^{ (\underline {m}) \, \mu \dag} = &\frac{g}{{\sqrt 2 }}{\gamma ^\mu } \Big[\Big(  V_{\ell}^{(\underline{m}) \dag} V_{\nu_\ell}^{(\underline{m})}{P_R}    + {P_L}    \Big) {\bf \bar 1}\Big]_{ab} \, .
\end{align}
In the case of leptonic currents, the matrix $K$ must be replaced by the identity matrix.

\begin{equation} \label{RFWmum+u0dm}
\begin{minipage}[r]{0.3\linewidth}
\includegraphics[width=2.0in]{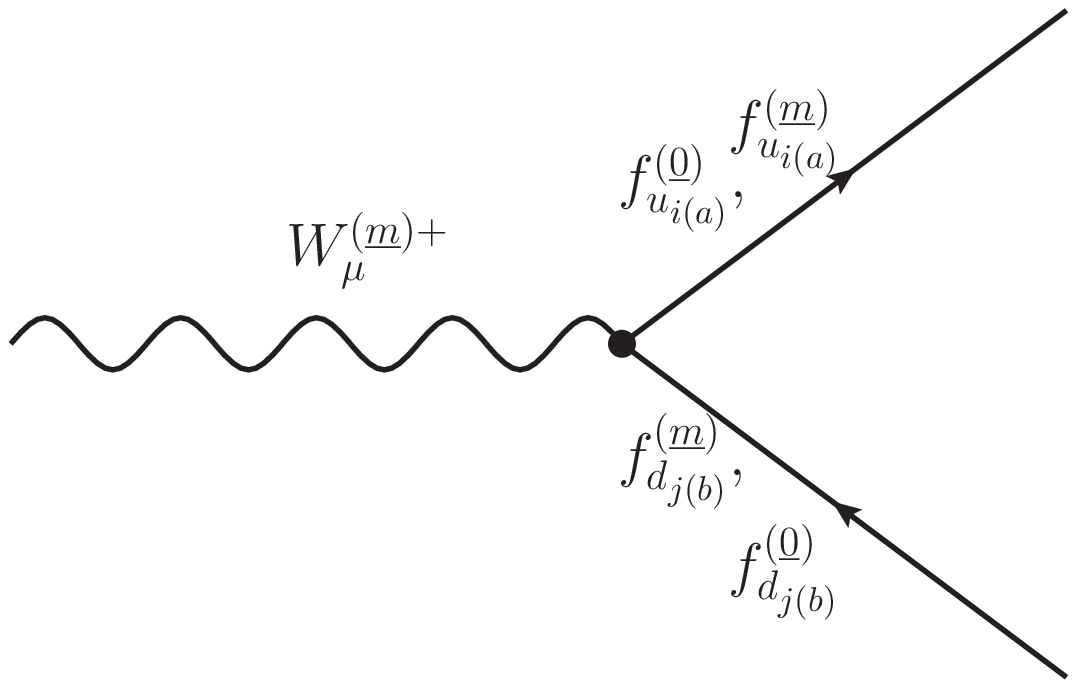}
\end{minipage}
\begin{minipage}[l]{0.45 \textwidth}
\begin{equation*}
:=  {ig \over {\sqrt 2 } }  K_{ij}\gamma^{\mu}P_L \delta_{ab} \, ,
\end{equation*}
\end{minipage}	
\end{equation}

\begin{equation} \label{RFWmum+u0dmdag}
\begin{minipage}[r]{0.3\linewidth}
\includegraphics[width=2.0in]{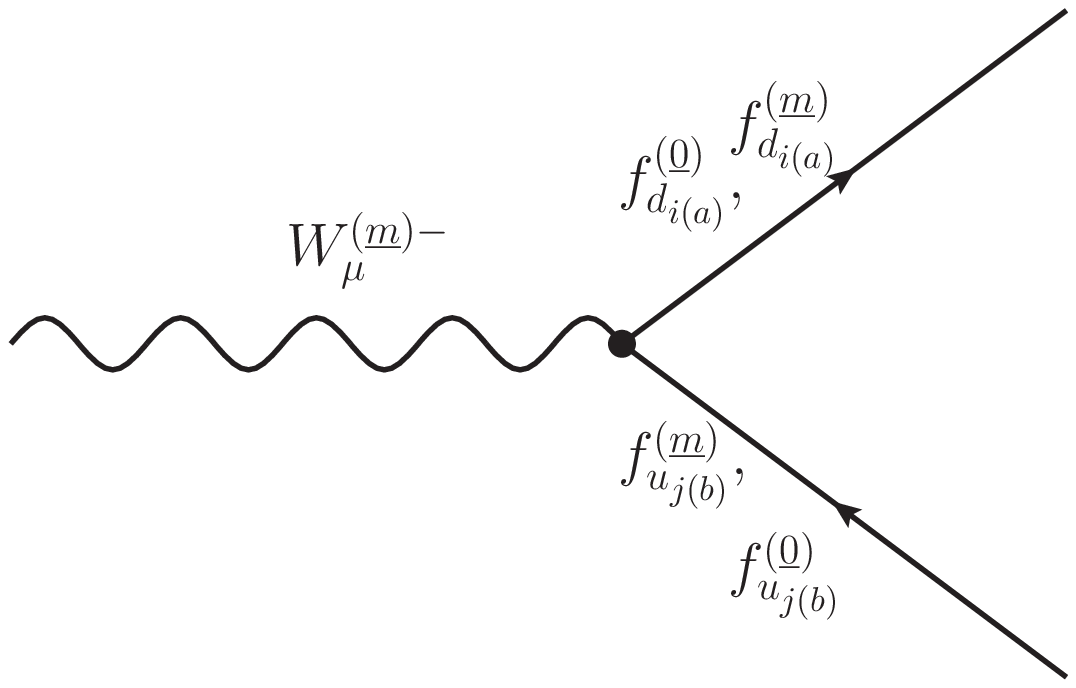}
\end{minipage}
\begin{minipage}[l]{0.45 \textwidth}
\begin{equation*}
:=  {ig \over {\sqrt 2 } }  K^\dag_{ij}\gamma^{\mu}P_L \delta_{ab}  \,  ,
\end{equation*}
\end{minipage}	
\end{equation}
Note that the above Feynman rules coincide with those generated by the SM. \\

\begin{equation*}
\begin{minipage}[r]{0.3\linewidth}
\includegraphics[width=2.0in]{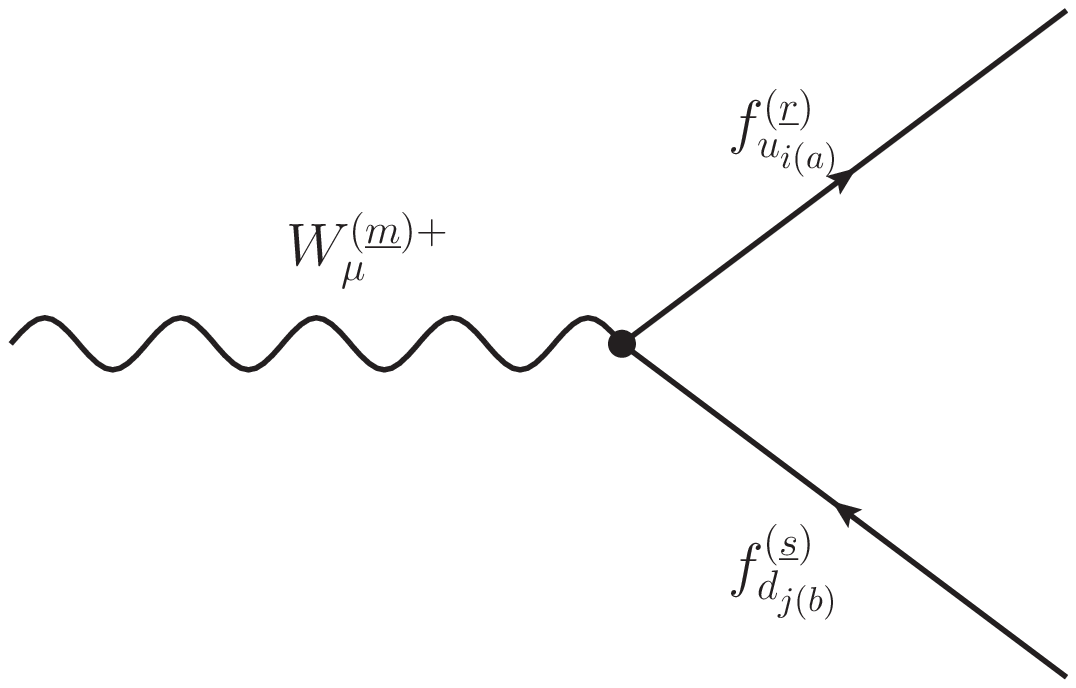}
\end{minipage}
\begin{minipage}[l]{0.45 \textwidth}
\begin{equation*}
:= i {\rm W}_{(ab)}^{(\underline {mrs}) \, \mu} \,  ,
\end{equation*}
\end{minipage}	
\end{equation*}
where
\begin{align}
{\rm W}_{ij(ab)}^{(\underline {mrs}) \, \mu}=& {g \over {\sqrt 2 }}  K_{ij} {\gamma ^\mu } \Big [      \Delta {'_{(\underline{rsm})}}    {\hat V}_{u_i}^{(\underline{r})  \dag}  {\hat V}_{d_j  }^{(\underline{s})}   {P_R}    + {1\over 2}{\Delta _{(\underline{rsm})}}    {P_L}   \Big  ]_{ab}  \, , \\
{\rm W}_{\ell (ab)}^{ (\underline {mrs}) \, \mu} =& {g \over {\sqrt 2 }}   {\gamma ^\mu }\Big[ {\bf \bar 1}\Big ( \Delta {'_{(\underline {rsm})}}       V_{\nu_\ell}^{(\underline {r}) \dag}   V_{\ell }^{(\underline{s})}  {P_R}  +{\Delta _{(\underline{rsm})}}  {P_L} \Big )\Big]_{ab} \, .
\end{align}

\begin{equation*}
\begin{minipage}[r]{0.3\linewidth}
\includegraphics[width=2.0in]{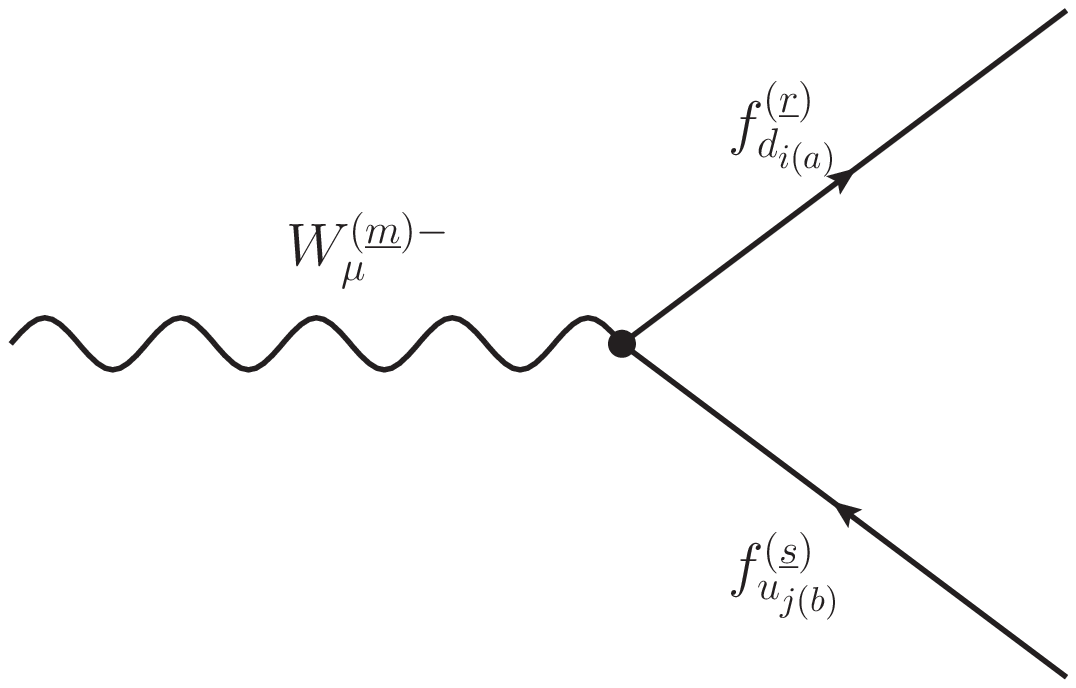}
\end{minipage}
\begin{minipage}[l]{0.45 \textwidth}
\begin{equation*}
:=i {\rm W}_{(ab)}^{(\underline {mrs}) \, \mu \dag} \, ,
\end{equation*}
\end{minipage}	
\end{equation*}
where
\begin{align}
{\rm W}_{ij(ab)}^{(\underline {mrs}) \, \mu \dag}=& {g \over {\sqrt 2 }}  K^\dag_{ij} {\gamma ^\mu } \Big [      \Delta {'_{(\underline{rsm})}}    {\hat V}_{d_i}^{(\underline{r})  \dag}  {\hat V}_{u_j  }^{(\underline{s})}   {P_R}    + {1\over 2}{\Delta _{(\underline{rsm})}}    {P_L}   \Big  ]_{ab}  \, , \\
{\rm W}_{\ell (ab)}^{ (\underline {mrs}) \, \mu \dag} =& {g \over {\sqrt 2 }}   {\gamma ^\mu }\Big[\Big ( \Delta {'_{(\underline {rsm})}}V_{\ell }^{(\underline{r})\dag} V_{\nu_\ell}^{(\underline {s}) }  {P_R}  +{\Delta _{(\underline{rsm})}}  {P_L} \Big ) {\bf \bar 1}\Big]_{ab} \, .
\end{align}

\noindent \textit{Scalar charged currents}

\begin{equation}\label{wbarf0fm}
\begin{minipage}[r]{0.3\linewidth}
\includegraphics[width=2.0in]{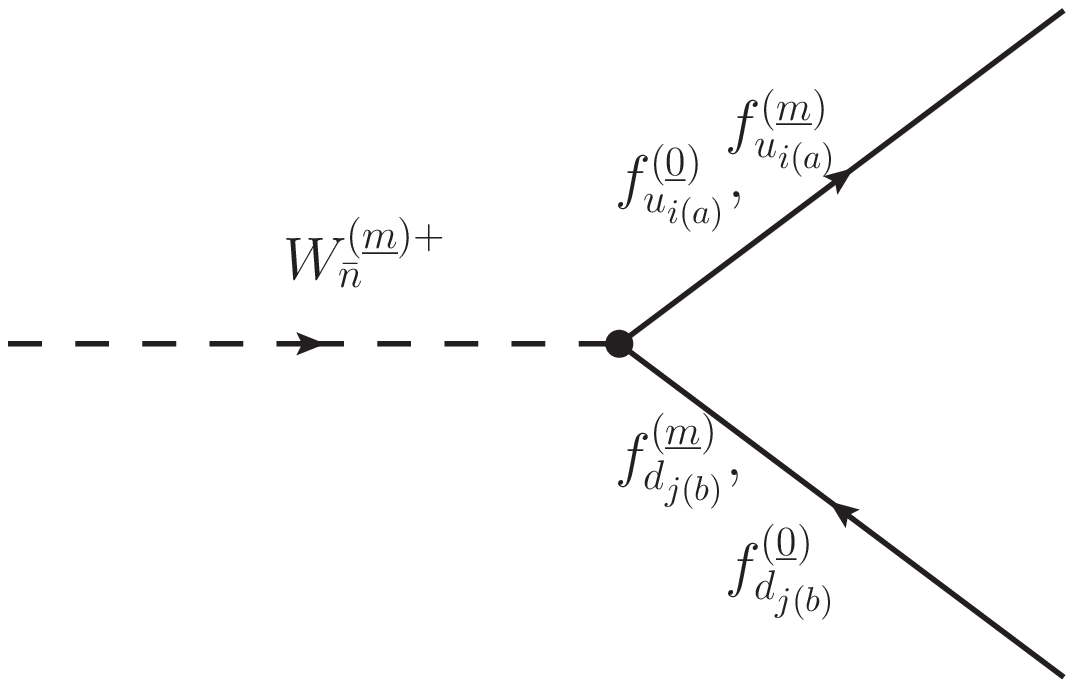}
\end{minipage}
\begin{minipage}[l]{0.55 \textwidth}
\begin{equation*}
 :=  \left\{ {\begin{array}{*{20}{c}}
{ -{ig \over {\sqrt 2 } }      \mathcal{R}^{(\underline{m})}_{\bar{\mu}\bar{n}} K_{ij}    \big[  \Pi ^{\bar \mu} V_{d_j}^{(\underline{m})} \big ]_{ab} {P_R} \, ,  }\\
{}\\
{ -{ig \over {\sqrt 2 } }      \mathcal{R}^{(\underline{m})}_{\bar{\mu}\bar{n}} K_{ij}    \big[   V_{u_i}^{(\underline{m}) \dag} \Pi ^{\bar \mu\dag } \big ]_{ab} {P_L}
 }
\end{array}} \right.
  \end{equation*}
\end{minipage}	
\end{equation}

\begin{equation}
\begin{minipage}[l]{0.3\linewidth}
\includegraphics[width=2.0in]{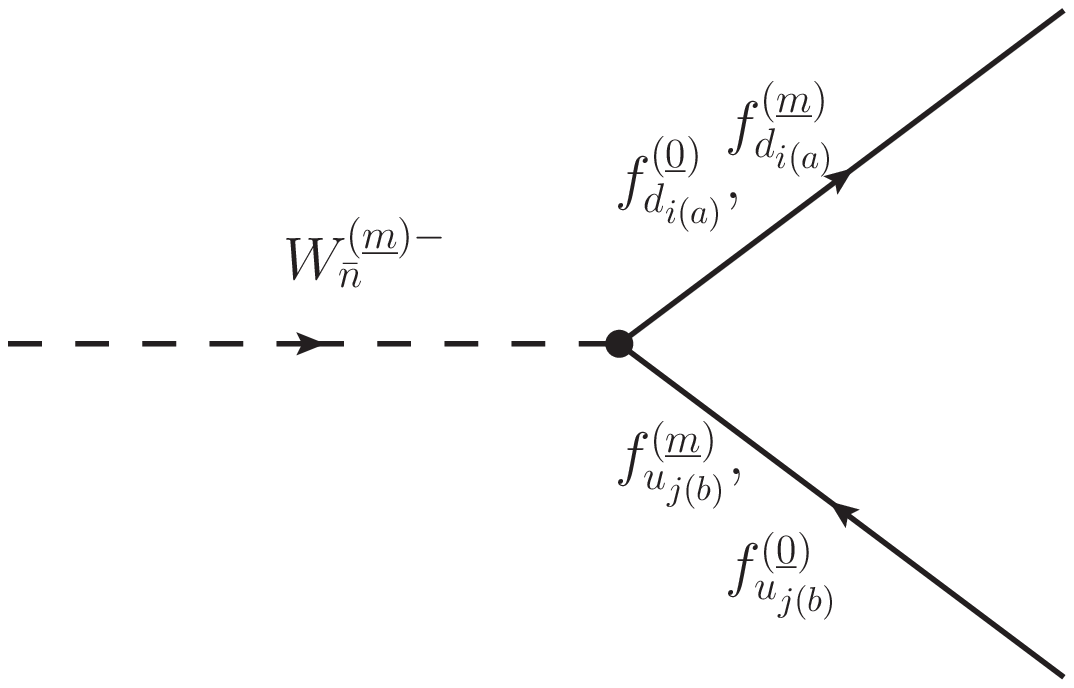}
\end{minipage}
\begin{minipage}[l]{0.55 \textwidth}
 \begin{equation*}
:= \left\{ {\begin{array}{*{20}{c}}
{ -{ig \over {\sqrt 2 } }      \mathcal{R}_{\bar{n}\bar{\mu}} ^{(\underline{m}) \dag} K_{ij}^{\dag}     \big[  \Pi ^{\bar \mu} V_{u_j}^{(\underline{m})} \big ]_{ab} {P_R}   \, , }\\
{}\\
{ -{ig \over {\sqrt 2 } }      \mathcal{R}_{\bar{n}\bar{\mu}} ^{(\underline{m}) \dag} K_{ij}^{\dag}    \big[   V_{d_i}^{(\underline{m}) \dag} \Pi ^{\bar \mu\dag } \big ]_{ab} {P_L} \, . }
\end{array}} \right.
\end{equation*}
\end{minipage}	
\end{equation}
Note that
\begin{equation*} \label{PrdtPiVf}
\Pi ^{\bar \mu} V_{f}^{(\underline{m})} = {1\over m_{f^{(\underline m)}}}  \left( {\begin{array}{*{20}{c}}
{ 0 }&{0}\\
{m_{f^{(\underline 0)}}   \Theta^{\bar \mu \dag} }&{  -i \Theta^{\bar \mu \dag} \Lambda_{(\underline m)}}
\end{array}} \right) \, , \hspace{3mm} \Pi ^{\bar \mu} V_{\nu_{\ell}}^{(\underline{m})} = {1\over m_{\nu_{\ell}^{(\underline m)}}}  \left( {\begin{array}{*{20}{c}}
{ 0 }&{0}\\
{0 }&{  -i \Theta^{\bar \mu \dag} \Lambda_{(\underline m)}}
\end{array}} \right) \, ,
\end{equation*}
where $f=q, \ell$.

\begin{equation*}
\begin{minipage}[r]{0.3\linewidth}
\includegraphics[width=2.0in]{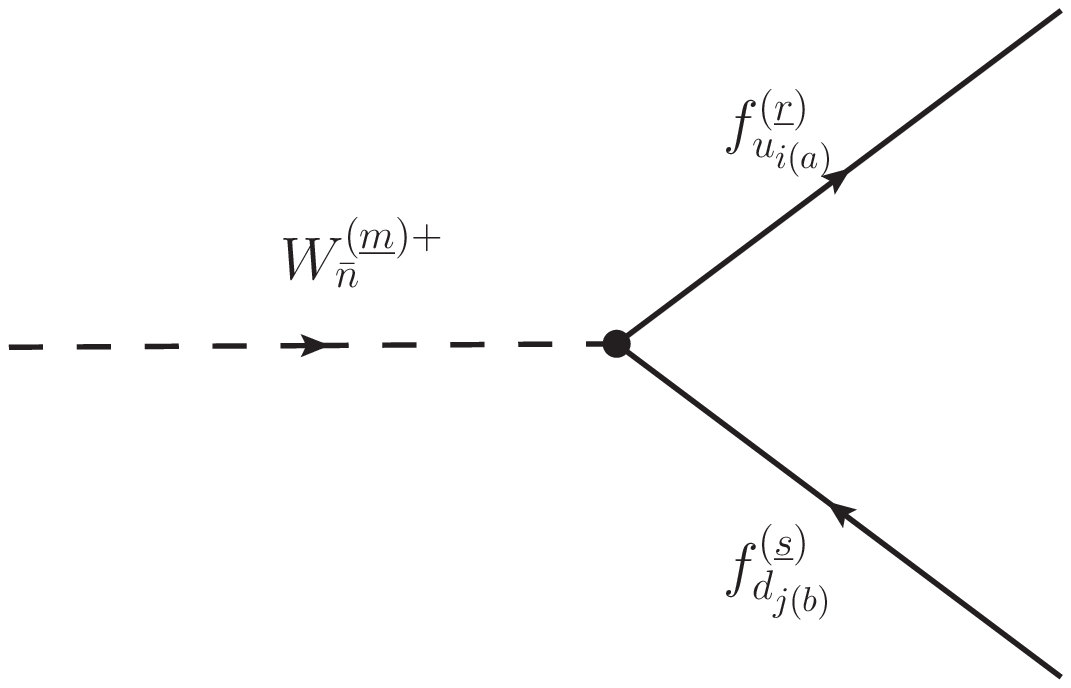}
\end{minipage}
\begin{minipage}[l]{0.50 \textwidth}
\begin{equation*}
:= i {\rm W}_{(ab)}^{\bar n \,(\underline {mrs})} \,  ,
\end{equation*}
\end{minipage}
\end{equation*}
where
\begin{align}
 {\rm W}_{ij(ab)}^{\bar n \,(\underline {mrs})} =&- {g \over {\sqrt 2 }}  K_{ij}     \mathcal{R}^{(\underline{m})}_{\bar{\mu}\bar{n}}   \big [ \Delta {'_{(\underline m \underline s  \underline r)}}  \Pi ^{\bar \mu} V_{d_j}^{(\underline{s})} {P_R}  +  \Delta {'_{(\underline m  \underline r  \underline s)}}   V_{u_i} ^{(\underline{r}) \dag}\Pi ^{\bar \mu \dag} {P_L} \big ]_{ab}  \, ,\\
 {\rm W}_{\ell \, (ab)}^{\bar n \,(\underline {mrs})} =&- {g \over {\sqrt 2 }}      \mathcal{R}^{(\underline{m})}_{\bar{\mu}\bar{n}}    \big [  \Delta {'_{(\underline m \underline s \underline r)}}   \Pi ^{\bar \mu} V_{\ell}^{(\underline{s})} {P_R} +  \Delta {'_{(\underline m  \underline r  \underline s)}}  V _{\nu_\ell}^{(\underline{r}) \dag} \Pi ^{\bar \mu \dag} {P_L} \big ]_{ab} \, .
\end{align}

\begin{equation*}
\begin{minipage}[r]{0.3\linewidth}
\includegraphics[width=2.0in]{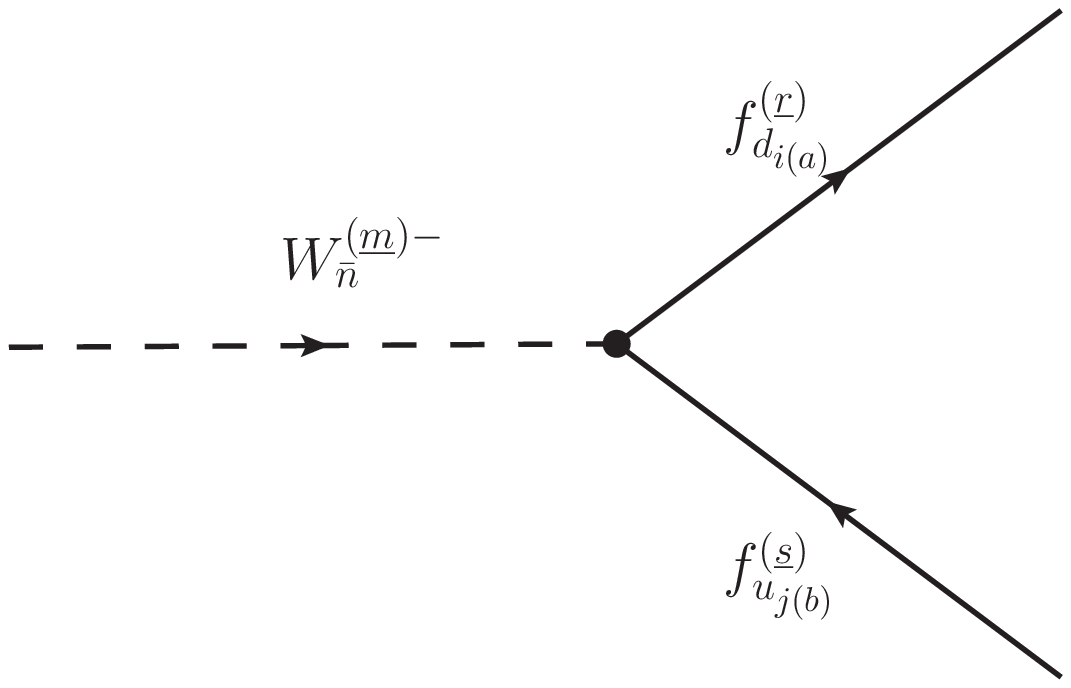}
\end{minipage}
\begin{minipage}[l]{0.50 \textwidth}
\begin{equation*}
:= i {\rm W}_{ (ab)}^{\bar n \,(\underline {mrs})\dag} \,  ,
\end{equation*}
\end{minipage}	
\end{equation*}
where
\begin{align}
 {\rm W}_{ij(ab)}^{\bar n \,(\underline {mrs})\dag} =&- {g \over {\sqrt 2 }}  K^\dag_{ij}     \mathcal{R}^{(\underline{m}) \dag}_{\bar{n}\bar{\mu}}   \big [ \Delta {'_{(\underline m \underline s  \underline r)}}  \Pi ^{\bar \mu} V_{u_j}^{(\underline{s})} {P_R}  +  \Delta {'_{(\underline m  \underline r  \underline s)}}   V_{d_i} ^{(\underline{r}) \dag}\Pi ^{\bar \mu \dag} {P_L} \big ]_{ab}  \, ,\\
 {\rm W}_{\ell \, (ab)}^{\bar n \,(\underline {mrs})\dag} =&- {g \over {\sqrt 2 }}      \mathcal{R}^{(\underline{m}) \dag}_{\bar{n}\bar{\mu}}    \big [  \Delta {'_{(\underline m \underline s \underline r)}}   \Pi ^{\bar \mu} V_{\nu_\ell}^{(\underline{s})} {P_R} +  \Delta {'_{(\underline m  \underline r \underline s)}}  V _{\ell}^{(\underline{r}) \dag} \Pi ^{\bar \mu \dag} {P_L} \big ]_{ab} \, .
\end{align}

\begin{equation*}\label{VWnpqmq0}
\begin{minipage}[r]{0.3\linewidth}
\includegraphics[width=2.0in]{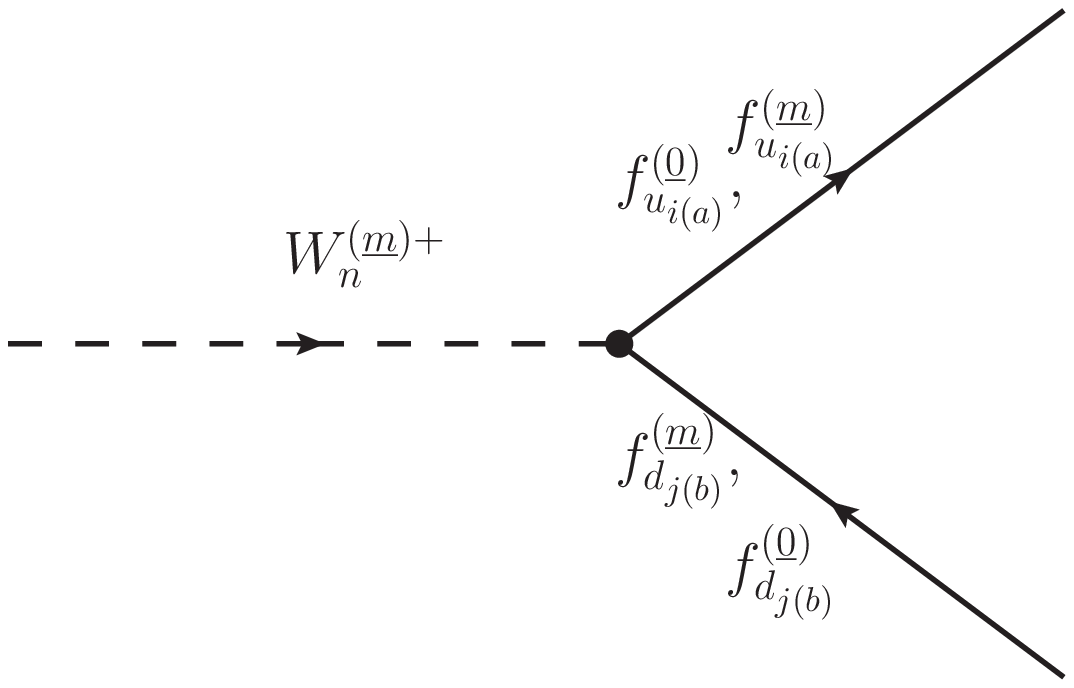}
\end{minipage}
\begin{minipage}[l]{0.55 \textwidth}
\begin{equation*}
:= i {\rm W}_{(ab)}^{(\underline {m})} \,   ,  \ \ i {\rm W}_{in\, (ab)}^{(\underline {m})} \,   ,
\end{equation*}
\end{minipage}
\end{equation*}

\begin{equation*}
\begin{minipage}[r]{0.3\linewidth}
\includegraphics[width=2.0in]{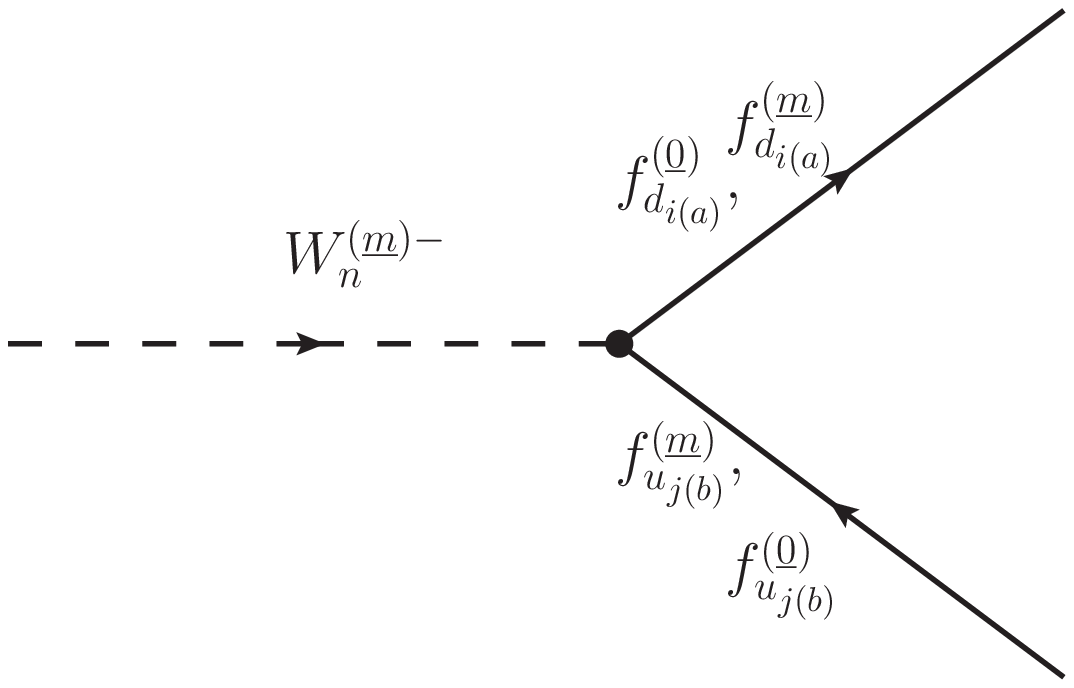}
\end{minipage}	
\begin{minipage}[l]{0.55 \textwidth}
\begin{equation*}
:= i \hat{{\rm {W}}}_{(ab)}^{(\underline {m})} \,   , \ \ i\hat{ {\rm W}}_{in\, (ab)}^{(\underline {m})} \,   ,
\end{equation*}
\end{minipage}	
\end{equation*}

\begin{align}
{\rm W}_{ij (ab)}^{(\underline {m})}=&  - {i g m_{W^{(\underline 0)}} \over{ {\sqrt 2} m_{W^{(\underline m)}}   } } K_{ij}     \Big [  \big (  \textstyle {m_{(\underline m)} \over m_{W^{(\underline 0)}}^2 }  m_{d_j^{(\underline 0)}}-  i   \Pi ^{(\underline m)}    \big) V_{d_j }^{(\underline{m})} {P_R}   - \textstyle {m_{(\underline m)} \over m_{W^{(\underline 0)}}^2 } m_{u_i^{(\underline 0)}} {P_L}  \Big ] _{ab} \, , \\
{\rm W}_{in\, ij (ab)}^{(\underline {m})} =&  - {ig m_{W^{(\underline 0)}} \over{ {\sqrt 2} m_{W^{(\underline m)}}   } }    K_{ij}  \Big [    \textstyle {m_{(\underline m)} \over m_{W^{(\underline 0)}}^2 }  m_{d_j^{(\underline 0)}} {P_R} -  V_{u_i}^{(\underline{m}) \dag} \big ( \textstyle {m_{(\underline m)} \over m_{W^{(\underline 0)}}^2 }   m_{u_i^{(\underline 0)}} +i  \Pi ^{(\underline m)\dag }   \big ){P_L}   \Big ]_{ab} \, ,
\end{align}

\begin{align}
{\hat {\rm W}}_{ij (ab)}^{(\underline {m})} =&   {i g m_{W^{(\underline 0)}} \over{ {\sqrt 2} m_{W^{(\underline m)}}   } }    K_{ij}^{\dag}  \Big [    \textstyle {m_{(\underline m)} \over m_{W^{(\underline 0)}}^2 }  m_{d_i^{(\underline 0)}} {P_L} -   \big ( \textstyle {m_{(\underline m)} \over m_{W^{(\underline 0)}}^2 }   m_{u_j^{(\underline 0)}} -i  \Pi ^{(\underline m) }   \big ) V_{u_j}^{(\underline{m}) } {P_R}   \Big ] _{ab}\, , \\
{\hat {\rm W}}_{in \, ij (ab)}^{(\underline {m})}=&   {i g m_{W^{(\underline 0)}} \over{ {\sqrt 2} m_{W^{(\underline m)}}   } } K_{ij}^{\dag}     \Big [ V_{d_i}^{(\underline{m}) \dag}  \big (  \textstyle {m_{(\underline m)} \over m_{W^{(\underline 0)}}^2 }  m_{d_i^{(\underline 0)}} +  i  \Pi ^{(\underline m)\dag}    \big) {P_L}   - \textstyle {m_{(\underline m)} \over m_{W^{(\underline 0)}}^2 } m_{u_j^{(\underline 0)}} {P_R}  \Big ] _{ab} \, ,
\end{align}

\begin{align}
{\rm W}_{\ell (ab)}^{(\underline {m})}=&  - {ig m_{W^{(\underline 0) }} \over{ {\sqrt 2 } m_{W^{(\underline m)}}   } }  \left[ {\bf {\bar 1}}\Big ( \textstyle {m_{(\underline m)} \over m_{W^{(\underline 0)}}^2 }   m_{\ell^{(\underline 0)}} - i \Pi ^{(\underline m)}   \Big ) V_{\ell}^{(\underline{m})}  \right]_{ab} {P_R}  \, ,\\
{\rm W}_{in\, \ell (ab)}^{(\underline {m})}=&  - {ig m_{W^{(\underline 0) }} \over{ {\sqrt 2 } m_{W^{(\underline m)}}   } }    \left[  {\bf {\bar 1}} \Big (\textstyle {m_{(\underline m)} \over m_{W^{(\underline 0)}}^2 } m_{\ell^{(\underline 0)}}  {P_R} -  i V_{\nu_\ell} ^{(\underline{m}) \dag}    \Pi ^{(\underline m)\dag }   {P_L}  \Big) \right]_{ab}
\end{align}

\begin{align}
{\hat {\rm W}}_{ \ell (ab)}^{(\underline {m})}=&  {ig m_{W^{(\underline 0) }} \over{ {\sqrt 2 } m_{W^{(\underline m)}}   } }     \left[ \Big (\textstyle {m_{(\underline m)} \over m_{W^{(\underline 0)}}^2 } m_{\ell^{(\underline 0)}}  {P_L} +  i  \Pi ^{(\underline m) }  V_{\nu_\ell} ^{(\underline{m}) }     {P_R}  \Big)    {\bf {\bar 1}} \right]_{ab} \, , \\
{\hat {\rm W}}_{in \, \ell (ab)}^{(\underline {m})}=&  {ig m_{W^{(\underline 0) }} \over{ {\sqrt 2 } m_{W^{(\underline m)}}   } }  \left[ V_{\ell}^{(\underline{m})\dag } \Big ( \textstyle {m_{(\underline m)} \over m_{W^{(\underline 0)}}^2 }   m_{\ell^{(\underline 0)}} + i  \Pi ^{(\underline m)\dag }   \Big )    {\bf {\bar 1}} \right]_{ab} {P_L}  \, ,
\end{align}

\begin{equation*}
\begin{minipage}[r]{0.3\linewidth}
\includegraphics[width=2.0in]{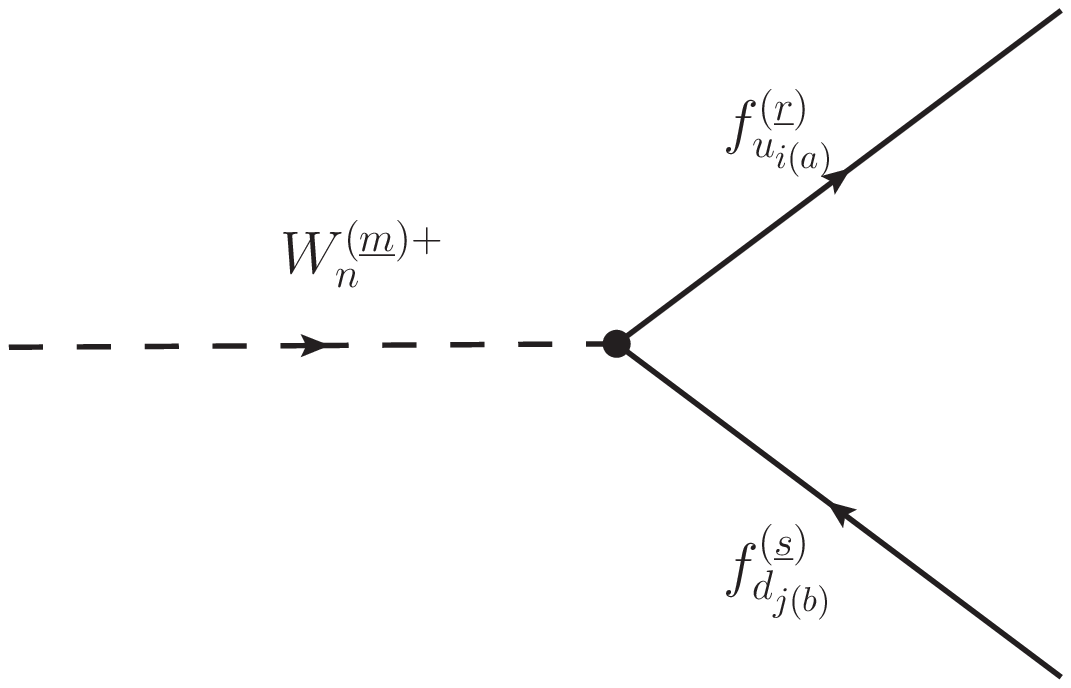}
\end{minipage}
\begin{minipage}[l]{0.55 \textwidth}
\begin{equation*}
:= i {\rm W}_{ij(ab)}^{(\underline {mrs})} \,  ,
\end{equation*}
\end{minipage}
\end{equation*}

\begin{equation*}
\begin{minipage}[r]{0.3\linewidth}
\includegraphics[width=2.0in]{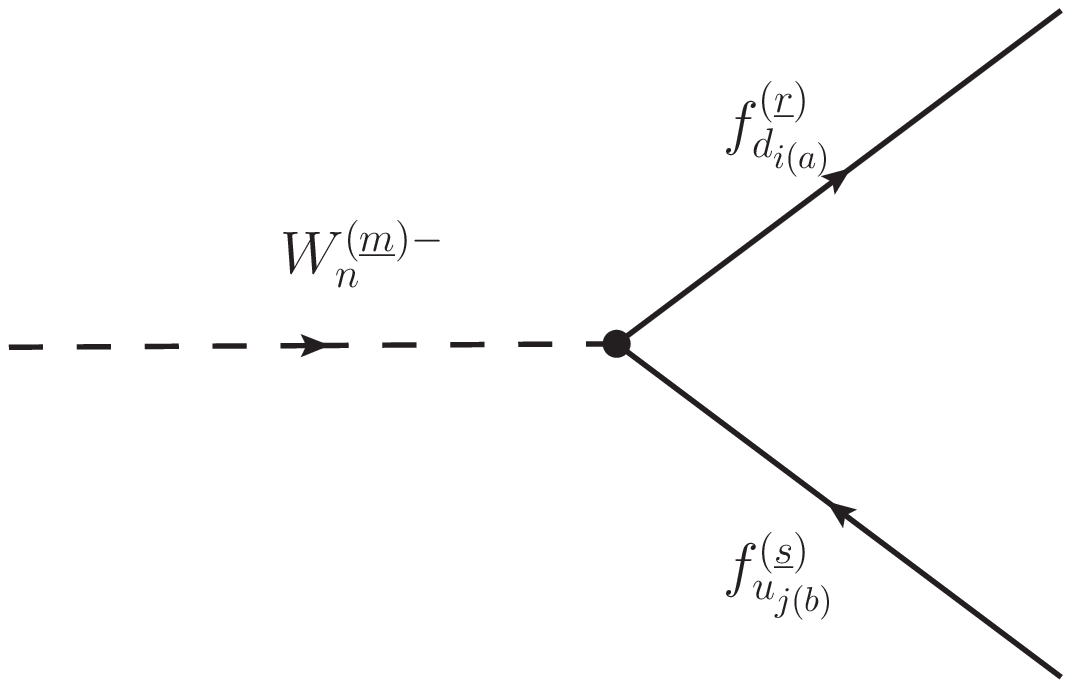}
\end{minipage}
\begin{minipage}[l]{0.55 \textwidth}
\begin{equation*}
:=i {\rm W}_{ij(ab)}^{(\underline {mrs})\dag} \, ,
\end{equation*}
\end{minipage}	
\end{equation*}

\begin{align}
{\rm W}_{ij(ab)}^{(\underline {mrs})}=& - {i g m_{W^{(\underline 0)}} K_{ij}   \over{ {\sqrt 2} m_{W^{(\underline m)}}  } }  \Big [ \nonumber \\
   &\Big (     {\Delta _{(\underline r  \underline s \underline m)}}    \textstyle {m_{(\underline m)} m_{d_j^{(\underline 0)}}  \over m_{W^{(\underline 0)}}^2 }   {\bar V}_{d_j }^{(\underline{s})}  - i\Delta {'_{(\underline m  \underline s  \underline r)}}  \Pi ^{(\underline m)}     V_{d_j }^{(\underline{s})} -  \Delta' _{(\underline r  \underline s  \underline m)}   \textstyle {m_{(\underline m)}m_{u_i^{(\underline 0)}}  \over m_{W^{(\underline 0)}}^2 }   {\hat V}_{d_j }^{(\underline{s})}  \Big )   {P_R}   \nonumber \\
    + & \Big (  \Delta' _{(\underline r  \underline s  \underline m)}  \textstyle {m_{(\underline m)}m_{d_j^{(\underline 0)}}  \over m_{W^{(\underline 0)}}^2 }    {\hat V}_{u_i} ^{(\underline{r}) \dag} -   {\Delta _{(\underline r  \underline s  \underline m)}}   \textstyle {m_{(\underline m)}m_{u_i^{(\underline 0)}} \over m_{W^{(\underline 0)}}^2 }     {\bar V}_{u_i}^{(\underline{r}) \dag}  -i\Delta {'_{(\underline m  \underline r  \underline s)}}  V_{u_i} ^{(\underline{r}) \dag}    \Pi ^{(\underline m) \dag}      \Big ) {P_L}   \Big ]_{ab}\, ,
\end{align}

\begin{align}
{\rm W}_{\ell (ab)}^{(\underline {mrs})}=&  - {i g m_{W^{(\underline 0) }} \over{ {\sqrt 2 } m_{W^{(\underline m)}}   } }  \Big[ {\bf {\bar 1}}  \Big \{   \Big ( {\Delta _{(\underline r  \underline s  \underline m)}}  \textstyle {m_{(\underline m)} m_{\ell^{(\underline 0)}} \over m_{W^{(\underline 0)}}^2 }    - i\Delta {'_{(\underline m  \underline s  \underline r)}}     \Pi ^{(\underline m)}    \Big ) V_{\ell }^{(\underline{s})}  {P_R}          \nonumber \\
&+ \,    V_{\nu_\ell}^{(\underline r) \dag}\Big (  \Delta {'_{(\underline r  \underline s  \underline m)}}   \textstyle {m_{(\underline m)}  m_{\ell^{(\underline 0)}}  \over m_{W^{(\underline 0)}}^2 }  -i\Delta {'_{(\underline m  \underline r  \underline s)}}    \Pi ^{(\underline m) \dag}  \Big ) {P_L}    \Big \} \Big]_{ab} \, , \label{Wnlnu}
\end{align}

\begin{align}
{\rm W}_{ij(ab)}^{(\underline {mrs}) \dag }=&  {i g m_{W^{(\underline 0)}} K_{ij}^{\dag}   \over{ {\sqrt 2} m_{W^{(\underline m)}}  } }  \Big [ \nonumber \\
   &\Big (     {\Delta _{(\underline s  \underline r \underline m)}}    \textstyle {m_{(\underline m)} m_{d_i^{(\underline 0)}}  \over m_{W^{(\underline 0)}}^2 }   {\bar V}_{d_i }^{(\underline{r})\dag}  + i\Delta {'_{(\underline m  \underline r  \underline s)}} V_{d_i }^{(\underline{r}) \dag}  \Pi ^{(\underline m) \dag }      -  \Delta' _{(\underline s  \underline r  \underline m)}   \textstyle {m_{(\underline m)}m_{u_j^{(\underline 0)}}  \over m_{W^{(\underline 0)}}^2 }   {\hat V}_{d_i }^{(\underline{r}) \dag}  \Big )   {P_L}   \nonumber \\
    + & \Big (  \Delta' _{(\underline s  \underline r  \underline m)}  \textstyle {m_{(\underline m)}m_{d_i^{(\underline 0)}}  \over m_{W^{(\underline 0)}}^2 }    {\hat V}_{u_j} ^{(\underline{s}) } -   {\Delta _{(\underline s  \underline r  \underline m)}}   \textstyle {m_{(\underline m)}m_{u_j^{(\underline 0)}} \over m_{W^{(\underline 0)}}^2 }     {\bar V}_{u_j}^{(\underline{s}) }  + i\Delta {'_{(\underline m  \underline s  \underline r)}}    \Pi ^{(\underline m) }  V_{u_j} ^{(\underline{s}) }       \Big ) {P_R}   \Big ]_{ab}\, ,
\end{align}

\begin{align}
{\rm W}_{\ell (ab)}^{(\underline {mrs}) \dag}=&   {i g m_{W^{(\underline 0) }} \over{ {\sqrt 2 } m_{W^{(\underline m)}}   } }  \Big[   \Big \{   V_{\ell }^{(\underline{r}) \dag}   \Big ( {\Delta _{(\underline s  \underline r  \underline m)}}  \textstyle {m_{(\underline m)} m_{\ell^{(\underline 0)}} \over m_{W^{(\underline 0)}}^2 }    + i\Delta {'_{(\underline m  \underline r  \underline s)}}     \Pi ^{(\underline m) \dag}    \Big )  {P_L}          \nonumber \\
&+ \,  \Big (  \Delta {'_{(\underline s  \underline r  \underline m)}}   \textstyle {m_{(\underline m)}  m_{\ell^{(\underline 0)}}  \over m_{W^{(\underline 0)}}^2 }  +i\Delta {'_{(\underline m  \underline s  \underline r)}}    \Pi ^{(\underline m)}  \Big )   V_{\nu_\ell}^{(\underline s) } {P_R}    \Big \}  {\bf {\bar 1}} \Big]_{ab} \, ,
\end{align}

\noindent \textbf{Z-mediated neutral currents}\\

\noindent \textit{Vector neutral currents}

\begin{equation*}
\begin{minipage}[r]{0.35\linewidth}
\includegraphics[width=2.0in]{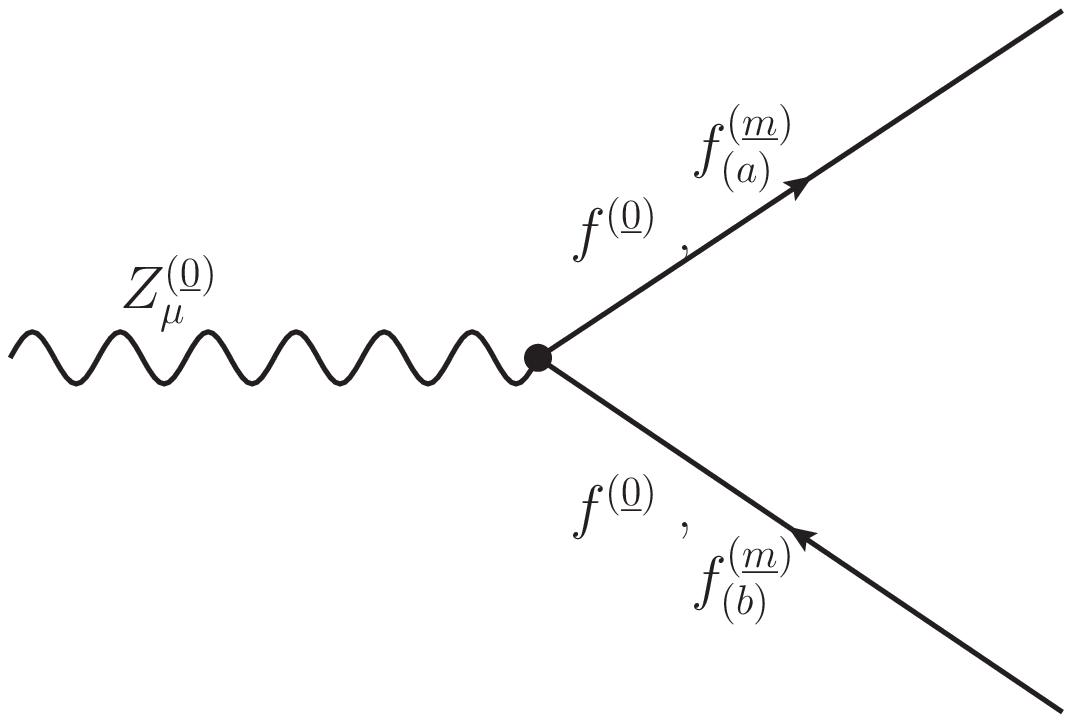}
\end{minipage}
\begin{minipage}[l]{0.6 \textwidth}
\begin{equation*}
:= \left\{ {\begin{array}{*{20}{c}}
{\frac{ig}{{2{c_W}}} {\gamma ^\mu }\big(  \mbox{g}_V^{f} -\mbox{g}_A^{f}{\gamma ^5} \big)  \, ,  \, i  {\rm Z}_{(ab)}^{(\underline{m}) \, \mu}  , \,  \mbox{ $f=q, \ell ,$}} \\{} \\
\frac{i g}{{2{c_W}}} {\gamma ^\mu }P_L \, ,\hspace{1.4 cm} \, i  {\rm Z}_{(ab)}^{\prime \, (\underline{m}) \, \mu}  , \,   \mbox{$f=\nu_\ell$,}
\end{array}} \right.   \,
\end{equation*}
\end{minipage}	
\end{equation*}
where
\begin{equation}
{\rm Z}_{(ab)}^{(\underline{m}) \, \mu}=\frac{g}{{2{c_W}}} \gamma^{\mu}\Big[ \big (  \mbox{g}_{+}^{f}   {\hat V}_f^{(\underline{m}) \dag} {\hat V}_{f}^{(\underline{m})}   + \mbox{g}_{-}^{f}   {\bar V}_f^{(\underline{m}) \dag} {\bar V}_{f}^{(\underline{m})} \big )P_R   + \mbox{g}_{V}^{f}P_L \Big]_{ab} \,
\end{equation}
\begin{equation}
{\rm Z}_{(ab)}^{\prime \, (\underline{m}) \, \mu} =\frac{g}{{2{c_W}}} \gamma^{\mu} {\delta_{ab}} \, .
\end{equation}

Neutral currents mediated by KK $Z$ excitations are given by
\begin{equation*}
\begin{minipage}[r]{0.3\linewidth}
\includegraphics[width=2.0in]{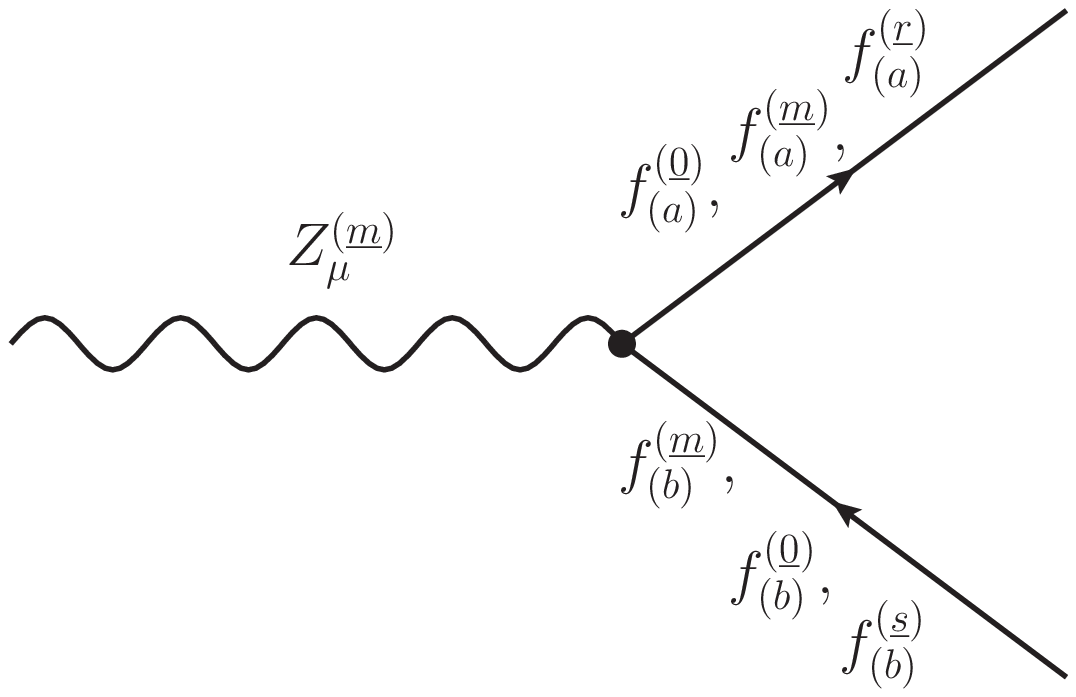}
\end{minipage}
\begin{minipage}[l]{0.65 \textwidth}
\begin{equation*}
:= \left\{ {\begin{array}{*{20}{c}}
{i [ {\rm Z}_{0\, (ab)}^{(\underline{m}) \, \mu}, \   {\rm Z}_{0in\, (ab)}^{(\underline{m}) \, \mu}, \  {\rm Z}_{(ab)}^{(\underline{mrs}) \, \mu}], \, \mbox{$f=q, \ell ,$}} \\ {}\\
{  i[Z'^{(\underline{m})\mu}_{0\, (ab)}, \,  Z'^{(\underline{m})\mu}_{0\, (ab)}, \   {\rm Z}_{(ab)}^{\prime \, (\underline{mrs}) \, \mu}] ,\,  \mbox{$f=\nu_\ell$,}\,  \, }
\end{array}} \right.   \,
\end{equation*}
\end{minipage}	
\end{equation*}
where
\begin{align}
{\rm Z}_{0 (ab)}^{(\underline{m}) \, \mu}=& {g \over {2{c_W}}}    {\gamma ^\mu }  \Big [ \mbox{g}_{-}^{f}   V_{f }^{(\underline{m})} {P_R} + \mbox{g}_{+}^{f}  {P_L} \Big ]_{ab} \, , \\
{\rm Z}_{0in (ab)}^{(\underline{m}) \, \mu}=& {g \over {2{c_W}}}    {\gamma ^\mu }  \Big [ \mbox{g}_{-}^{f}   V_{f }^{(\underline{m})\dag} {P_R} + \mbox{g}_{+}^{f}  {P_L} \Big ]_{ab} \, ,\\
{\rm Z}_{(ab)}^{(\underline{mrs})\, \mu}=&{{g \over {2{c_W}}}}  {\gamma ^\mu }\Big [      \Big (   \Delta {'_{(\underline{rsm})}}  \mbox{g}_{+}^{f}    {\hat V}_f^{(\underline{r}) \dag} {\hat V}_{f }^{(\underline{s})} \nonumber\\
 &+{\Delta _{(\underline{rsm})}}  \mbox{g}_{-}^{f}    {\bar V}_f^{(\underline{r}) \dag}  {\bar V}_{f}^{(\underline{s})}   \Big )   {P_R} +\,{1\over 2} \big(\Delta {'_{(\underline{rsm})}}   \mbox{g}_{-}^{f}  +{\Delta _{(\underline{rsm})}} \mbox{g}_{+}^{f}  \big )  {P_L}     \Big ]_{ab} \, ,
\end{align}

\begin{align}
Z'^{(\underline{m})\mu}_{0\, (ab)}=&\frac{g}{2c_W}\gamma^\mu P_L\, \delta_{ab}\, , \\
{\rm Z}_{(ab)}^{\prime \, (\underline{mrs}) \, \mu}=&{{g \over {2{c_W}}}}  {\gamma ^\mu } \left[ {\bf \bar 1}\Big(  \Delta {'_{(\underline{rsm})}}     V_{\nu_\ell}^{(\underline{r}) \dag}  V_{\nu_\ell }^{(\underline{s})}     {P_R}   +  {\Delta _{(\underline{rsm})}}   {P_L}    \Big ) \right]_{ab}\, .
\end{align}

\noindent \textit{Scalar neutral currents}

\begin{equation*}
\begin{minipage}[r]{0.3\linewidth}
\includegraphics[width=2.0in]{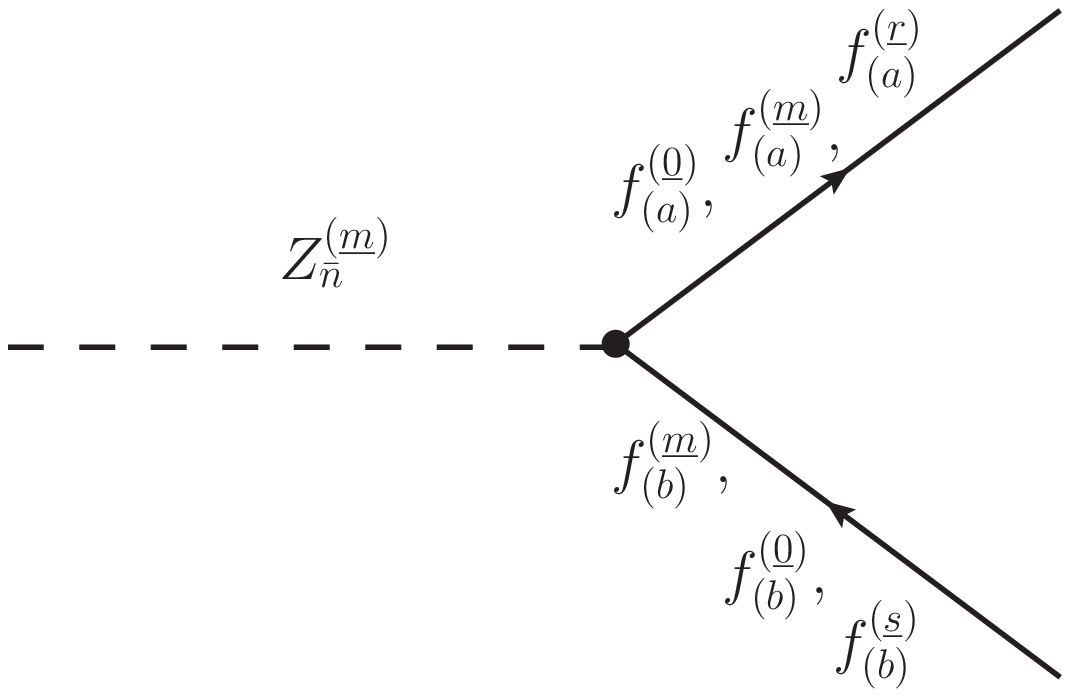}
\end{minipage}
\begin{minipage}[l]{0.65 \textwidth}
\begin{equation*}
:= \left\{ {\begin{array}{*{20}{l}}
{i [ {\rm Z}_{(ab)}^{\bar n (\underline{m})}, \  {\rm Z}_{in\, (ab)}^{\bar n (\underline{m})}, \   {\rm Z}_{ (ab)}^{\bar n (\underline{mrs})} ], \ \mbox{$f=q,\ell,$}} \\ {} \\
 {i [Z'^{\bar n (\underline{m})}_{(ab)}, \ Z'^{\bar n (\underline{m})}_{in\, (ab)}, \  {\rm Z}_{\nu\, (ab)}^{\bar n (\underline{mrs})} ], \  \mbox{ $f=\nu_\ell$,}}
\end{array}} \right.   \,
\end{equation*}
\end{minipage}	
\end{equation*}
where
\begin{align}
{\rm Z}_{(ab)}^{\bar n (\underline{m})}=&  - {g \over {2 {c_W}}}    \mathcal{R}^{(\underline{m})}_{\bar{\mu}\bar{n}}   \left[ \Pi^{\bar \mu} \big (  \mbox{g}_{+}^{f} V_{f}^{(\underline{m})} {P_R} - \mbox{g}_{-}^{f}
 {P_L}  \big ) \right]_{ab}   \, , \\
{\rm Z}_{in(ab)}^{\bar n (\underline{m})}=&- {g \over {2 {c_W}}}    \mathcal{R}_{\bar{n} \bar{\mu}}^{(\underline{m}) \dag}   \left[ \big (  \mbox{g}_{+}^{f} V_{f}^{(\underline{m}) \dag} {P_L} - \mbox{g}_{-}^{f}
 {P_R}  \big )  \Pi^{\bar \mu \dag} \right]_{ab} \, , \\
{\rm Z}_{(ab)}^{\bar n (\underline{mrs})} =& - {g \over {2 c_W} }      \mathcal{R}^{(\underline{m})}_{\bar{\mu}\bar{n}}     \Big [ \Big ( \Delta {'_{(\underline m \underline s  \underline r)}}   \mbox{g}_{+}^{f}    \Pi^{\bar \mu} - \Delta {'_{(\underline m  \underline r  \underline s)}}  \mbox{g}_{-}^{f}   \Pi^{\bar \mu \dag} \Big ) V_{f }^{(\underline{s})} {P_R} \nonumber\\
&+ \, V_f^{(\underline{r}) \dag} \Big(  \Delta {'_{(\underline m  \underline r  \underline s)}} \mbox{g}_{+}^{f}  \Pi^{\bar \mu \dag}    - \Delta {'_{(\underline m  \underline s  \underline r)}} \mbox{g}_{-}^{f}  \Pi^{\bar \mu}    \Big ){P_L} \Big ]_{ab}
\end{align}

\begin{align}
Z'^{\bar n (\underline{m})}_{(ab)}=&- {g \over {2 {c_W}}}    \mathcal{R}^{(\underline{m})}_{\bar{\mu}\bar{n}} \big[ \Pi^{\bar \mu} V_{\nu_\ell}^{(\underline{m})} \big]_{ab} {P_R}\, , \\
Z'^{\bar n (\underline{m})}_{in\, (ab)}=&- {g \over {2 {c_W}}}    \mathcal{R}_{\bar{n} \bar{\mu}}^{(\underline{m}) \dag}  \big[V_{\nu_\ell}^{(\underline{m})\dag}  \Pi^{\bar \mu\dag}\big] _{ab}{P_L}\, , \\
{\rm Z}_{\nu (ab)}^{\bar n (\underline{mrs})} =&  - {g \over {2 c_W} }      \mathcal{R}^{(\underline{m})}_{\bar{\mu}\bar{n}} \Big [  \Delta {'_{(\underline m  \underline s  \underline r)}}   \Pi^{\bar \mu}V_{\nu_\ell }^{(\underline{s})} {P_R}  +\Delta {'_{(\underline m  \underline r  \underline s)}}   V_{\nu_\ell}^{(\underline{r}) \dag}\Pi^{\bar \mu \dag}{P_L} \Big ]_{ab}
\end{align}

\begin{equation*}
\begin{minipage}[r]{0.3\linewidth}
\includegraphics[width=2.0in]{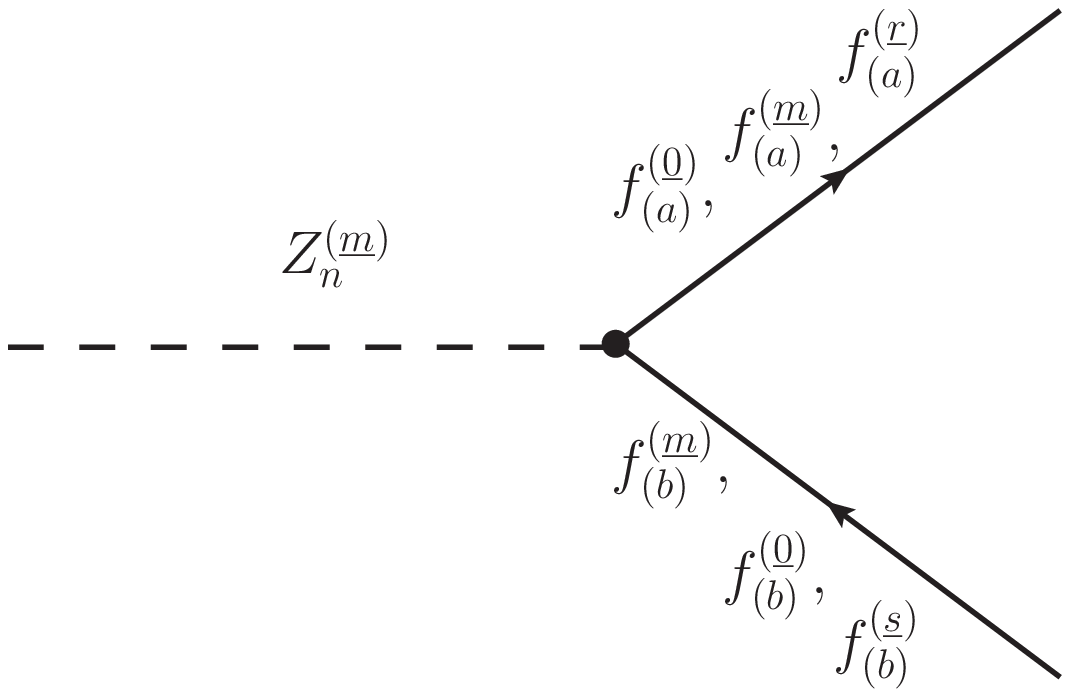}
\end{minipage}
\begin{minipage}[l]{0.65 \textwidth}
\begin{equation*}
:= \left\{ {\begin{array}{*{20}{l}}
{i[  {\rm Z}_{(ab)}^{(\underline{m})}, {\rm Z}_{in(ab)}^{(\underline{m})}, {\rm Z}_{(ab)}^{(\underline{mrs})}],  \ \mbox{$f=q,\ell,$}} \\ {} \\
{i[{\rm Z}_{(ab)}^{\prime \, (\underline{m})}, {\rm Z}^{\prime \, (\underline{m})}_{in(ab)},  {\rm Z}_{(ab)}^{\prime \,(\underline{mrs})}],    \  \mbox{ $f=\nu_\ell$,}}
\end{array}} \right.   \,
\end{equation*}
\end{minipage}	
\end{equation*}
where
\begin{align}
{\rm Z}_{(ab)}^{(\underline{m})} =&  { ig m_{Z^{(\underline 0)}} \over {2 c_W  m_{Z^{(\underline m)}}   } } \Big [  \Big ( \epsilon_f     \textstyle {m_{(\underline m)} \over m_{Z^{(\underline 0)}}^2 }  m_{f^{(\underline 0)}}  + i \mbox{g}_{+}^{f} \Pi ^{(\underline m)}  \Big ) V_{f}^{(\underline{m})} {P_R}\nonumber \\
&- \Big(  \epsilon_f     \textstyle {m_{(\underline m)} \over m_{Z^{(\underline 0)}}^2 }   m_{f^{(\underline 0)}}   +i\mbox{g}_{-}^{f}  \Pi ^{(\underline m)}  \Big ){P_L}\Big ]_{ab}    \, , \\
{\rm Z}_{in(ab)}^{(\underline{m})}=&  - { ig m_{Z^{(\underline 0)}} \over {2 c_W  m_{Z^{(\underline m)}}  } }  \Big [  V_{f}^{(\underline{m})\dag} \Big ( \epsilon_f     \textstyle {m_{(\underline m)} \over m_{Z^{(\underline 0)}}^2 }  m_{f^{(\underline 0)}}  - i \mbox{g}_{+}^{f}  \Pi ^{(\underline m)\dag}  \Big )  {P_L} \nonumber \\
 &- \Big(  \epsilon_f     \textstyle {m_{(\underline m)} \over m_{Z^{(\underline 0)}}^2 }   m_{f^{(\underline 0)}}   -i\mbox{g}_{-}^{f} \Pi ^{(\underline m)\dag}  \Big ){P_R}\Big ]_{ab}    \, , \\
 {\rm Z}_{(ab)}^{(\underline{mrs})} &={ ig m_{Z^{(\underline 0)}} \over {2 c_W  m_{Z^{(\underline m)}}   } } \Big [ \mbox{g}_{Z_n }^{F_{ R }^{rsm}} {P_R}+  \mbox{g}_{Z_n }^{F_{ L }^{rsm}} {P_L}    \Big ]_{ab}  \, , \label{Znfrfsq}
\end{align}

\begin{align}
{\rm Z}_{(ab)}^{\prime \, (\underline{m})} =&{ -g m_{Z^{(\underline 0)}} \over {2 c_W  m_{Z^{(\underline m)}}   } }  \big [ \Pi ^{(\underline m)} V_{\nu_\ell}^{(\underline{m})}  \big ]_{ab}{P_R} ,\\
{\rm Z}^{\prime \, (\underline{m})}_{in(ab)} =&{-g m_{Z^{(\underline 0)}} \over {2 c_W  m_{Z^{(\underline m)}}  }}  \big [V_{\nu_\ell }^{(\underline{m}) \dag}  \Pi ^{(\underline m)\dag}   \big ]_{ab}{P_L}, \\
{\rm Z}_{(ab)}^{\prime \,(\underline{mrs})} =&{ -g m_{Z^{(\underline 0)}} \over {2 c_W  m_{Z^{(\underline m)}}   } }  \left[ {\bf \bar 1} \Big (  \Delta {'_{(\underline m  \underline s  \underline r)}}   \Pi ^{(\underline m)}   V_{\nu_\ell}^{(\underline{s})} {P_R}  +\Delta {'_{(\underline m  \underline r  \underline s)}}   V_{\nu_\ell}^{(\underline{r}) \dag }  \Pi ^{(\underline m)\dag}  {P_L} \Big ) \right] _{ab}   \label{Znfrfsl}\, .
\end{align}
In the above expressions,
\begin{align}
\mbox{g}_{Z_n }^{F_{ R }^{rsm}} &=  \epsilon_f  \textstyle {m_{(\underline m)}m_{f^{(\underline 0)}} \over m_{Z^{(\underline 0)}}^2 }  \big( {\Delta _{(\underline r  \underline s  \underline m)}}  {\bar V}_{f}^{(\underline{s})}  -   \Delta {'_{(\underline r  \underline s  \underline m)}}  {\hat V}_{f }^{(\underline{s})}  \big)\nonumber \\
  &+  i \big( \Delta {'_{(\underline m  \underline s  \underline r)}}   \mbox{g}_{+}^{f}   \Pi ^{(\underline m)}     - \Delta {'_{(\underline m  \underline r  \underline s)}}  \mbox{g}_{-}^{f}   \Pi ^{(\underline m)\dag}    \big) { V}_{f }^{(\underline{s})}  \, ,  \\
\mbox{g}_{Z_n}^{F_{ L }^{rsm}} &= \epsilon_f  \textstyle {m_{(\underline m)}m_{f^{(\underline 0)}} \over m_{Z^{(\underline 0)}}^2 }    \big (\Delta {'_{(\underline r  \underline s  \underline m)}} {\hat V}_f^{(\underline{r}) \dag}  -  {\Delta _{(\underline r  \underline s  \underline m)}} {\bar V}_f^{(\underline{r}) \dag}     \big) \nonumber \\
       &+  i { V}_f ^{(\underline{r}) \dag} \big( \Delta {'_{(\underline m  \underline r  \underline s)}} \mbox{g}_{+}^{f}   \Pi ^{(\underline m)\dag} -\Delta {'_{(\underline m \underline s  \underline r)}} \mbox{g}_{-}^{f}    \Pi ^{(\underline m)}     \big) \,  ,
\end{align}
\begin{equation}
\epsilon_f = \left\{ {\begin{array}{*{20}{l}}
{- 1, \, f =u, }\\
{  1,  \hspace{3.5mm} f =d,e,}
\end{array}} \right.
\end{equation}
note that $\epsilon_f=-2\mbox{g}_A^{f}$.

\noindent \textbf{Electromagnetic currents}\\

\noindent \textit{Vector currents}

\begin{equation*}
\begin{minipage}[r]{0.3\linewidth}
\includegraphics[width=2.0in]{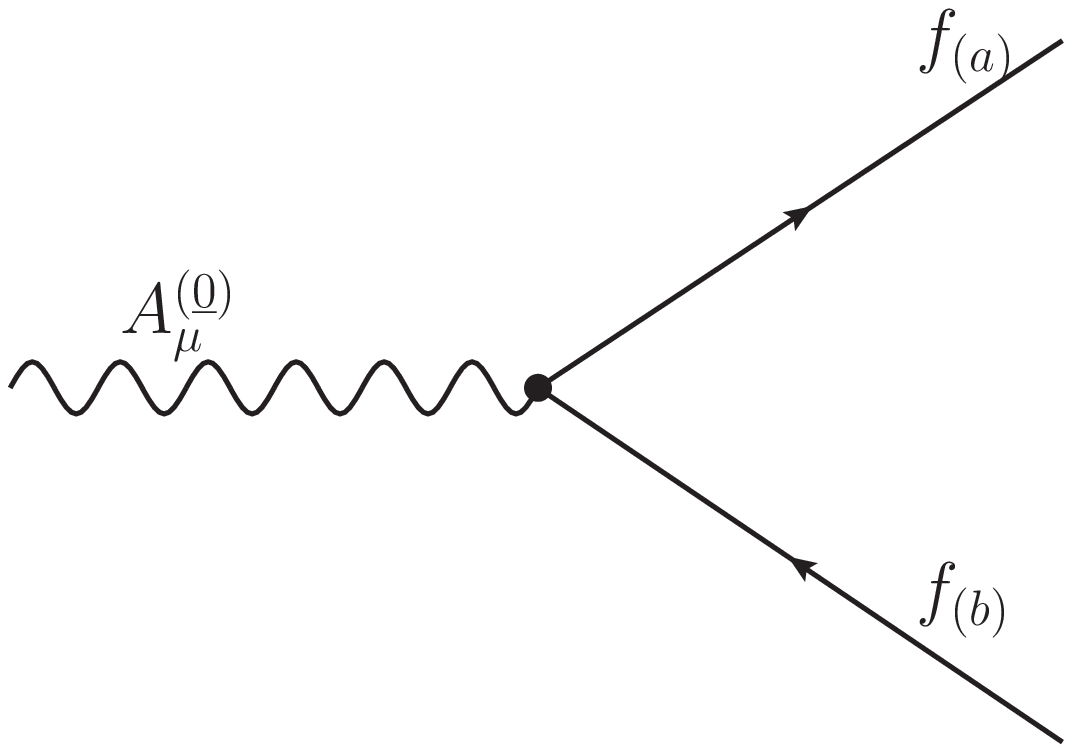}
\end{minipage}
\begin{minipage}[l]{0.60 \textwidth}
\begin{equation*}
:=i eQ_f \gamma^{\mu} \delta_{ab}  \, ,\hspace{3mm}  f=\{f^{(\underline 0)},f^{(\underline m)} \} \, ,
\end{equation*}
\end{minipage}	
\end{equation*}

\begin{equation*}
\begin{minipage}[r]{0.3\linewidth}
\includegraphics[width=2.0in]{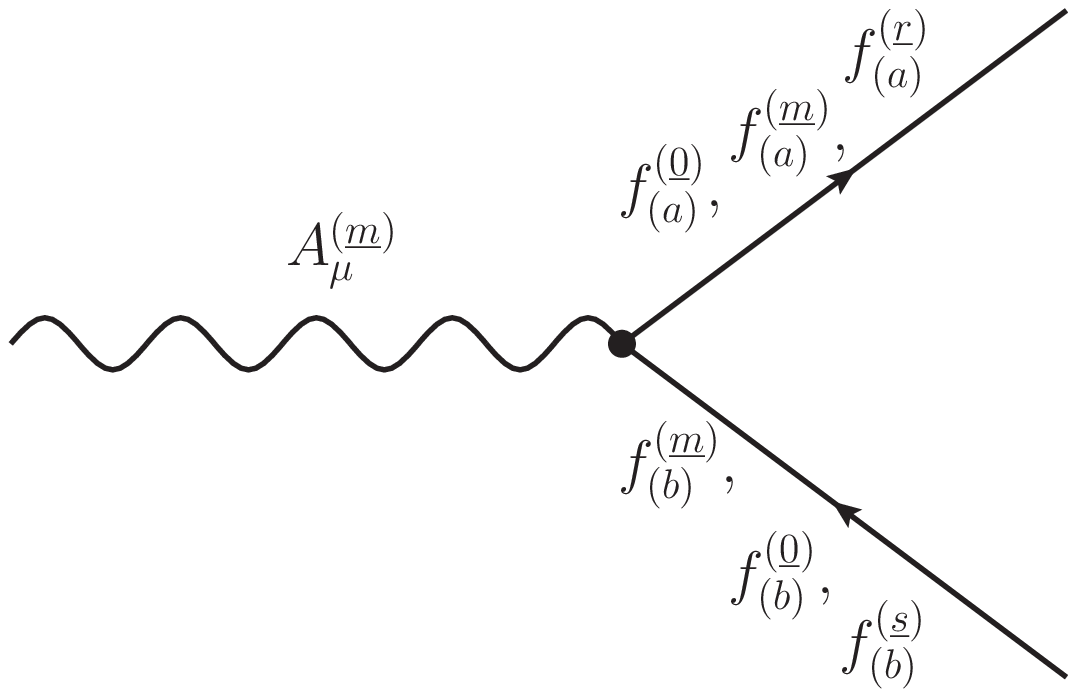}
\end{minipage}
\begin{minipage}[l]{0.55 \textwidth}
\begin{equation*}
:=i [{\Gamma}_{ (ab)}^{ (\underline {m}) \, \mu}, \Gamma_{in \,  (ab)}^{ (\underline {m}) \, \mu}, \Gamma_{ (ab)}^{ (\underline {mrs}) \, \mu}] \, ,
\end{equation*}
\end{minipage}
\end{equation*}
where
\begin{align}
{\Gamma}_{(ab)}^{ (\underline {m}) \, \mu}=&e     {Q_f}  {\gamma ^\mu }  \big [ V_{f }^{(\underline{m})} {P_R} +{P_L} \big ] _{ab}   \, , \\
{\Gamma}_{in(ab)}^{ (\underline {m}) \, \mu}  = &e    {Q_f}  {\gamma ^\mu }  \big [ V_{f}^{(\underline{m})\dag} {P_R} +{P_L} \big ]_{ab}     \, , \\
{\Gamma}_{(ab)}^{ (\underline {mrs}) \, \mu} =&e      {Q_f}  {\gamma ^\mu }   \Big [   \Big (   \Delta {'_{(\underline{rsm})}}  {\hat V}_f^{(\underline{r}) \dag } {\hat V}_{f}^{(\underline{s})}   +  {\Delta _{(\underline {rsm})}} {\bar V}_f^{(\underline{r}) \dag} {\bar V}_{f}^{(\underline{s})}  \Big )     {P_R}\nonumber \\
&+  {1\over 2}\big( \Delta {'_{(\underline{rsm})}}  + {\Delta _{(\underline {rsm})}} \big ) {P_L}   \Big ]_{ab} \, .
\end{align}

\noindent \textit{Scalar currents}

\begin{equation*}
\begin{minipage}[r]{0.3\linewidth}
\includegraphics[width=2.0in]{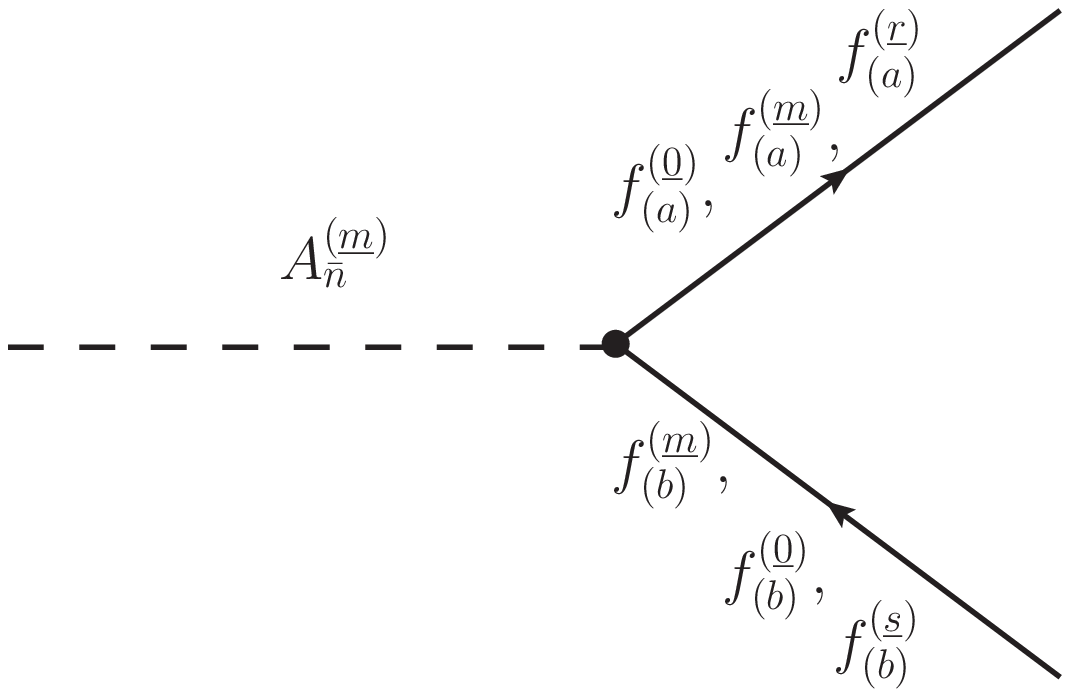}
\end{minipage}
\begin{minipage}[l]{0.55 \textwidth}
\begin{equation*}
:=i [\Gamma_{(ab)}^{\bar n\, (\underline m)},  \Gamma_{in(ab)}^{\bar n\, (\underline m)}, \Gamma_{(a b)}^{(\underline {m  r s})} ] \, ,
\end{equation*}
\end{minipage}	
\end{equation*}
where
\begin{align}
\Gamma_{(ab)}^{\bar n\, (\underline m)} =&  -e {Q_f}  \mathcal{R}^{(\underline{m})}_{\bar{\mu}\bar{n}} \left[ \Pi^{\bar \mu} \left( V_{f}^{(\underline{m})} {P_R} - {P_L} \right) \right]_{ab}  \, , \\
\Gamma_{in(ab)}^{\bar n\, (\underline m)}=&    -e {Q_f}  \mathcal{R}^{(\underline{m}) \dag}_{\bar{n} \bar{\mu}} \left[ \left( V_{f}^{(\underline{m})\dag} {P_R} - {P_L} \right) \Pi^{\bar \mu \dag} \right]_{ab}\, ,\\
\Gamma_{(ab)}^{(\underline {m  r s})}=&-e  {Q_f}  \mathcal{R}^{(\underline{m})}_{\bar{\mu}\bar{n}}   \Big [ \big( \Delta {'_{(\underline {msr})}}     \Pi^{\bar \mu}  -  \Delta {'_{(\underline{mrs})}}     \Pi^{\bar \mu  \dag }   \big ) V_{f}^{(\underline{s})} {P_R} \nonumber \\
 &+ \, V _{f}^{(\underline{r}) \dag} \big ( \Delta {'_{(\underline {mrs})}}  \Pi^{\bar \mu \dag }  - \Delta {'_{(\underline {msr})}}    \Pi^{\bar \mu} \big)  {P_L}    \Big ]_{ab}  \, .
\end{align}

\noindent \textbf{Higgs couplings}

\begin{equation}
\begin{minipage}[r]{0.3\linewidth}
\includegraphics[width=2.0in]{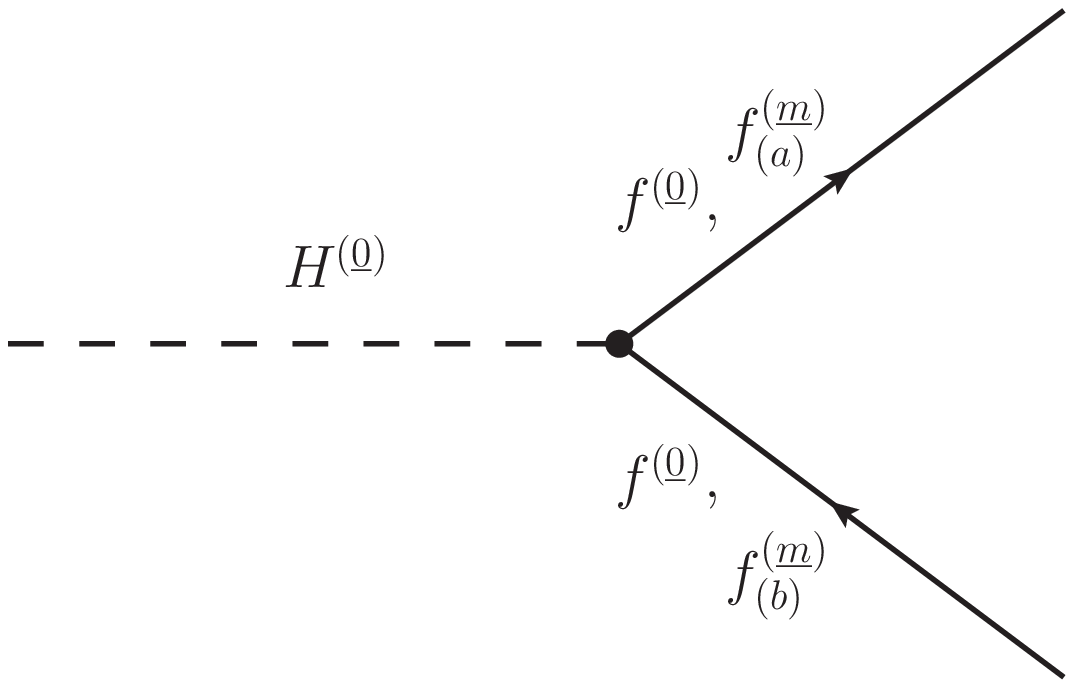}
\end{minipage}
\begin{minipage}[l]{0.55 \textwidth}
\begin{equation*}
:=  -  {igm_{f^{(\underline 0)}} \over {2 m_{W^{(\underline 0)}} } } , \,  \hspace{2mm} i  \Gamma^{(\underline{m})}_{(ab)} ,
\end{equation*}
\end{minipage}	
\end{equation}
where
\begin{equation}
\Gamma_{(ab)}^{(\underline{m})}=-  {gm_{f^{(\underline 0)}} \over {2 m_{W^{(\underline 0)}}} m_{f^{(\underline m)}} }    \Big [ m_{f^{(\underline 0)}}- i\Omega^{(\underline m)}  \gamma^5 \Big]_{ab}
\end{equation}

\begin{equation*}
\begin{minipage}[r]{0.3\linewidth}
\includegraphics[width=2.0in]{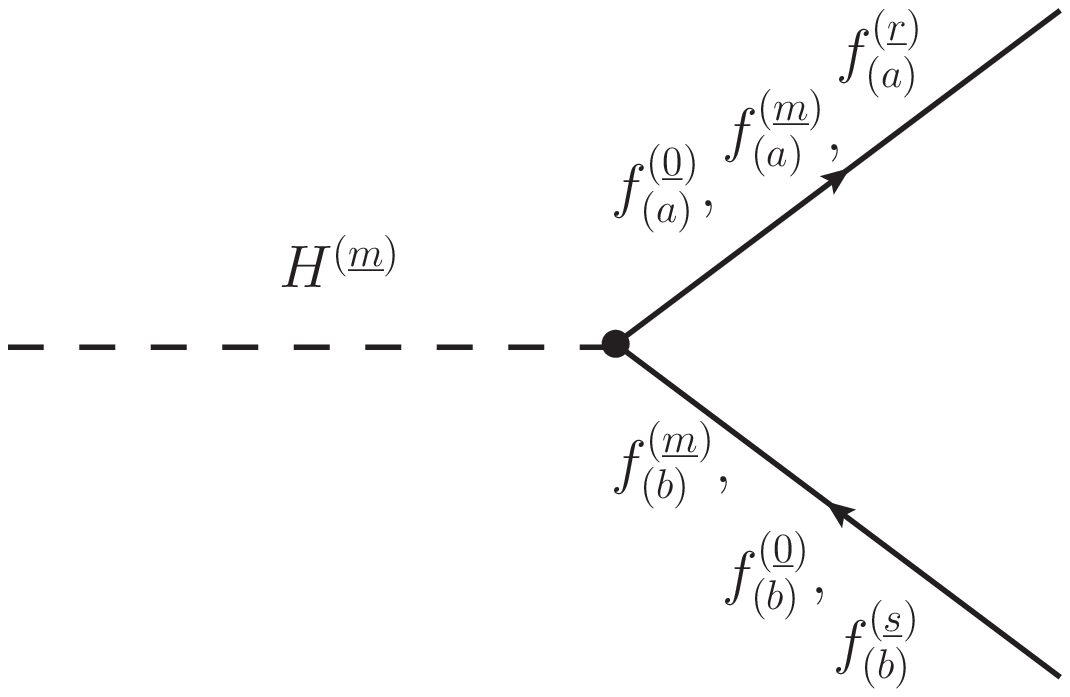}
\end{minipage}
\begin{minipage}[l]{0.55 \textwidth}
\begin{equation*}
:= i [{\cal H}_{(ab)}^{(\underline {m})}, {\cal H}_{in \ (ab)}^{(\underline {m})}, {\cal H}_{(ab)}^{(\underline {mrs})} ]  \, ,
\end{equation*}
\end{minipage}	
\end{equation*}
where
\begin{align}
{\cal H}_{(ab)}^{(\underline {m})}=&- {gm_{f^{(\underline 0)}}  \over {2 m_{W^{(\underline 0)}} } }        \big [ V_{f}^{(\underline{m})} {P_R}+ {P_L}  \big ]_{ab}  \, , \\
{\cal H}_{in(ab)}^{(\underline {m})} =&- {g m_{f^{(\underline 0)}} \over {2 m_{W^{(\underline 0)}} } }       \big [ V_{f}^{(\underline{m})\dag +} {P_L}+ {P_R}  \big ]_{ab} \, , \\
{\cal H}_{(ab)}^{(\underline {mrs})}=&- {gm_{f^{(\underline 0)}} \over {2 m_{W^{(\underline 0)}} } } \Big [    \Big (  {\Delta _{(\underline{rsm})}}      {\bar V}_{f }^{(\underline{s})} +\Delta {'_{(\underline {rsm})}} {\hat V}_{f }^{(\underline{s})}  \Big)  {P_R}\nonumber \\
  &+ \Big (   {\Delta _{(\underline{rsm})}} {\bar V}_f^{(\underline{r}) \dag}  + \Delta {'_{(\underline{rsm})}} {\hat V}_f^{(\underline{r}) \dag}   \Big) {P_L}  \Big ]_{ab} .
\end{align}

\end{document}